\documentclass{SciPost}

\hypersetup{
    colorlinks,
    linkcolor={red!50!black},
    citecolor={blue!50!black},
    urlcolor={blue!80!black}
}
\usepackage{braket}
\usepackage[bitstream-charter]{mathdesign}
\usepackage{subcaption}
\DeclareSymbolFont{usualmathcal}{OMS}{cmsy}{m}{n}
\DeclareSymbolFontAlphabet{\mathcal}{usualmathcal}
\DeclareSymbolFont{cmletters}{OML}{cmm}{m}{it}
\DeclareMathSymbol{\cmg}{\mathalpha}{cmletters}{`g}
\fancypagestyle{SPstyle}{
\fancyhf{}
\lhead{\colorbox{scipostblue}{\bf \color{white} ~SciPost Physics }}
\rhead{{\bf \color{scipostdeepblue} ~Submission }}

\fancyfoot[C]{\textbf{\thepage}}
}

\begin{document}

\pagestyle{SPstyle}

\begin{center}{\Large \textbf{\color{scipostdeepblue}{
Boundary critical phenomena in the quantum Ashkin-Teller model\\
}}}\end{center}

\begin{center}\textbf{
Yifan Liu\textsuperscript{1$\star$},
Natalia Chepiga\textsuperscript{2},
Yoshiki Fukusumi\textsuperscript{3,4}
 and
Masaki Oshikawa\textsuperscript{1,5,6}
}\end{center}

\begin{center}
{\bf 1} Institute for Solid State Physics, University of Tokyo, Kashiwa 277-8581, Japan\\
{\bf 2} {Kavli Institute of Nanoscience, Delft University of Technology, Lorentzweg 1, 2628 CJ Delft, the Netherlands}\\
{\bf 3} Physics Division, National Center for Theoretical Sciences,
 National Taiwan University, Taipei 106319, Taiwan
\\
{\bf 4} Center for Theory and Computation, National Tsing Hua University, Hsinchu 300044, Taiwan
\\
{\bf 5} Kavli Institute for the Physics and Mathematics of the Universe (WPI),
The University of Tokyo, Kashiwa, Chiba 277-8583, Japan
\\
{\bf 6} Trans-scale Quantum Science Institute, University of Tokyo, Bunkyo-ku, Tokyo 113-0033, Japan
\\[\baselineskip]
$\star$ \href{mailto:yifan@issp.u-tokyo.ac.jp}{\small yifan@issp.u-tokyo.ac.jp}
\end{center}

\section*{\color{scipostdeepblue}{Abstract}}
\textbf{\boldmath{%
We investigate the boundary critical phenomena of the one-dimensional quantum Ashkin-Teller model using boundary conformal field theory and density matrix renormalization group (DMRG) simulations. Based on the $\mathbb{Z}_2$-orbifold of the $c=1$ compactified boson boundary conformal field theory, we construct microscopic lattice boundary terms that renormalize to the stable conformal boundary conditions, utilizing simple current extensions and the underlying $\mathrm{SU}(2)$ symmetry to explicitly characterize the four-state Potts point. We validate these theoretical identifications via finite-size spectroscopy of the lattice energy spectra, confirming their consistency with $D_4$ symmetry and Kramers-Wannier duality. Finally, we discuss the boundary renormalization group flows among these identified fixed points to propose a global phase diagram for the boundary criticality.
}}

\vspace{\baselineskip}

\noindent\textcolor{white!90!black}{%
\fbox{\parbox{0.975\linewidth}{%
\textcolor{white!40!black}{\begin{tabular}{lr}%
  \begin{minipage}{0.6\textwidth}%
    {\small Copyright attribution to authors. \newline
    This work is a submission to SciPost Physics. \newline
    License information to appear upon publication. \newline
    Publication information to appear upon publication.}
  \end{minipage} & \begin{minipage}{0.4\textwidth} 
    {\small Received Date \newline Accepted Date \newline Published Date}%
  \end{minipage} 
\end{tabular}}
}}
}


\vspace{10pt}
\noindent\rule{\textwidth}{1pt}
\tableofcontents
\noindent\rule{\textwidth}{1pt}
\vspace{10pt}


\section{Introduction}
\label{sec:intro}

In recent decades, there have been extensive studies on systems with impurities and boundaries~\cite{PhysRevB.46.15233,grimm1993,GRIMM1990633,UweGrimm_2002,saleur1998lecturesnonperturbativefield,affleck2009quantumimpurityproblemscondensed}. One of the most fundamental quests in this area, particularly for systems at criticality, is to determine how different boundary conditions (BCs) modify the properties of the system. This question spans from traditional aspects of boundary critical phenomena—such as the excitation spectra~\cite{affleckBoundaryCriticalPhenomena1998,affleckLogarithmicCorrectionsQuantum1999,Cardy:2004hm,campostriniQuantumIsingChains2015,Chepiga:2021tkj} and surface exponents~\cite{alcarazSurfaceExponentsQuantum1987}—to modern quantum information metrics like entanglement entropy controlling the complexity of numerical simulations on classical computers~\cite{affleckEntanglementEntropyQuantum2009,royEntanglementEntropyIsing2022,royTopologicalInterfacesLuttinger2024,chepigaResilientInfiniteRandomness2024}. The profound influence of boundaries has been further highlighted in emerging fields like non-Hermitian physics, where the properties of the system become rather sensitive when open boundary conditions are applied~\cite{PASQUIER1990523,PhysRevX.13.021007,b55l-7tbc,Edvardsson2019, Koch2020, Edvardsson2022, Yao2018,chouPTSymmetryenrichedNonunitary2025,10.21468/SciPostPhys.7.5.069,PhysRevB.103.085428,10.21468/SciPostPhys.12.6.194,tangBoundaryCriticality2025,ioNonHermitianFreefermionCritical2026}.

Across various fields such as condensed matter physics, high energy physics, and quantum information, boundary conformal field theory (BCFT) has proven to be a powerful tool for deriving exact results for such systems at criticality~\cite{CARDY1984514,affleckBoundaryCriticalPhenomena1998,PETKOVA2001157,Picco_2010,calabreseEntanglementEntropyQuantum2004}. It can be applied directly to systems with boundaries, or to those with defects and interfaces by performing the folding trick~\cite{wongTunnelingQuantumWires1994,oshikawaBoundaryConformalField1996,karchUniversalityEffectiveCentral2023}. In the context of BCFT, the "fundamental quest" mentioned above amounts to classifying the conformally invariant boundary conditions of the low-energy field theory and determining how to realize them through the boundary renormalization group flow induced by boundary perturbations. For minimal models, such identifications can be easily made due to the rationality of the CFT, as a finite number of primary fields are involved, and the Cardy states that respect the bulk extended chiral symmetry can be constructed directly from the modular S-matrix~\cite{Cardy:1989ir,Cardy:2004hm}.

However, similar problems for irrational CFTs, like the $c=1$ CFT, turn out to be more subtle. A complete classification of the boundary conditions, for instance, remains elusive, due to the infinite number of Virasoro primaries in a generic $c=1$ theory (e.g., the compactified boson)~\cite{Ginsparg:1987eb}. Furthermore, even when candidate conformally invariant boundary conditions are known, it is often unclear whether they are stable fixed points that can be realized through concrete lattice counterparts.

In this work, we aim to address these issues in the context of the
$\mathbb{Z}_2$-orbifold of free boson with $c=1$~\cite{Ginsparg:1987eb}.
The Ashkin-Teller model, a well-known generalization of the Ising model, provides a canonical lattice realization for this universality class.
It is renowned for its rich phase diagram featuring a line of critical points described by the $\mathbb{Z}_2$-orbifold CFT with a varying compactification radius~\cite{Ginsparg:1988ui}.
A key feature is that some universal properties, such as the thermal critical exponents, vary continuously along this line, while others, like the scaling of the order parameter and the central charge, keep
the same~\cite{henkelConformalInvarianceCritical1999,delfinoUniversalRatiosLine2004}.
This makes the model an ideal theoretical laboratory for investigating the subtleties of irrational CFTs.

The Ashkin-Teller model includes, as a special case,
two decoupled Ising models.
The system of two decoupled critical Ising models is indeed described
by the $\mathbb{Z}_2$-orbifold of free boson with
the compactification radius $r=1$.
The boundary of the two decoupled critical Ising models can be
identified with a defect line in the single critical Ising model.
This motivated the study of the boundary conformal field theory
for the $\mathbb{Z}_2$-orbifold of free boson in
Refs.~\cite{oshikawaIsingDefectPRL,oshikawaBoundaryConformalField1996}
where continuous families and
several discrete sets of the orbifold versions
of Dirichlet and Neumann boundary conditions were constructed.
While these constructions are valid beyond the special radius
discussed in Refs.~\cite{oshikawaIsingDefectPRL,oshikawaBoundaryConformalField1996} for the Ising defect problem,
the rich boundary phase structures of the critical Ashkin-Teller model
have not been fully explored.
It is our aim in this work to do so, based on the boundary conformal field theory
of the $\mathbb{Z}_2$-orbifold of free boson at general radii.

In particular, at another special point, the Ashkin-Teller model
is equivalent to the four-state Potts model with a higher symmetry.
The four-state Potts model is one of the simplest and most canonical examples of a system possessing an integer spin simple current~\cite{Schellekens:1990ys,Fuchs:1996dd}. This simple current structure is not just a mathematical curiosity; it plays a central role in contemporary condensed matter physics~\cite{cappelliModularInvariantPartition1997,milovanovicEdgeExcitationsPaired1996,schoutensSimplecurrentAlgebraConstructions2016}, appearing in studies of topological order~\cite{Moore:1991ks,Fuchs:2004dz,Lu_2010} and the connectivity of quantum spin chains. For instance, it underlies the physics of $\{ SU(N)\}^{K}$ Wess-Zumino-Witten (WZW) models and their relation to generalizations of the Haldane conjecture~\cite{Furuya:2015coa,Lecheminant:2015iga,Fukusumi_2022_c,kikuchi2022rgflowswzwmodels}.

While the generic $c=1$ theory has infinite primaries, the four-state Potts model's high degree of symmetry (specifically $D_4$) constrains the theory~\cite{KEDEM1993263}. It is known that one can extract a finite set of eleven primary fields that form a "diagonal" modular invariant partition function~\cite{Dijkgraaf:1989hb,Cappelli:2002wq}. A key open question, which bridges the formal CFT understanding with condensed matter applications, is how boundary states constructed from these bulk fields relate to stable, physical boundary conditions on a lattice. Although several conformally invariant boundary conditions were proposed~\cite{jacobsenConformalBoundaryLoop2008,Chepiga:2021tkj}, a careful study of their consistency and the RG flows connecting them is still lacking, especially since the numerical study of CFTs with simple currents is still under development.

In this work, we present a concrete identification of the stable boundary conditions in this model and propose the corresponding boundary renormalization group (RG) flows. To provide strong, non-perturbative evidence for our analytical claims, we supplement our study with large-scale density matrix renormalization group (DMRG) simulations. This powerful numerical method allows us to directly visualize the RG flows on the lattice and confirm the stability of the proposed boundary fixed points. Furthermore, we leverage DMRG to explore the particularly subtle phenomena at the four-state Potts point, where a complete analytical understanding remains challenging and peculiar behaviors emerge. Our numerical results not only demonstrate the validity of our analytical identifications but also provide new insights into this special critical point.

The paper is organized as follows. Section~\ref{sec:boundary_states} begins with a brief overview of the quantum Ashkin-Teller model. We then detail our primary finding: an explicit identification of the lattice boundary terms that renormalize to the stable conformally invariant boundary conditions of the corresponding BCFT. In Secs.~\ref{sec:D4} and \ref{sec:KW}, we test this identification for consistency with the model's internal symmetries, specifically the $D_4$ group symmetry and the non-invertible Kramers-Wannier duality, respectively. In Sec.~\ref{sec:four-state}, we focus on the identification of boundary conditions in a special case of the AT model: the four-state Potts model. The boundary renormalization group flows among these identified boundary conditions are discussed in detail in Sec.~\ref{sec:boundary_RG}. To further verify our analytical results, in Sec.~\ref{sec:FSS}, we perform the finite-size scaling analysis on the energy spectrum of the lattice model. We conclude this paper by discussing the possible future directions based on this work in Sec.~\ref{sec:conclusion}.

\section[Conformally invariant boundary conditions of generic Ashkin-Teller model]{Conformally invariant boundary conditions of generic\\Ashkin-Teller model}
\label{sec:boundary_states}
The 1+1d quantum Ashkin-Teller (AT) model can be defined as two transverse-field Ising models coupled through a bulk interaction. 
To be specific, the Hamiltonian on an open chain of length $L$ with the Free boundary condition at both ends is given as
\begin{align}
    \label{eq:HFree-Free}
    H_\text{AT}^\text{Free-Free}=
    -h \sum_{j=1}^L\left(\sigma_j^x+\tau_j^x+\lambda \sigma_j^x \tau_j^x\right)-J \sum_{j=1}^{L-1}\left(\sigma_j^z \sigma_{j+1}^z+\tau_j^z \tau_{j+1}^z+\lambda \sigma_j^z \tau_j^z \sigma_{j+1}^z \tau_{j+1}^z\right),
\end{align}
where $\sigma^{x,z},\tau^{x,z}$ are Pauli matrices~\cite{henkelConformalInvarianceCritical1999}.
This model possesses a $D_4$ (the dihedral group of order 8)
symmetry, which is the symmetry of a square.
It is generated by two $\mathbb{Z}_2$ actions flipping each Ising spin and another $\mathbb{Z}_2$ action exchanging them. This model is critical at $h=J$ for all $0\leq \lambda\leq1$. By varying the strength of the bulk interaction $\lambda$, it presents a line of criticality that is described by a $c=1$ CFT. As a first step to study the boundary critical phenomena in this model, in this section, we would like to determine the conformally invariant boundary conditions (CBCs) of the underlying BCFT and identify them with explicit realization on the lattice.

This model can also be described in terms of the four-state Potts variable.
The four spin states of the Potts model, labeled as $A,B,C,D$ correspond to
the four states of the pair of Ising spins.
We can identify
\begin{align}
    A = \uparrow \uparrow,\quad B = \uparrow \downarrow,\quad C = \downarrow \downarrow,\quad D = \downarrow \uparrow .
\label{eq:Potts_vs_AT}
\end{align}
At $\lambda=1$, the Hamiltonian~\eqref{eq:HFree-Free} has the enhanced $S_4$ (permutation group)
symmetry corresponding to arbitrary exchanges among the four states $A,B,C,D$;
the model is thus equivalent to the four-state Potts model at this point.
Correspondingly, the known moduli space of CBCs is enlarged due to enhanced symmetry~\cite{recknagelBoundaryDeformationTheory1999a}.
Generically, on the other hand, the Ashkin-Teller model has the $D_4$ symmetry
generated by the spin flips of each Ising spin and by the exchange between the two Ising spins.
Let us describe this symmetry in terms of the Potts variable introduced above.
The spin flips of the first and second Ising spins give, respectively,
\begin{align}
    A \leftrightarrow D, & \quad B \leftrightarrow C,
    \notag \\
    A \leftrightarrow B, & \quad C \leftrightarrow D .
\end{align}
The exchange of the two Ising spins gives
\begin{align}
    B \leftrightarrow D, & \quad A \leftrightarrow A, \quad C \leftrightarrow C.
\end{align}
All these transformations preserve the ``antipodal pairs'' $\{A,C\}$ and $\{B,D\}$.
Therefore, the $D_4$ symmetry can be regarded as the subset that preserves the antipodal pairs,
of the full permutation group
$S_4$ exchanging the four states $A,B,C,D$.
As the four-state Potts model is a special case of the generic AT model,
most of the results derived for the AT model remain valid.
Consequently, we focus on the properties of the generic AT model in this section,
and defer the discussion of the specific features of the Potts point $\lambda=1$ to a later section.

For a 1+1d critical spin chain with system size $L$ and open ends, the spectrum of the system should read
\begin{align}\label{eq:FSS}
E_n^{a b}(L) = & \mathfrak{E}_a+\mathfrak{E}_b+\varepsilon_0 \tilde{L} + \frac{\pi v}{\tilde{L}}\left(x_n-\frac{c}{24}\right) + o(\tilde{L}^{-1}),
\end{align}
where $\mathfrak{E}_{a/b}$ are boundary energy associated with corresponding boundary conditions, $\varepsilon_0$ is the bulk energy density, $v$ is the spin-velocity, $c$ is the central charge, $x_n$ is the scaling dimension of the operators generating the corresponding states, $\tilde{L}=L+\delta L_a+\delta L_b$ is the "effective system size" modified due to the open boundary conditions~\cite{10.21468/SciPostPhys.17.4.099} and $o(\tilde{L}^{-1})$ represents higher-order corrections arising from irrelevant perturbations.
The $O(\tilde{L}^{-1})$ term is universal, and can be read from the so-called open-string partition function of BCFT:
\begin{align}
    Z_{ab}(q):=\mathrm{Tr}q^{L_0-\frac{c}{24}}=\mathrm{Tr}\exp[-\beta H^{ab}(\tilde{L})],
    \label{eq:Zalpha_beta}
\end{align}
where $q \equiv e^{-\frac{\pi v \beta}{\tilde{L}}}$.
In BCFT, it is often convenient to rotate the space-(imaginary) time so that the 
boundaries are regarded as the ``boundary states'' (initial/final states of the time evolution).
In this picture, the same partition function is obtained as 
\begin{align}
 Z_{ab}(\tilde{q}) = \langle\alpha|\exp[-\tilde{L}H^{PBC}(\beta)]|\beta\rangle=\langle\alpha|\tilde{q}^{H_p}|\beta\rangle,
 \label{eq:Zalpha_beta_closed}
\end{align}
where $\tilde{q} \equiv e^{-\frac{4\pi \tilde{L}}{v \beta}}$. 
The two forms, Eqs.~\eqref{eq:Zalpha_beta} and \eqref{eq:Zalpha_beta_closed}, are related by the modular transformation
$S: \tau \to -1/\tau$~\cite{Caldeira:2004jy}.

These allow us to identify the conformally invariant boundary conditions
by comparing the finite-size energy spectrum of the lattice model with the BCFT predictions.

\subsection{Boundary conditions and their BCFT identification}
\label{sec:bs_orbifold}

The low-energy theory of this model, in the scaling limit, corresponds to the 1+1d $\mathbb{Z}_2$ orbifold of the compactified boson CFT with a continuously variable compactification radius $r$, defined by the Lagrangian density:
\begin{align}
    \mathcal{L}=\frac{1}{2\pi}(\partial_\mu\varphi)^2,
\end{align}
where $\varphi$ is the bosonic field with a $\mathrm{U}(1)$ compactification $\varphi\sim\varphi+2\pi r$ and a $\mathbb{Z}_2$ identification $\varphi\sim-\varphi$~\cite{yangModularInvariantPartition1987}. By matching the exact thermal exponent of the Ashkin-Teller model~\cite{kohmotoHamiltonianStudies$d2$1981}, 
\begin{align}
    \nu = \frac{1}{2 - \frac{\pi}{2}[\arccos(-\lambda)]^{-1}},
\end{align}
with the CFT prediction for correlation length critical exponent $\nu = (2 - \frac{1}{r^2})^{-1}$ , 
one can establish the mapping between the compactification radius $r$ and the lattice coupling $\lambda$ in \eqref{eq:HFree-Free}:
\begin{align}
    r=\sqrt{2\left[1-\frac{\cos^{-1}(\lambda)}{\pi}\right]},\quad r\in[1,\sqrt{2}].
    \label{eq:r_vs_lambda}
\end{align}

Throughout this work, we adopt the boundary state formalism, in which conformally invariant boundary conditions are represented by bulk CFT states satisfying the Cardy condition~\cite{Cardy:1989ir,recknagelBoundaryConformalField2013}. For this $\mathbb{Z}_2$ orbifold model, one can construct boundary states corresponding to the Dirichlet and Neumann boundary conditions (BCs) for the bosonic field by symmetrizing the corresponding boundary states of the compactified boson and resolving the degeneracy at the fixed point of the $\mathbb{Z}_2$ orbifolding.

Let us summarize the conformally invariant boundary states of the $\mathbb{Z}_2$ orbifold of
the free boson, originally constructed for the special radius $r=1$ in Refs.~\cite{oshikawaIsingDefectPRL,oshikawaBoundaryConformalField1996}, but are valid for general compactification radii.
They consist of two continuous families of BCs:
\begin{align}
    &\left|D_O(\theta)\right\rangle,\quad\theta \in(0,\pi),\label{eq:Dirichlet}\\
    &\left|N_O(\theta)\right\rangle,\quad \theta \in(0,\pi) ,
\end{align}
and eight discrete endpoint BCs:
\begin{align}
    \left|D_O(0, \pi )\pm\right\rangle,\quad \ket {N_O(     0, \pi )\pm}.
\end{align}
Here and throughout this paper, we use the simplified notations for the Dirichlet/Neumann 
boundary states parametrized by the boundary values as 
\begin{align}\label{eq:notation_D/N}
    \ket{D_O(\theta)}:=\ket{D_O(\varphi = \theta r)},\quad
    \ket{N_O(\theta)}:=\ket{N_O(\tilde{\varphi}=\frac{\theta}{2r})},
\end{align}
so that $\theta\in [0,\pi]$ for any compactification radius $r$. 
The right-hand side of Eq.~\eqref{eq:notation_D/N} corresponds to the convention in Ref.~\cite{oshikawaBoundaryConformalField1996}.
In this convention, $\theta=0,\pi$ correspond to the fixed points of the $\mathbb{Z}_2$ orbifolding;
there are two distinct boundary Dirichlet/Neumann boundary states at each of these fixed points  
labeled by $\pm$ correspondingly to the sign of the twisted sector
contribution\cite{oshikawaBoundaryConformalField1996}.

In Ref.~\cite{oshikawaBoundaryConformalField1996}, it is noticed that at $r=1$, which corresponds to the decoupled point $\lambda=0$, i.e., a doubled Ising model, the above eight discrete endpoint boundary states, together with the midpoint of the continuous family of the Dirichlet boundary state $\ket{D_O(\frac{\pi}{2})}$
can be identified with the $3\times3$ tensor products of the Ising boundary states:
\begin{gather}
|\uparrow \uparrow\rangle  =\left|D_O(0)+\right\rangle,\quad |\uparrow \downarrow\rangle =\left|D_O(\pi)-\right\rangle, \quad
|\downarrow \downarrow\rangle  =\left|D_O(0)-\right\rangle,\quad|\downarrow \uparrow\rangle  =\left|D_O(\pi)+\right\rangle, \notag\\
|\uparrow f\rangle  =\left|N_O(0)+\right\rangle,\quad |f \uparrow\rangle  =|N_O(\pi)+\rangle,\quad
|\downarrow f\rangle  =|N_O(0)-\rangle,\quad |f \downarrow\rangle  =|N_O(\pi)-\rangle,\notag \\
|f f\rangle  =|D_O(\frac{\pi}{2})\rangle,
\label{eq:orbifoldBC_vs_AT}
\end{gather}
where for each independent Ising spin ($\sigma,\tau$) one can apply free ($f$) or fixed ($\uparrow/\downarrow$) boundary conditions.
Although this identification was originally made at the decoupled point,
we can naturally generalize these identifications between the boundary conditions of the AT model and
the $\mathbb{Z}_2$ orbifold free boson boundary states
to the whole AT criticality line with generic $\lambda$ and $r$ related by Eq.~\eqref{eq:r_vs_lambda}.

While the Hamiltonian~\eqref{eq:HFree-Free} has the $S_4$ symmetry of the Potts model only
at the special point $\lambda=1$, the Ashkin-Teller model with a general $\lambda$
can be described in terms of the Potts variable $S=A,B,C,D$ introduced in Eq.~\eqref{eq:Potts_vs_AT}.
The deviation $\lambda-1$ from the four-state Potts point can be regarded as 
a perturbation which explicitly breaks the $S_4$ symmetry.

The four boundary conditions corresponding to the fixed Ising spins are naturally identified with 
the four boundary conditions corresponding to the fixed Potts spin. 
On the other hand, the boundary condition $\uparrow f$ in terms of the Ising spins
may be regarded as a ``resonant state'' between $\uparrow \uparrow = A$ and $\uparrow \downarrow =B$.
Thus it can be identified with the ``mixed'' $AB$ boundary condition in terms of the Potts variable.

Generalizing this, the set of nine boundary conditions
in terms of the Ising spins can be rewritten in terms of the Potts variables as
\begin{gather}
    ff=\text{Free},\ \uparrow\uparrow=A,\ \uparrow\downarrow=B,\ \downarrow\downarrow=C,\ \downarrow\uparrow=D,\notag\\
    \uparrow f=AB,\ \downarrow f=CD,\ f\uparrow=AD,\ f\downarrow=BC .
    \label{eq:BC_AT_vs_Potts}
\end{gather}

On the lattice, such boundary conditions can be realized by defining the corresponding spin projectors:
\begin{align}
    P^{\{\mu\}}_j:=\sum_{\mu}\ket{\mu}\bra{\mu}.
\end{align}
where $\{\mu\} \subseteq \{A,B,C,D\}$ is a subset of possible spin directions that is allowed.
We refer to these as "blob" boundary conditions corresponding to the terminology used in the Q-state Potts model~\cite{jacobsenConformalBoundaryLoop2008,dubailConformalTwoboundaryLoop2009,robertsonConformallyInvariantBoundary2019a}. In practice, these projectors are represented by the spin operators. We then define the Hamiltonian with boundary conditions $a$ and $b$ on the left and right sides by acting such projectors on the left-most and right-most sites, respectively\footnote{This realization can also be seen as a boundary perturbation, see Sec.~\ref{sec:boundary_RG}}:
\begin{align}
    H^{\alpha-\beta}:=P^a_1P^b_LHP^a_1P^b_L.
\end{align}
We can choose the subset $\{\mu\}$ to consist of one, two, or three spin states.
When $|\{\mu\}|=1$, we have the fixed boundary conditions $A,B,C,D$.
When $|\{\mu\}|=2$, we have the mixed boundary conditions $AB,AC,AD,BC,BD,CD$.
When $|\{\mu\}|=3$, we have the three-state mixed boundary conditions $ABC,ABD,ACD,BCD$.
(When $|\{\mu\}|=4$, the projection does nothing, resulting in the free boundary condition.)
We studied the correspondence between these blob boundary conditions and
the boundary states of the $\mathbb{Z}_2$ orbifold.
Some of these correspondences can be explicitly identified based on Eqs.~\eqref{eq:orbifoldBC_vs_AT} and ~\eqref{eq:BC_AT_vs_Potts}.

For the free boundary condition ($|\{\mu\}|=0$ or $4$), we have 
\begin{align}
    \ket{\text{Free}}=\ket{D_O(\frac{\pi}{2})} .
\end{align}
The diagonal partition function for the free boundary condition is 
\begin{align}
    Z_\text{Free-Free}=& \bra{D_O(\frac{\pi}{2})}\tilde{q}^{H_p}\ket{D_O(\frac{\pi}{2})}
    \notag \\ 
    = & \frac{1}{ \eta(q)} \left[\sum_{n=-\infty}^{\infty} q^{\frac{r^2}{2}(2n+1)^2}+\sum_{n=-\infty}^{\infty} q^{2r^2n^2}\right] ,
\label{eq:Zfree-free}
\end{align}
where 
\begin{align}
    \eta(q)=q^{\frac{1}{24}}\prod_{n=1}^\infty (1-q^n)
\end{align}
is the Dedekind eta function.
The scaling dimensions of the boundary operators generated by the vertex operators in the sum are
\begin{align}
0,\frac{r^2}{2}, 2r^2,\frac{9r^2}{2},\dots,
\end{align}
where the $h=0$ state corresponds to the identity operator.
Note that the full spectrum also includes integer descendants arising from the $1/\eta(q)$ factor.
The Affleck-Ludwig ground-state degeneracy is read off from the boundary state as  
\begin{align}
g_{\text{Free}} = \frac{1}{\sqrt{r}}. 
\end{align}

For the fixed boundary conditions ($\{\mu\}|=1$), we have
\begin{align}
    \ket{A}=\ket{D_O(0)+},\ \ket{B}=\ket{D_O(\pi )-},\ \ket{C}=\ket{D_O(0)-},\ \ket{D}=\ket{D_O(\pi)+} .
    \label{eq:fixed_DO}
\end{align}
The diagonal partition function for these fixed boundary conditions reads
\begin{align}
    Z_{a-a}= & \bra{D_O(0/\pi)\pm}\tilde{q}^{H_p}\ket{D_O(0/\pi)\pm} \notag \\
    =& \frac{1}{2 \eta(q)} \sum_{n=-\infty}^{\infty}\left[q^{2 r^2 n^2}+(-1)^n q^{n^2}\right],\quad
\text{with}\ a\in\{A,B,C,D\}.
\end{align}
Similarly, for the fixed boundary conditions, expanding the partition function reveals the primary scaling dimensions:
\begin{equation}
0, 2r^2, 8r^2, \dots.
\end{equation}
From the boundary state, we can read off the Affleck-Ludwig ground-state degeneracy   
\begin{align} 
g_A =\dots= \frac{1}{2\sqrt{r}} .
\end{align}

Concerning the ``mixed'' boundary conditions of two Potts spin states ($|\{\mu\}|=2$),
we can distinguish between the ``adjacent'' mixed boundary conditions ($AB,CD,\dots$) and the ``antipodal'' mixed boundary conditions ($AC,BD$) due to their different behavior under the $D_4$ group action.
For the adjacent mixed boundary conditions, we have
\begin{align}
    \ket{AB}=\ket{N_O(0)+},\ \ket{CD}=\ket{N_O(0 )-},\ \ket{AD}=\ket{N_O(\pi)-},\ \ket{BC}=\ket{N_O(\pi)-} .
\end{align}
The diagonal partition function for these adjacent mixed boundary conditions are   
\begin{align}
Z_{a-a}= & \bra{N_O(0/\pi)\pm}\tilde{q}^{H_p}\ket{N_O(0/\pi)\pm} \notag \\ 
= & \frac{1}{2 \eta(q)} \sum_{n=-\infty}^{\infty}\left[q^{ \frac{n^2}{2r^2} }+(-1)^n q^{n^2}\right],\quad
\text{with}\ a\in\{AB,BC,CD,AD\},
\label{eq:AB-AB}
\end{align}
The scaling dimensions of the boundary operators for these boundary conditions are thus  
\begin{align} 
 0,\frac{1}{2r^2}, \frac{2}{r^2}, \frac{9}{2 r^2}, \ldots ,
\end{align}
and the Affleck-Ludwig ground-state degeneracy is 
\begin{align} 
 g_{AB} = \dots = \sqrt{\frac{r}{2}} .
\end{align}

On the other hand, the antipodal mixed boundary conditions should belong to
the continuous Dirichlet family of boundary states~\eqref{eq:Dirichlet} 
of the $\mathbb{Z}_2$ orbifold. 
In the microscopic model, the antipodal mixed boundary condition may be 
imposed by introducing the boundary field $h_{\sigma \tau} \sigma^z \tau^z$. 
In general, the boundary value $\theta$ that parametrizes the Dirichlet boundary
state~\eqref{eq:Dirichlet} is a non-universal continuous function of the 
boundary field $h_{\sigma \tau}$.
In the limit $h_{\sigma \tau}\rightarrow \pm \infty$, the boundary spin is completely 
projected to the antipodal pair of states $AC$ or $BD$.
We argue that, these limits correspond to special values of $\theta$,
namely 
\begin{align}
|AC\rangle=|D_O(\theta_c)\rangle,\quad|BD\rangle=|D_O(\pi-\theta_c)\rangle,
    \label{eq:AC_BD-theta_c}
\end{align}
where
\begin{align}
\theta_c(\lambda)=\frac{1}{\frac{1}{\arcsin(\lambda)}+\frac{2}{\pi}} .
\label{eq:theta_c-lambda}
\end{align}
In terms of the compactification radius $r$, 
\begin{align}
    \theta_c(r) = \frac{\pi}{2} \left( 1 - \frac{1}{r^2} \right) .
\label{eq:theta_c-r}
\end{align}

Although the boundary state belongs to the same ``continuous Dirichlet'' family
of the boundary states, the diagonal partition function is different from 
that for the free boundary condition because of the orbifolding~\cite{oshikawaBoundaryConformalField1996}.
\begin{align}\label{eq:AC-AC}
    Z_{D_O(\theta)-D_O(\theta)} = &
    \bra{D_O(\theta)}\tilde{q}^{H_p}\ket{D_O(\theta)} \notag \\
    = & \frac{1}{ \eta(q)} \left[\sum_{n=-\infty}^{\infty} q^{2r^2 (n+\frac{\theta}{\pi})^2}+
    \sum_{n=-\infty}^{\infty} q^{2r^2 n^2}\right] .
\end{align}
The Affleck-Ludwig boundary degeneracy, on the other hand, is identical to that of the Free boundary condition, 
reflecting the fact that these boundary conditions are related by exactly marginal boundary deformations:
\begin{align} 
    g_{AC} = g_{BD} = & g_{\text{Free}} = \frac{1}{\sqrt{r}} .
\end{align}
The detailed discussions and justifications of the identification~\eqref{eq:AC_BD-theta_c}
will be given later in Sec.~\ref{sec:boundary_RG}.

Finally, let us discuss the ``three-state mixed'' boundary conditions ($|\{\mu\}|=3$) in 
the microscopic model.
Actually there are no boundary states of the $\mathbb{Z}_2$ orbifold which specifically correspond  
to the ``three-state mixed'' boundary conditions. 
Instead, the ``three-state mixed'' boundary conditions in the microscopic model are renormalized 
into one of the boundary states discussed above, in the infrared limit.
For the generic critical AT model with $\lambda <1$,
we expect that the boundary is renormalized eventually 
to one of the fixed or two-state mixed boundary conditions in the infrared limit.
A detailed discussion of which boundary condition will be the IR fixed point is provided in Sec.~\ref{sec:RG_Free_AT}.
At the four-state Potts point, on the other hand, a totally symmetric three-state mixed perturbation will be normalized into the free boundary condition, and any non-symmetric one will end up with one of the fixed boundary conditions in the infrared limit. 
The justification of this claim will be given later in Sec.~\ref{sec:RGflow_four-state_3-state}.

\subsection{Off-diagonal partition functions and \texorpdfstring{$D_4$}{D4} symmetry in AT model}\label{sec:D4}

The $D_4$ symmetry of the AT model was already evident in the diagonal partition functions for various boundary conditions 
discussed in Sec.~\ref{sec:bs_orbifold}. 
Here let us check that the $D_4$ symmetry is also respected in the off-diagonal partition functions, namely 
the partition functions with different boundary conditions on the two sides.

First, let us check the off-diagonal partition functions among the fixed ($A,B,C,D$) boundary conditions.
The $D_4$ symmetry leaves two distinct possibilities for the off-diagonal partition functions: 
adjacent and antipodal pairs of fixed spin directions\footnote{We do not distinguish between $Z_{\alpha-\beta}$ and $Z_{\beta-\alpha}$ as the spatial inversion symmetry is manifested.}.
\begin{align}
Z_{A-B/D}= & \bra{D_O(0)+}\tilde{q}^{H_p}\ket{D_O(\pi)\mp}=\frac{1}{2 \eta(q)} \sum_{n=-\infty}^{\infty}q^{\frac{r^2}{2}(2n+1)^2}\notag\\&\quad\quad\quad\ =\bra{D_O(0)-}\tilde{q}^{H_p}\ket{D_O(\pi)\mp}=Z_{C-B/D},\\
Z_{A-C}=Z_{B-D}= & \bra{D_O(0/\pi)+}\tilde{q}^{H_p}\ket{D_O(0/\pi)-}=\frac{1}{2 \eta(q)} \sum_{n=-\infty}^{\infty}\left[q^{2r^2 n^2}-(-1)^n q^{n^2}\right] .
\end{align}
Here and in the following, we employ a compact slash (/) notation in the subscripts, which corresponds to the compact notations (/ or $\pm$) in the bra-kets. For example, $Z_{A-B/D}$ is shorthand for the equal functions $Z_{A-B}$ and $Z_{A-D}$, corresponding to the $\mp$ sign in $\ket{D_O(\pi)\mp}$. As expected, the fixed-fixed partition function depends only on whether the two boundary conditions are the identical (diagonal partition function), 
an adjacent pair, or an antipodal pair.
This is consistent with the $D_4$ symmetry of the Ashkin-Teller model.

For adjacent two-state mixed boundary conditions $\{AB,BC,CD,AD\}$,
partition functions among them read the same as those
with fixed boundary conditions by replacing $r$ with $\frac{1}{2r}$ due to the T-duality, which maps between Dirichlet and Neumann BCs:
\begin{align}
    Z_{AB-BC/AD}= & \bra{N_O(0)+}\tilde{q}^{H_p}\ket{N_O(\pi)\mp}=\frac{1}{2 \eta(q)} \sum_{n=-\infty}^{\infty}q^{\frac{1}{8r^2}(2n+1)^2}\notag\\&\quad\quad\quad\quad\quad\,\,=\bra{N_O(0)-}\tilde{q}^{H_p}\ket{N_O(\pi)\mp}=Z_{CD-BC/AD},\\
    Z_{AB-CD}=Z_{BC-AD}= & 
    \bra{N_O(0/\pi)+}\tilde{q}^{H_p}\ket{N_O(0/\pi)-}=\frac{1}{2 \eta(q)} \sum_{n=-\infty}^{\infty}\left[q^{\frac{n^2}{2r^2} }-(-1)^n q^{n^2}\right].
\end{align}
For the partition functions between fixed and adjacent two-state mixed BC,
we will consider the amplitudes between Dirichlet and Neumann boundary states.
The contribution from the $\mathrm{U}(1)$ boundary states is always the same,
and the only difference comes from the amplitudes between twisted sectors. So there are only two independent partition functions, which read
\begin{align}
    Z_{A-CD/BC}=&\bra{D_O(0)+}\tilde{q}^{H_p}\ket{N_O(0/\pi)-}=\bra{D_O(0)-}\tilde{q}^{H_p}\ket{N_O(0/\pi)+}=Z_{C-AB/AD}\notag\\=&Z_{B-CD/AD}=\bra{D_O(\pi)-}\tilde{q}^{H_p}\ket{N_O(0)-/N_O(\pi)+}\notag\\=&\bra{D_O(\pi)+}\tilde{q}^{H_p}\ket{N_O(0)+/N_O(\pi)-}=Z_{D-AB/BC}=\frac{1}{ \eta(q)}\sum_{n=-\infty}^\infty q^{\frac{(8n+3)^2}{16}}
\end{align}
and
\begin{align}
    Z_{A-AB/AD}=&\bra{D_O(0)+}\tilde{q}^{H_p}\ket{N_O(0/\pi)+}=\bra{D_O(0)-}\tilde{q}^{H_p}\ket{N_O(0/\pi)-}=Z_{C-CD/BC}\notag\\=&Z_{B-AB/BC}=\bra{D_O(\pi)-}\tilde{q}^{H_p}\ket{N_O(0)+/N_O(\pi)-}\notag\\=&\bra{D_O(\pi)+}\tilde{q}^{H_p}\ket{N_O(0)-/N_O(\pi)+}=Z_{D-CD/AD}=\frac{1}{ \eta(q)}\sum_{n=-\infty}^\infty q^{\frac{(8n+1)^2}{16}}.
\end{align}
These correspond exactly to the two distinct ways of pairing the fixed and adjacent two-state mixed boundary conditions under the $D_4$ symmetry, 
namely whether the spin direction of the fixed boundary condition is included in the mixed boundary condition or not.

The partition function with the two distinct antipodal two-state mixed boundary conditions $\{AC,BD\}$ reads
\begin{align}
    Z_{AC-BD}=\bra{D_O(\theta_c)}\tilde{q}^{H_p}\ket{D_O(\pi-\theta_c)}=\frac{1}{ \eta(q)} \left[\sum_{n=-\infty}^{\infty} q^{2r^2(n+\frac{1}{2})^2}+\sum_{n=-\infty}^{\infty} q^{2r^2(n+\frac{1}{2}+\frac{\theta_c}{\pi})^2}\right].
\end{align}
The partition function with an antipodal two-state mixed boundary condition and a fixed boundary condition differs only by the relative value of the bosonic field, so there are again only two possibilities:
\begin{align}\label{eq:AC-A}
    Z_{AC-A/C}&=\bra{D_O(\theta_c)}\tilde{q}^{H_p}\ket{D_O(0)\pm}=
    \frac{1}{\eta(q)}
    \sum_{n=-\infty}^\infty q^{2r^2(n+\frac{\theta_c}{2\pi})^2}\notag\\
    &=\bra{D_O(\pi-\theta_c)}\tilde{q}^{H_p}\ket{D_O(\pi)\mp}=Z_{BD-B/D},\\
    \label{eq:AC-B}
    Z_{AC-B/D}&=\bra{D_O(\theta_c)}\tilde{q}^{H_p}\ket{D_O(\pi)\mp}=\frac{1}{ \eta(q)}\sum_{n=-\infty}^\infty q^{2r^2(n+\frac{\pi-\theta_c}{2\pi})^2}\notag\\&=\bra{D_O(\pi-\theta_c)}\tilde{q}^{H_p}\ket{D_O(0)\pm}=Z_{BD-A/C}.
\end{align}
The partition functions between an adjacent and an antipodal two-state mixed boundary conditions are all the same:
\begin{align}\label{eq:AC-AB}
    Z_{AC/BD-a}=\bra{D_O(\theta_c)}\tilde{q}^{H_p}\ket{N_O(0/\pi)\pm}=\frac{1}{ \eta(q)} \sum_{n=0}^{\infty}q^{\frac{(2n+1)^2}{16} },\ \text{with}\ a\in\{AB,BC,CD,AD\},
\end{align}
due to the amplitudes between U(1) Dirichlet and Neumann boundary conditions being the same for all values of (dual) fields.

Finally,
the off-diagonal partition functions between the Free and fixed boundary conditions are all the same,
since the free boundary state sits at the midpoint of the Dirichlet boundary condition family:
\begin{align}
    Z_{\text{Free-}A/B/C/D}=\bra{D_O(\frac{\pi}{2})}\tilde{q}^{H_p}\ket{D_O(0/\pi)\pm}=\frac{1}{\eta(q)}\sum_{n=0}^{\infty}q^{\frac{r^2}{8}(2n+1)^2}.
\end{align}
Similar to the antipodal two-state mixed boundary condition case~\eqref{eq:AC-AB}, the partition functions between Free and adjacent two-state mixed boundary conditions are all the same:
\begin{align}
    Z_{\text{Free}-a}=\bra{D_O(\frac{\pi}{2})}\tilde{q}^{H_p}\ket{N_O(0/\pi)\pm}=\frac{1}{ \eta(q)} \sum_{n=0}^{\infty}q^{\frac{(2n+1)^2}{16} },\ \text{with}\ a\in\{AB,BC,CD,AD\}.
\end{align}
For the two partition functions with antipodal two-state mixed BC, they are once again the same due to the symmetric structure:
\begin{align}
    Z_{\text{Free}-AC/BD}=\bra{D_O(\frac{\pi}{2})}\tilde{q}^{H_p}\ket{D_O(\frac{\pi}{4}/\frac{3\pi}{4})}=\frac{1}{ \eta(q)} \sum_{n=0}^{\infty}q^{\frac{r^2}{32} (2n+1)^2}.
\end{align}
With the above results, we confirm that the 
$D_4$ symmetry of the Ashkin-Teller model is respected in all off-diagonal partition functions as well.

\subsection{Kramers-Wannier transformation on boundary conditions}\label{sec:KW}
Since the quantum Ashkin-Teller model can be represented as two TF-Ising chains with mutual interactions, we can also apply the Kramers-Wannier transformation~\cite{kramersStatisticsTwoDimensionalFerromagnet1941,liNoninvertibleDualityTransformation2023b} to these Ising chains:
\begin{align}\label{eq:KW}
\sigma_j^x & = \begin{cases}\tilde{\sigma}_{j-1}^z \tilde{\sigma}_j^z, & j=2, \ldots, L, \\
\tilde{\sigma}_1^z, & j=1,\end{cases} & \qquad \sigma_j^z & =\prod_{k=j}^L \tilde{\sigma}_k^x ,\\
\tau_j^x & = \begin{cases}\tilde{\tau}_{j-1}^z \tilde{\tau}_j^z, & j=2, \ldots, L, \\
\tilde{\tau}_1^z, & j=1,\end{cases} & \qquad \tau_j^z & =\prod_{k=j}^L \tilde{\tau}_k^x.
\end{align}
For the AT model with periodic boundary conditions, this is a symmetry of the model at the critical point. With OBC, this transformation keeps the bulk invariant but alters the boundary conditions on each side. In fact, this can be viewed as moving the non-invertible Kramers-Wannier topological defect (TD) from one side to the other side \cite{PhysRevLett.93.070601,Kevin_Graham_2004,schweigert2007kramerswannierdualitieswzwtheories,FROHLICH2007354,PhysRevB.104.125418,doi:10.1142/9789814304634_0056}. Effectively, this transforms one boundary condition into another. In this section, we discuss how the blob boundary conditions we define transform under this duality transformation.

\subsubsection{Duality between free and fixed boundary conditions}
For the critical Ashkin-Teller chain with free boundary conditions, the Hamiltonian after the transformation reads
\begin{align}\label{eq:dual_FreeFree}
    \widetilde{H^\text{Free-Free}}=&- \sum_{j=1}^{L-1}\left(\tilde{\sigma}_j^x+\tilde{\tau}_j^x+\lambda \tilde{\sigma}_j^x \tilde{\tau}_j^x\right)- \sum_{j=1}^{L-1}\left(\tilde{\sigma}_j^z \tilde{\sigma}_{j+1}^z+\tilde{\tau}_j^z \tilde{\tau}_{j+1}^z+\lambda \tilde{\sigma}_j^z \tilde{\tau}_j^z \tilde{\sigma}_{j+1}^z \tilde{\tau}_{j+1}^z\right)\notag\\
    &-(\tilde{\sigma}_1^z+\tilde{\tau}_1^z+\lambda\tilde{\sigma}_1^z\tilde{\tau}_1^z) .
\end{align}
We note that there is no transverse magnetic field acting on the rightmost spins $\tilde{\sigma}_L$ and $\tilde{\tau}_L$.
As a consequence, $\tilde{\sigma}_L^{z}$ and $\tilde{\tau}_L^{z}$ commute with this Hamiltonian.
The Hilbert space thus breaks down into four independent sectors,
each of which corresponds to a fixed boundary condition on the right end of the chain.
On the other hand, the last term in Eq.~\eqref{eq:dual_FreeFree} can be viewed as a longitudinal
boundary field acting on the leftmost spins $\tilde{\sigma}_1$ and $\tilde{\tau}_1$.
This will drive the boundary condition on the left end to the fixed boundary condition to the spin state $A$.
Overall, this means
\begin{align}\label{eq:KW_lat_Free_A}
    Z^\text{lat}_{\text{Free}-\text{Free}}(L) = \sum_{\mathcal{B}=A,B,C,D} Z^\text{lat}_{A-\mathcal{B}}(L+1),
\end{align}
where $Z^\text{lat}_{a-b}(L)$ is the partition function of an AT chain with $L$ sites and $a,b$ boundary conditions applied on the left and right-hand side, respectively.
This shows that the free boundary condition is mapped to the fixed boundary condition.
We may interpret the boundary condition of the dual model after the KW transformation as 
the result of the TD acting on the original boundary state. 
Note that our KW transformation, as defined in Eq.~\eqref{eq:KW}, is not inversion symmetric.
As a consequence, the boundary conditions are transformed differently at the left and right ends.
In the following, we focus on the boundary condition on the right end\footnote{For general boundary conditions on the left-hand side, the KW transformation may introduce a string operator along the whole chain. As we are discussing conformally invariant boundary conditions here, we fix the LHS to be free so that the boundary condition after the KW transformation is still conformally invariant (fixed).}.
Then we have 
\begin{align}
    D_{KW}\ket{\text{Free}}=\ket{A}+\ket{B}+\ket{C}+\ket{D},\quad D_{KW}\ket{A/B/C/D}=\ket{\text{Free}}
    \label{eq:KW_free}
\end{align}
or in terms of Dirichlet boundary states:
\begin{align}
    \label{eq:KW_free_DO}
    D_{KW}\ket{D_O(\frac{\pi}{2})}=\sum_{\theta={0,\pi},\pm}\ket{D_O(\theta),\pm},\quad D_{KW}\ket{D_O(0/\pi),\pm}=\ket{D_O(\frac{\pi}{2})},
\end{align}
where we denote the non-invertible Kramers-Wannier TD as $D_{KW}$. Note that the free boundary condition is mapped to a superposition of four fixed BCs, which is called a spontaneous symmetry breaking (SSB) BC\footnote{Historically, such SSB boundary conditions first appeared in the studies of BCFT for the tricritical Ising model~\cite{Chim1996}, though they remained less familiar in broader contexts. Later, specific types were investigated by Graham and Watts~\cite{Graham:2003nc}, and more recently, general cases have been connected to boundary zero modes and qubits~\cite{PhysRevB.104.125418, Fukusumi2021, Fukusumi2025}.}. On the lattice, this superposition implies that the boundary is not pinned to a specific direction but is allowed to fluctuate among all four possible symmetry-breaking configurations, consistent with the duality mapping from a disordered (free) phase.
We note that, the free boundary conditions on the both ends in the Hamiltonian~\eqref{eq:HFree-Free} is 
mapped to the fixed boundary condition $A$ on the left end and the SSB boundary condition
on the right end in the dual Hamiltonian~\eqref{eq:dual_FreeFree}\footnote{The choice of the fixed boundary condition here is not unique since one can combine $D_4$ action to the KW transformation to define a new KW transformation. Nevertheless, the boundary condition on the right-hand side is always the SSB boundary condition, so the partition function is not changed.}.
This reflects the lack of inversion symmetry of the KW transformation~\eqref{eq:KW} as discussed above.

\subsubsection{Duality among two-state mixed boundary conditions}
We can also apply this procedure to two-state mixed boundary conditions.
For instance, let us keep both sides of the $\tau$-Ising chain free but
fix the right-hand side of the $\sigma$-Ising chain to $\left|\uparrow\right\rangle$,
by replacing the transverse field $-\sigma^x_L - \lambda \sigma^x_L \tau^x_L$
by the longitudinal field $-\sigma^z_L$
on the rightmost site $L$, in the original Hamiltonian~\eqref{eq:HFree-Free}.
This realizes the ``mixed'' boundary condition between two adjacent states AB
on the right end and keeps the Free boundary condition on the left end.
The resulting Hamiltonian is
\begin{align}
    H_\text{AT}^{\text{Free}-AB}=&- \sum_{j=1}^{L-1}\left(\sigma_j^x+\tau_j^x+\lambda \sigma_j^x \tau_j^x\right)-\tau_L^x-\sigma_L^z \notag\\
    & - \sum_{j=1}^{L-1}\left(\sigma_j^z \sigma_{j+1}^z+\tau_j^z \tau_{j+1}^z+\lambda \sigma_j^z \tau_j^z \sigma_{j+1}^z \tau_{j+1}^z\right).
\end{align}
After the Kramers-Wanier transformation, the dual Hamiltonian reads:
\begin{align}\label{eq:dual_free_2}
    \widetilde{H_\text{AT}^{\text{Free}-AB}}=&
    - \sum_{j=1}^{L-1}\left(\tilde{\sigma}_j^x+\tilde{\tau}_j^x+\lambda \tilde{\sigma}_j^x \tilde{\tau}_j^x\right)
    - \sum_{j=1}^{L-2}\left(\tilde{\sigma}_j^z \tilde{\sigma}_{j+1}^z+\tilde{\tau}_j^z \tilde{\tau}_{j+1}^z+\lambda \tilde{\sigma}_j^z \tilde{\tau}_j^z \tilde{\sigma}_{j+1}^z \tilde{\tau}_{j+1}^z\right)\notag\\
    &-(\tilde{\sigma}_1^z+\tilde{\tau}_1^z+\lambda\tilde{\sigma}_1^z\tilde{\tau}_1^z)
    -\tilde{\sigma}^x_L - \tilde{\tau}^z_{L-1} \tilde{\tau}^z_L .
\end{align}
The bulk and the left boundary remain the same as in Eq.~\eqref{eq:dual_FreeFree}.
The difference appears at the right boundary. 
The $\tilde{\sigma}$ spin now obeys the Free boundary condition on the right end, thanks to the transverse field $-\tilde{\sigma}^x_L$
on the rightmost site.
On the other hand, because of the lack of the transverse field $\tilde{\tau}^x_L$ on the same site, $\tilde{\tau}^z_L$ commutes with the Hamiltonian.
This allows us to separate the Hilbert space into two sectors labeled by the eigenvalue of $\tilde{\tau}_L^z$.
If we choose $\tilde{\tau}_L^z=1$, then the $\tilde{\tau}$ spin is polarized to $\ket{\uparrow}$ while $\tilde{\sigma}$ spin remains free on the right end. 
This is equivalent to imposing the adjacent two-state mixed boundary condition $AD$ on the right end.
Similarly, choosing $\tilde{\tau}_L^z=-1$ corresponds to the $BC$ boundary condition.

Overall, the KW transformation on the adjacent two-state mixed boundary condition may be schematically written as
\begin{align}\label{eq:KW-2}
    AB\to AD+BC
\end{align}
and the relation between the partition function at the continuum limit can be verified easily:
\begin{align}
    Z_{\text{Free}-AB}&=\frac{1}{2\eta(q)}\sum_{n=-\infty}^\infty q^{\frac{(2n+1)^2}{16}}\notag\\
    &=\frac{1}{\eta(q)}\sum_{n=-\infty}^\infty q^{\frac{(8n+1)^2}{16}}+\frac{1}{\eta(q)}\sum_{n=-\infty}^\infty q^{\frac{(8n+3)^2}{16}}=Z_{A-AD}+Z_{A-BC}.
\end{align}
The above transformations are a natural generalization of the decouple point $\lambda=0\ (r=1)$,
where the same relations can be obtained by performing the tensor product of each independent Ising spin~\cite {oshikawaBoundaryConformalField1996}.

\subsubsection{Duality between antipodal two-state mixed boundary conditions}

The transformation of the antipodal two-state mixed boundary states $\ket{AC}$ and $\ket{BD}$ 
turns out to be more complicated.
To realize the $AC$ boundary condition, for example, we apply the projector
\begin{align}
    P^{AC}_L=\frac{1+\sigma^z_L\tau^z_L}{2}\stackrel{\text{KW}}{=}\frac{1+\tilde{\sigma}^x_L\tilde{\sigma}^x_L}{2},
\end{align}
on the rightmost site of the original Hamiltonian~\eqref{eq:HFree-Free}.
Alternatively, we can introduce the boundary field $- h_{\sigma\tau} \sigma^z_L \tau^z_L$ at the site $L$
and take the limit $h_{\sigma\tau}\to +\infty$ to enforce $\sigma^z_L \tau^z_L=1$ in the ground state. 
This can also be emulated by keeping the field $h_{\sigma \tau}$ finite but removing the transverse field terms
$-\sigma^x_L - \tau^x_L$ at the same site.
We nevertheless keep $- \lambda \sigma^x_L \tau^x_L$ which commutes with the boundary field
$- h_{\sigma\tau} \sigma^z_L \tau^z_L$. (The $- \lambda \sigma^x_L \tau^x_L$ term survives either way,
under the projector or in the limit of $h_{\sigma\tau} \to \infty$).
The Hamiltonian is then given by
\begin{align} 
    H_\text{AT}^{\text{Free}-AC}=&- \sum_{j=1}^{L-1}\left(\sigma_j^x+\tau_j^x+\lambda \sigma_j^x \tau_j^x\right)
    - \sum_{j=1}^{L-1}\left(\sigma_j^z \sigma_{j+1}^z+\tau_j^z \tau_{j+1}^z+\lambda \sigma_j^z \tau_j^z \sigma_{j+1}^z \tau_{j+1}^z\right)\notag\\
    &
    - \lambda \sigma^x_L \tau^x_L
    - h_{\sigma\tau} \sigma_L^z \tau_L^z .
\end{align}
After the Kramers-Wannier transformation, the dual Hamiltonian reads
\begin{align}
    \label{eq:H_dual_Free_AC}
    \widetilde{H_\text{AT}^{\text{Free}-AC}}=&
    - \sum_{j=1}^{L-1}\left(\tilde{\sigma}_j^x+\tilde{\tau}_j^x+\lambda \tilde{\sigma}_j^x \tilde{\tau}_j^x\right)
    - \sum_{j=1}^{L-2}\left(\tilde{\sigma}_j^z \tilde{\sigma}_{j+1}^z+\tilde{\tau}_j^z \tilde{\tau}_{j+1}^z+\lambda \tilde{\sigma}_j^z \tilde{\tau}_j^z \tilde{\sigma}_{j+1}^z \tilde{\tau}_{j+1}^z\right)\notag\\
    &-(\tilde{\sigma}_1^z+\tilde{\tau}_1^z+\lambda\tilde{\sigma}_1^z\tilde{\tau}_1^z)
    - \lambda \tilde{\sigma}_{L-1}^z \tilde{\sigma}_{L}^z \tilde{\tau}_{L-1}^z \tilde{\tau}_{L}^z
    -  h_{\sigma \tau} \tilde{\sigma}_L^x \tilde{\tau}_L^x .
\end{align}
On the rightmost site $L$, only the last term exists. 
Thus we can project to the sector with $\tilde{\sigma}_L^x \tilde{\tau}_L^x=1$.
This still allows the two possibility $\tilde{\sigma}_L^z \tilde{\tau}_L^z= \pm 1$.
Since $\tilde{\sigma}_L^z \tilde{\tau}_L^z$ commutes with the Hamiltonian,
we can discuss each sector $\tilde{\sigma}_L^z \tilde{\tau}_L^z = \pm 1$ separately.
The dual Hamiltonian can thus be reduced to
\begin{align}
    \label{eq:H_dual_Free_AC-2}
    \widetilde{H_\text{AT}^{\text{Free}-AC}} \sim &
    - \sum_{j=1}^{L-1}\left(\tilde{\sigma}_j^x+\tilde{\tau}_j^x+\lambda \tilde{\sigma}_j^x \tilde{\tau}_j^x\right)
    - \sum_{j=1}^{L-2}\left(\tilde{\sigma}_j^z \tilde{\sigma}_{j+1}^z+\tilde{\tau}_j^z \tilde{\tau}_{j+1}^z+\lambda \tilde{\sigma}_j^z \tilde{\tau}_j^z \tilde{\sigma}_{j+1}^z \tilde{\tau}_{j+1}^z\right)\notag\\
    &-(\tilde{\sigma}_1^z+\tilde{\tau}_1^z+\lambda\tilde{\sigma}_1^z\tilde{\tau}_1^z)
    \mp \lambda \tilde{\sigma}_{L-1}^z \tilde{\tau}_{L-1}^z ,
\end{align}
which is defined on the shorter chain with the sites $j=1,2,\ldots,L-1$.
The left boundary is subject to the fixed (A) boundary condition, as in the previous cases.
The rightmost site (now the site $L-1$) has the field $\tilde{h}_{\sigma\tau} = \pm \lambda$
coupled to the product $\tilde{\sigma}_{L-1}^z \tilde{\tau}_{L-1}^z$,
which tends to drive the boundary to the antipodal two-state mixed boundary condition,
competing with the standard transverse field on the same site
$-\tilde{\sigma}_{L-1}^x - \tilde{\tau}_{L-1}^x - \lambda \tilde{\tau}_{L-1}^x \tilde{\sigma}_{L-1}^x$.
As we will discuss details in Sec.~\ref{sec:RG_Free_AT}, the boundary condition then belongs to the
continous Dirichlet family $\ket{D_O(\theta)}$ of the $\mathbb{Z}_2$ orbifold,
with the boundary value $\theta$
determined by the dual magnetic field $\tilde{h}_{\sigma\tau} = \pm \lambda$.

Based on
the KW duality transformation of the bosonic field~\cite{oshikawaBoundaryConformalField1996,delfinoUniversalRatiosLine2004}\footnote{We note that the Kramers-Wannier transformation we discuss here can be viewed as a superposition of the two self-dualities $D_\pm$ in this model~\cite{delfinoUniversalRatiosLine2004}, i.e., $D_{KW}=D_++D_-$.},
we generally expect 
\begin{align}
    D_{KW} \ket{D_O(\theta)} = \ket{D_O\left(\frac{\pi}{2}-\theta\right)} + \ket{D_O\left(\frac{\pi}{2}+\theta\right)}. 
\label{eq:KW_DO_theta}
\end{align}
As a special case, the antipodal two-state mixed boundary conditions~\eqref{eq:AC_BD-theta_c} transform as 
\begin{align}
    D_{KW} \ket{D_O(\theta_c / \pi-\theta_c)} = \ket{D_O\left(\frac{\pi}{2}-\theta_c\right)} + \ket{D_O\left(\frac{\pi}{2}+\theta_c\right)}. 
\end{align}
The transformation~\eqref{eq:KW_DO_theta} also covers the KW duality transformation of the
free boundary condition~\eqref{eq:KW_free} as a special case, considering that
\begin{align}
    \ket{D_O(\theta\to 0)} &= \ket{D_O(0)+} + \ket{D_O(0)-} .
    \label{eq:DO_0_decomp}
\end{align}

Similarly, for the Neumann BC, we have
\begin{align}
    D_{KW}\ket{N_O(\theta)}=2\ket{N_O(\pi-\theta)},
    \label{eq:KW-Neumann}
\end{align}
which is consistent with Eq.~\eqref{eq:KW-2}.

We also observe that the $g$-factor is universally scaled by a factor of 2 under the action of the Kramers-Wannier duality defect:
\begin{align}
\frac{g_{A+B+C+D}}{g_\text{Free}}=\frac{g_\text{Free}}{g_A}=\frac{2g_{AB}}{g_{AB}}=\frac{2g_{D_O(\theta)}}{g_{D_O(\theta)}}=\frac{2g_{N_O(\theta)}}{g_{N_O(\theta)}}=2.
\end{align}
This result aligns with the quantum dimension $\sqrt{2}$ of the duality defect in the transverse-field Ising model, reflecting the fact that the Ashkin-Teller model consists of two coupled Ising copies ($\sqrt{2} \times \sqrt{2} = 2$).
\section{Extended structure at the four-state Potts point (\texorpdfstring{$\lambda=1$}{lambda})}
\label{sec:four-state}
The line of criticality of the AT model ends at $\lambda=1$. At this point, the Ashkin-Teller Hamiltonian~\eqref{eq:HFree-Free} can be organized into another form:
\begin{align}
    H_\text{AT}(\lambda=1)=&-J \sum_{j=1}^{L-1}\left(\sigma_j^z \sigma_{j+1}^z+\tau_j^z \tau_{j+1}^z+ \sigma_j^z \tau_j^z \sigma_{j+1}^z \tau_{j+1}^z\right)-h \sum_{j=1}^L\left(\sigma_j^x+\tau_j^x+ \sigma_j^x \tau_j^x\right)\notag\\
    =&-4J\sum_{j=1}^{L-1}\left(\frac{1+\sigma_j^z\sigma_{j+1}^z}{2}\frac{1+\tau_j^z\tau_{j+1}^z}{2}-\frac{1}{4}\right)-4h\sum_{j=1}^L\left(\frac{1+\sigma_j^x}{2}\frac{1+\tau_j^x}{2}-\frac{1}{4}\right)\notag\\
    =&-4J\sum_{j=1}^{L-1}\sum _{\mu=1}^4P_j^\mu P_{j+1}^\mu-4h\sum_{j=1}^L P_j,
\end{align}
where $P_i^\mu = \ket{\mu}_{ii} \! \bra{\mu} - \frac{1}{4}$ tends to project the spin at site $i$ along the $\mu\in\{A,B,C,D\}$ direction while $P_i=\ket{\eta_0}_{ii} \! \bra{\eta_0}-\frac{1}{4}$ tends to align spins along the direction $\ket{\eta_0}:=\frac{1}{2}(\ket{A}+\ket{B}+\ket{C}+\ket{D})$. This is exactly the four-state Potts model~\cite{affleckBoundaryCriticalPhenomena1998,Chepiga:2021tkj} that possesses $S_4$ symmetry that permutes all possible four AT spin directions. For this model, the blob boundary conditions we defined in Sec.~\ref{sec:bs_orbifold} are still conformally invariant, so our analysis applies here as well, and it is easy to check that one recovers the full $S_4$ symmetry for the partition functions at this special point by taking $r=\sqrt{2}$.

It was known that the $\mathbb{Z}_2$-orbifold theory is rational at some special points of the radius $r=\sqrt{\frac{M}{N}}r_\text{sd}$, where $M,N\in\mathbb{Z}$ and the self-dual radius is taken to be $r_{sd}=\frac{1}{\sqrt{2}}$ in our convention~\cite{GINSPARG1988153}. Four-state Potts model falls in this case, and we expect a richer structure of the boundary states here. In this section, we will first show how our $\mathbb{Z}_2$-orbifold boundary conditions can be identified with the extended W-symmetry preserving boundary conditions that can be easily obtained by Cardy's states for a rational CFT (RCFT) in Sec.~\ref{sec:W}. Then, we will use another formalism, which emphasizes the $\mathrm{SU}(2)$ symmetry in the model, to obtain a more comprehensive picture of boundary states in this model in Sec.~\ref{sec:SU(2)/D_2}.
\subsection{W-symmetry preserving boundary conditions from RCFT formalism}\label{sec:W}
It was shown that the four-state Potts model is equivalent to a $D_2$ orbifold of the $\mathrm{SU}(2)_1$ WZW model~\cite{KEDEM1993263, 10.21468/SciPostPhys.7.5.069,Cappelli:2002wq}, featuring an extended conformal symmetry known in the literature as the $\mathrm{W}(2,4,4)$ chiral algebra~\cite{BOUWKNEGT1993183,eholzer1993}. This notation explicitly denotes the spins of its generators: the standard spin-2 energy-momentum tensor and two distinct spin-4 currents. The physical origin of these higher-spin currents can be traced back to the generic $\mathbb{Z}_2$ orbifold branch~\cite{Thorngren:2019iar}. While the $\mathbb{Z}_2$ orbifold projection removes the original spin-1 $\mathrm{U}(1)$ current, the first spin-4 current generically survives along the entire branch. At the special radius corresponding to the four-state Potts point, the symmetry is further enhanced as the second spin-4 current emerges. For our practical purposes, this $\mathrm{W}(2,4,4)$ symmetry organizes the spectrum into eleven chiral primaries $1,j_{1,2,3},\phi,\sigma_{1,2,3},\tau_{1,2,3}$, whose scaling dimensions and characters are provided in Tab.~\ref{tab:W(2,4,4)}.
\begin{table}[tb]
    \centering
    \renewcommand{\arraystretch}{2.0}
    \begin{tabular}{c|c|c}
        \textbf{Field} & $h$ & \textbf{Characters} \\ \hline
        $1$ & 0 & $\displaystyle\chi_1=\frac{1}{\eta(q)}\left[\sum_{n=-\infty}^\infty q^{4n^2}-\sum_{n=0}^\infty q^{(2n+1)^2}\right]$ \\ 
        $j_a$ & 1 & $\displaystyle\chi_{j_a}=\frac{1}{\eta(q)}\sum_{n=0}^\infty q^{(2n+1)^2}$\\ 
        
        $\phi$ & $\frac{1}{4}$ & $\displaystyle\chi_\phi=\frac{1}{\eta(q)}\sum_{n=0}^\infty q^\frac{(2n+1)^2}{4}$\\ 
        $\sigma_a$ & $\frac{1}{16}$ & $\displaystyle\chi_{\sigma_a}=\frac{1}{\eta(q)}\sum_{n=0}^\infty q^\frac{(8n+1)^2}{16}$\\ 
        $\tau_a$ & $\frac{9}{16}$ & $\displaystyle\chi_{\tau_a}=\frac{1}{\eta(q)}\sum_{n=0}^\infty q^\frac{(8n+3)^2}{16}$ \\ 
    \end{tabular}
    \caption{Principle fields for the $\mathrm{W}(2,4,4)$ theory.}
    \label{tab:W(2,4,4)}
\end{table}
\begin{table}[tb]
  \centering
  \begin{tabular}{|l|ccccc|}
    \hline
    & $1$ & $j_b$ & $\phi$ & $\sigma_b$ & $\tau_b$ \\
    \hline
    $1$ & 1 & 1 & 2 & 2 & 2 \\
    $j_a$ & 1 & 1 & 2 & $2\varepsilon_{ab}$ & $2\varepsilon_{ab}$ \\
    $\phi$ & 2 & 2 & $-4$ & 0 & 0 \\
    $\sigma_a$ & 2 & $2\varepsilon_{ab}$ & 0 & $\delta_{ab}\sqrt{8}$ & $-\delta_{ab}\sqrt{8}$ \\
    $\tau_a$ & 2 & $2\varepsilon_{ab}$ & 0 & $-\delta_{ab}\sqrt{8}$ & $\delta_{ab}\sqrt{8}$ \\
    \hline
  \end{tabular}
  \caption{The modular S-matrix $S_{ij}$ (up to an overall factor $\frac{1}{4\sqrt{2}}$) for the $\mathrm{W}(2,4,4)$ theory~\cite{Dijkgraaf:1989hb}.}
    \label{tab:S_matrix}
\end{table}
The torus partition function of this theory can be organized into a diagonal form:
\begin{align}
    Z(\tau, \bar{\tau})=\left|\chi_1\right|^2+\sum_{a=1,2,3}\left|\chi_{j_a}\right|^2+\left|\chi_{\phi}\right|^2+\sum_{a=1,2,3}\left|\chi_{\sigma_a}\right|^2+\sum_{a=1,2,3}\left|\chi_{\tau_a}\right|^2.
\end{align}
This allows us to construct Cardy states that preserve the extended symmetry at the boundary directly through the W-Ishibashi states $|i\rangle\rangle$~\cite{Ishibashi:1988kg,recknagelBoundaryConformalField2013}, where $i$ denotes one of the W-primaries. Through the Verlinde formula, Cardy states can be constructed in the following forms:
\begin{align}
    |\tilde{i}\rangle=\sum_j\frac{S_{ij}}{\sqrt{S_{1j}}}|j\rangle\rangle,
\end{align}
where $S_{ij}$ is the modular S-matrix (Tab.~\ref{tab:S_matrix})~\cite{Cardy:1989ir}. Therefore, for each W-primary $i$, we can associate it with a boundary state $\ket{\tilde{i}}$ satisfying the consistency conditions by construction.

It is straightforward to identify these boundary states with those we obtained in Sec.~\ref{sec:bs_orbifold}, using the form of the partition function with a fixed boundary condition applied on the left-hand side:
\begin{align}
    Z_{1i}(q)=\chi_i=Z_{A-\mathcal{B}_i},
\end{align}
where $\mathcal{B}_i$ is the boundary condition corresponding to the primary field $i$. The resulting identifications are
\begin{align}
    &\ket{\tilde{1}}:A,\quad \ket{\tilde{j}_{1,2,3}}:B,C,D,\quad \ket{\tilde{\phi}}:\text{Free},\quad\ket{\tilde{\sigma}_{1,2,3}}:CD,BD,BC,\quad \ket{\tilde{\tau}_{1,2,3}}:AB,AC,AD.
\end{align}
Therefore, the boundary states we obtained from the Dirichlet and Neumann boundary conditions through the $\mathbb{Z}_2$-orbifold approach exhaust all the W-symmetry preserving boundary conditions of the four-state Potts model.
\subsection{W-symmetry breaking boundary conditions in the \texorpdfstring{$\mathrm{SU}(2)/D_2$}{SU(2)/D2} orbifold}\label{sec:SU(2)/D_2}
In this section, we further consider boundary conditions that explicitly break the enhanced $\mathrm{W}$-symmetry, yet can be constructed from the $\mathrm{SU}(2)$-preserving boundary states of the parent theory. We briefly review this construction for the $\mathrm{SU}(2)_1/D_2$ WZW model, which was first given in Ref.~\cite{ishikawaTwistedBoundaryStates2003}. Generic boundary states in this formalism are given by symmetrizing the original $\mathrm{SU(2)}$ boundary states, which correspond to the orbifold action. The $\mathrm{SU}(2)_1$ WZW model is equivalent to the $S_1$ compactified boson with $r=1$. Due to the $\mathrm{SU}(2)$ symmetry, its conformal boundary states are parametrized by the group element $g\in\mathrm{SU(2)}$:
\begin{align}
    \left.|\cmg\right\rangle=2^{-1 / 4} \sum_{j, m, n} D_{m, n}^j(\cmg)|j ; m, n\rangle\rangle,
\end{align}
where $ D_{m, n}^j(\cmg)$ is the Wigner D-matrix\cite{sakuraiModernQuantumMechanics2017} and $|j ; m, n\rangle\rangle$ are the $\mathrm{SU}(2)$-Ishibashi states for the degenerate representation $(j;m,n)$ of the Virasoro algebra with $c=1$~\cite{gaberdielConformalBoundaryStates2001,gaberdielConformalBoundaryStates2002}. These Ishibashi states are normalized as follows:
\begin{align}
    \langle\langle j ; m, n| q^{H_c} \mid j^{\prime} ; m^{\prime}, n^{\prime}\rangle\rangle=\delta_{j j^{\prime}} \delta_{m m^{\prime}} \delta_{n n^{\prime}} \chi_j(q), \quad \chi_j(q)=\frac{1}{\eta(q)}\left(q^{j^2}-q^{(j+1)^2}\right).
\end{align}
The $D_2$ orbifold of $\mathrm{SU}(2)$ is defined by the identification under a $D_2$ action:
\begin{align}
    \cmg\sim\gamma \cmg\gamma^{-1},\quad \gamma\in \Gamma,
\end{align}
where $\Gamma\subset \mathrm{SU}(2)/\{1,-1\}$ is a subgroup of $\mathrm{SU}(2)$ and contains four elements. In the fundamental representation of $\mathrm{SU}(2)$, these four elements can be written as
\begin{align}
    \Gamma=\{1,\gamma_1=i\sigma_1,\gamma_2=i\sigma_2,\gamma_3=i\sigma_3\}.
\end{align}
As in the $\mathbb{Z}_2$ orbifold of the compactified boson theory, generic boundary states of this orbifold theory can be obtained by symmetrizing the $\mathrm{SU}(2)$ boundary states:
\begin{align}
    |\cmg\rangle_{\Gamma}=\frac{1}{\sqrt{|\Gamma|}} \sum_{\gamma \in \Gamma}\left|\gamma \cmg \gamma^{-1}\right\rangle, \quad \cmg \in \mathrm{SU}(2)
\end{align}
and then resolve the degeneracy at the fixed points:
\begin{align}
& F_1=\left\{\left.f_1(\tilde{\theta}) :=\left(\begin{array}{cc}
\cos \tilde{\theta} & i \sin \tilde{\theta} \\
i \sin \tilde{\theta} & \cos \tilde{\theta}
\end{array}\right) \right\rvert\, 0 \leq \tilde{\theta}<2 \pi\right\}, \\
& F_2=\left\{\left.f_2(\theta) :=\left(\begin{array}{cc}
\cos \theta & \sin \theta \\
-\sin \theta & \cos \theta
\end{array}\right) \right\rvert\, 0 \leq \theta<2 \pi\right\}, \\
& F_3=\left\{\left.f_3(\alpha) :=\left(\begin{array}{cc}
e^{i \alpha} & 0 \\
0 & e^{-i \alpha}
\end{array}\right) \right\rvert\, 0 \leq \alpha<2 \pi\right\},\\
&1\equiv\left(\begin{array}{cc}
1 & 0 \\
0 & 1
\end{array}\right),\quad-1\equiv\left(\begin{array}{cc}
-1 & 0 \\
0 & -1
\end{array}\right),
\end{align}
where $F_a,\ a=1,2,3$, are fixed points under the action of the subgroups $\{1,\gamma_a\} \cong \mathbb{Z}_2$ and $\cmg=\pm1$ are fixed point of the whole group $\Gamma$. Like in the case of $\mathbb{Z}_2$-orbifold, these fixed points are resolved by additional Ishibashi states from the twisted sectors.

It turns out that all the boundary states that can be inherited from the generic Ashkin-Teller model through the $\mathbb{Z}_2$-orbifold formalism can be identified with boundary states of $\mathrm{SU}(2)/D_2$. To start, let us first take a look at the fixed points $F_3$. Boundary states along this line can be resolved by the $\mathrm{U}(1)$-Ishibashi states at $r_{sd}=1/\sqrt{2}$,
\begin{align}
    \left.\left|\mathcal{T}_3(n)\right\rangle\right\rangle \equiv \frac{1}{\sqrt{2}}\left(|\mathcal{D}\left(n+\frac{1}{2}\right)\rangle\rangle+\mathcal{D}\left(-n-\frac{1}{2}\right)\rangle\rangle\right) ,\quad\left(n \in \mathbb{Z}_{\geq 0}\right).
\end{align}
The $F_3$ fixed point can then be resolved by:
\begin{align}\label{eq:F1}
\left|f_3(\alpha)\right\rangle_{\Gamma} & .=\frac{1}{2}\left(\left|f_3(\alpha)\right\rangle+\left|f_3(-\alpha)\right\rangle\right)+2^{1 / 4}\left|\mathcal{T}_3[\alpha]\right\rangle\rangle, \\
\left|\mathcal{T}_3[\alpha]\right\rangle\rangle & =\sum_{n \in \mathbb{Z}_{\geq 0}} \cos \left[\left(n+\frac{1}{2}\right) \alpha\right]|\mathcal{T}_3(n)\rangle\rangle, \quad(0 \leq \alpha<2 \pi),
\end{align}
which can be shown to coincide with the Dirichlet boundary states in $S_1/\mathbb{Z}_2$ at $r=\sqrt{2}$:
\begin{align}
    \ket{f_3(\alpha)}_\Gamma=\ket{D_O(\frac{\alpha}{2})}.
\end{align}
For $F_1$ and $F_2$, the resolved boundary states can be constructed in a similar way to Eq.~\eqref{eq:F1}:
\begin{align}
\left|f_a(\beta)\right\rangle_{\Gamma} & =\frac{1}{2}\left(\left|f_a(\beta)\right\rangle+\left|f_a(-\beta)\right\rangle\right)+2^{1 / 4}\left|\mathcal{T}_a[\beta]\right\rangle\rangle, \\
\left.\left|\mathcal{T}_a[\beta]\right\rangle\right\rangle & =\sum_{n \in \mathbb{Z}_{\geq 0}} \cos \left[\left(n+\frac{1}{2}\right) \beta\right]|\mathcal{T}_a(n)\rangle\rangle, \quad(0 \leq \beta<2 \pi),
\end{align}
where the Ishibashi states $\ket{\mathcal{T}_1(n)}\rangle$ and $\ket{\mathcal{T}_2(n)}\rangle$ are constructed by the Virasoro Ishibashi state from the twisted sector of $X=-X$. All these Ishibashi states, including $\ket{\mathcal{T}_3(n)}\rangle$, satisfy the orthogonal relation:
\begin{align}
    \langle\langle\mathcal{T}_a(m)| q^{H_c} | \mathcal{T}_b(n)\rangle\rangle=\delta_{a b} \delta_{m n} \frac{1}{\eta(q)} q^{\frac{1}{4}\left(n+\frac{1}{2}\right)^2}.
\end{align}
Among these states, we can find the four twisted Neumann boundary states:
\begin{align}
    \ket{f_1\left(\frac{\pi}{2}/\frac{3\pi}{2}\right)}_{\Gamma}=\ket{N_O(\pi)\pm},\quad\ket{f_2\left(\frac{\pi}{2}/\frac{3\pi}{2}\right)}_{\Gamma}=\ket{N_O(0)\pm}.
\end{align}
The $\mathbb{Z}_2$ Neumann boundary states are actually the line connecting these four points, two in a group:
\begin{align}
    \ket{N_O(\alpha)}=\ket{\cmg=\left(\begin{array}{cc}
0 & e^{i \alpha/2} \\
e^{-i \alpha/2} & 0
\end{array}\right)}_\Gamma,\quad \alpha\in(0,\pi).
\end{align}

\begin{figure}[tb]
  \centering
\includegraphics[keepaspectratio, width=86mm]{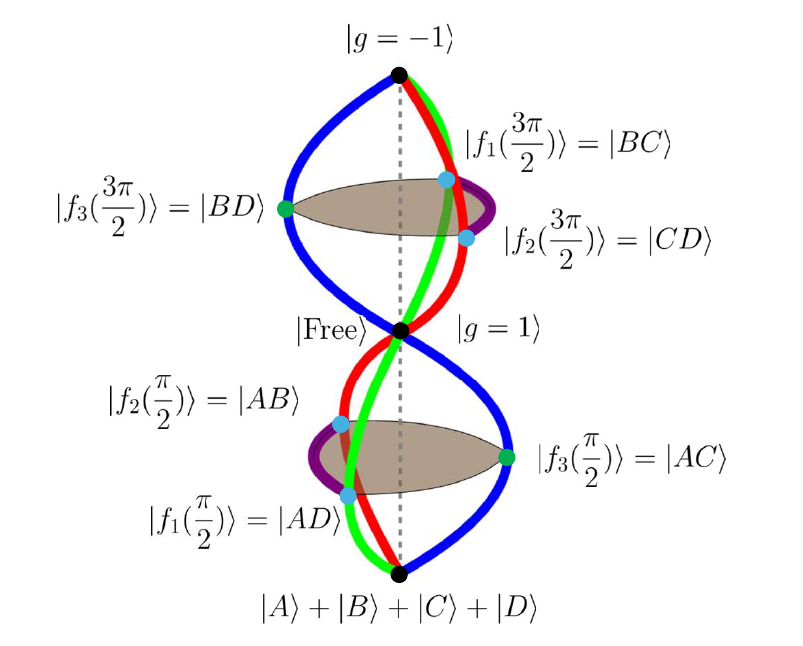}
  \caption{Illustration of the moduli space of boundary states in $\mathrm{SU}(2)/D_2$ model. Green, red, and blue lines are lines of fixed points $f_{1,2,3}$, respectively. Blob boundary conditions we identified in Sec.~\ref{sec:bs_orbifold} are denoted by separate (blue, green, and black) points. Note that in the orbifold space, the upper and lower halves of the three lines are identified by parity in this figure. Purple lines are also identified and correspond to the Neumann boundary states. The interiors of the upper and lower triangular spindles are not identified, although the boundary surfaces of them are identified through the parity.}
  \label{fig:SU(2)/D2}
\end{figure}

At $g=1$, the boundary states are resolved by linear combinations of these Ishibashi states:
\begin{align}
|D_O(0)\pm\rangle & =\frac{1}{2}|\cmg=1\rangle+2^{1 / 4} \frac{1}{2}\left[|\mathcal{T}_3[0]\rangle\rangle \pm\left(\left|\mathcal{T}_2[0]\right\rangle\right\rangle+\left|\mathcal{T}_1[0]\rangle\rangle\right)\right],\notag \\
|D_O(\pi)\pm\rangle & =\frac{1}{2}|\cmg=1\rangle+2^{1 / 4} \frac{1}{2}\left[-\left|\mathcal{T}_3[0]\rangle\rangle \pm\left(\left|\mathcal{T}_2[0]\right\rangle\right\rangle-\left|\mathcal{T}_1[0]\right\rangle\rangle\right)\right],
\end{align}
which correspond to fixed boundary conditions. On the other hand, the $g=-1$ point has no resolution and just corresponds to the midpoint Dirichlet boundary state:
\begin{align}
    \ket{D_O(\frac{\pi}{2})}=\ket{\cmg=-1}.
\end{align}
One can then assign the blob boundary states we find in the four-state Potts model to these $\mathrm{SU}(2)/D_2$-orbifold boundary states according to this. We illustrate this identification by Fig.~\ref{fig:SU(2)/D2}. For these boundary conditions, the partition functions are the same as those we obtained in Sec.~\ref{sec:D4}.

For those not in the region of the fixed point, the general partition functions among them are given as a combination of partition functions due to the symmetrizing:
\begin{align}
    Z_{\cmg_1-\cmg_2}=\bra{\cmg_1}_\Gamma\tilde{q}^{H_p}\ket{\cmg_2}_\Gamma=\frac{1}{\eta(q)}\sum_{\rho\in\Gamma}\sum_{n=-\infty}^\infty q^{(n+\frac{\alpha_\rho}{2\pi})^2},
\end{align}
where the phase $\alpha_\rho$ is given as
\begin{align}
    2\cos\alpha_\rho=\mathrm{Tr}(\cmg_1^\dagger\rho \cmg_2\rho^{-1}),\quad \rho\in\{1,\gamma_1,\gamma_2,\gamma_3\}
\end{align}
in the fundamental representation. The partition function between a generic orbifold boundary condition and a fixed point boundary condition is given by only considering the contribution from the $\mathrm{SU}(2)$ channel. For instance, for free BC:
\begin{align}
    Z_{\text{Free}-\cmg}=\bra{f_3(\pi)}_\Gamma\tilde{q}^{H_p}\ket{\cmg}_\Gamma=\bra{f_3(\pi)}\tilde{q}^{H_p}\ket{\cmg}_\Gamma=\frac{1}{\eta(q)}\sum_{n=-\infty}^\infty q^{(n+\frac{1}{2}+\frac{\alpha_\cmg}{2\pi})},
\end{align}
where $\alpha_\cmg$ is the phase of $\cmg$:~$ 2\cos(\alpha_\cmg)=\mathrm{Tr}(\cmg)$. For the diagonal partition function, there is always a non-changing part:
\begin{align}
    Z_{\cmg-\cmg}=\bra{\cmg}_\Gamma\tilde{q}^{H_p}\ket{\cmg }_\Gamma=\frac{1}{\eta(q)}\left[\sum_{n=-\infty}^\infty q^{n^2}+\sum_{\gamma=\gamma^{1,2,3}}\sum_{n=-\infty}^\infty q^{(n+\frac{\alpha_\gamma}{2\pi})^2}\right]
\end{align}
that contains three marginal operators. This is also consistent with the analysis made in boundary conformal perturbation theory, which says that the moduli space of boundary states in this model connected to the Dirichlet and Neumann boundary states has the following form~\cite{recknagelBoundaryDeformationTheory1999a}:
\begin{align}
    \hat{I}_{\sqrt{2}}\sqcup\hat{C}_{\frac{1}{2\sqrt{2}}}/\mathbb{Z}_2,
\end{align}
where $\hat{I}_{\sqrt{2}}$ denotes the disjoint union of an open interval corresponding to the Dirichlet line $\ket{D_O}$ ($\ket{f_3}$) and four points for the twisted boundary states $\ket{D_O(0/\pi r)\pm}$. The space $\hat{C}_{\frac{1}{2\sqrt{2}}}$ consists of five disjoint parts; one has the topology of an open ball, which is the interiors of the upper and lower triangular spindles connected by the parity in Fig.~\ref{fig:SU(2)/D2} containing Neumann line $\ket{N_O}$ (purple line), the four remaining components are open intervals that form a single circle (green and red lines ) were it not for the four Dirichlet-like points (black points).
\section{Boundary renormalization group flow}\label{sec:boundary_RG}

In this section, we discuss the boundary renormalization group (BRG) flows among
various conformally invariant boundary conditions, which correspond to the fixed points (and lines) 
of BRG. The BRG flows determine the global phase diagram of the boundary problem.
To understand the BRG flows, we analyze the boundary perturbations on the
conformally invariant boundary conditions.
The BRG flows are dictated by the scaling dimensions of boundary perturbations, depending
in particular, on whether the boundary perturbation is relevant, irrelevant, or marginal
under BRG.
The boundary perturbations are often realized in the lattice model as
boundary magnetic fields.
Thus, it is the first step towards understanding the BRG flows and the global phase diagram, to analyze the scaling of the boundary magnetic fields.

\subsection{Generic Ashkin-Teller model}\label{sec:RGflow_AT}
\subsubsection{Stability of the Free boundary condition}
\label{sec:RG_Free_AT}

Here we focus on the generic Ashkin-Teller model with $0 \leq \lambda<1$,
which is described by the $\mathbb{Z}_2$ orbifold with the compactification radius $1 \leq r < \sqrt{2}$.
Let us first discuss the boundary field perturbations to the free boundary condition.
Including the longitudinal boundary fields on the right-most site $L$
to the Hamiltonian with the Free--Free boundary conditions~\eqref{eq:HFree-Free},
the perturbed Hamiltonian reads
\begin{align}
    H_\text{AT}^{\text{Free-}\{h^z_\sigma,h^z_\tau,h^z_{\sigma\tau}\}}=&- \sum_{j=1}^L\left(\sigma_j^x+\tau_j^x+\lambda \sigma_j^x \tau_j^x\right)\notag- \sum_{j=1}^{L-1}\left(\sigma_j^z \sigma_{j+1}^z+\tau_j^z \tau_{j+1}^z+\lambda \sigma_j^z \tau_j^z \sigma_{j+1}^z \tau_{j+1}^z\right)\\
    &+h^z_\sigma\sigma^z_L+h^z_\tau\tau^z_L+h^{z}_{\sigma\tau}\sigma^z_L\tau^z_L,
    \label{eq:H_boundary-fields}
\end{align}
where $\{h^z_\sigma,h^z_\tau,h^z_{\sigma\tau}\}$ represent the magnitudes of the longitudinal boundary fields applied to the edge spins.

The scaling dimensions of the various boundary perturbations to the free boundary condition
can be read off from the diagonal partition function~\eqref{eq:Zfree-free}.
We see that there are always two relevant operators with the same scaling dimension $\frac{r^2}{2}$
and one marginal operator.
We can identify the two relevant operators with the boundary fields $h^z_\sigma$ and $h^z_\tau$, 
generalizing the identification at the decoupling point $\lambda=0$ ($r=1$) discussed in
Ref.~\cite{oshikawaBoundaryConformalField1996}.
Let us first analyze the BRG flows induced by these two boundary fields $h^z_\sigma$ and $h^z_\tau$,
setting $h^z_{\sigma\tau}=0$.
\begin{figure}[tb]
  \centering
\includegraphics[keepaspectratio, width=60mm]{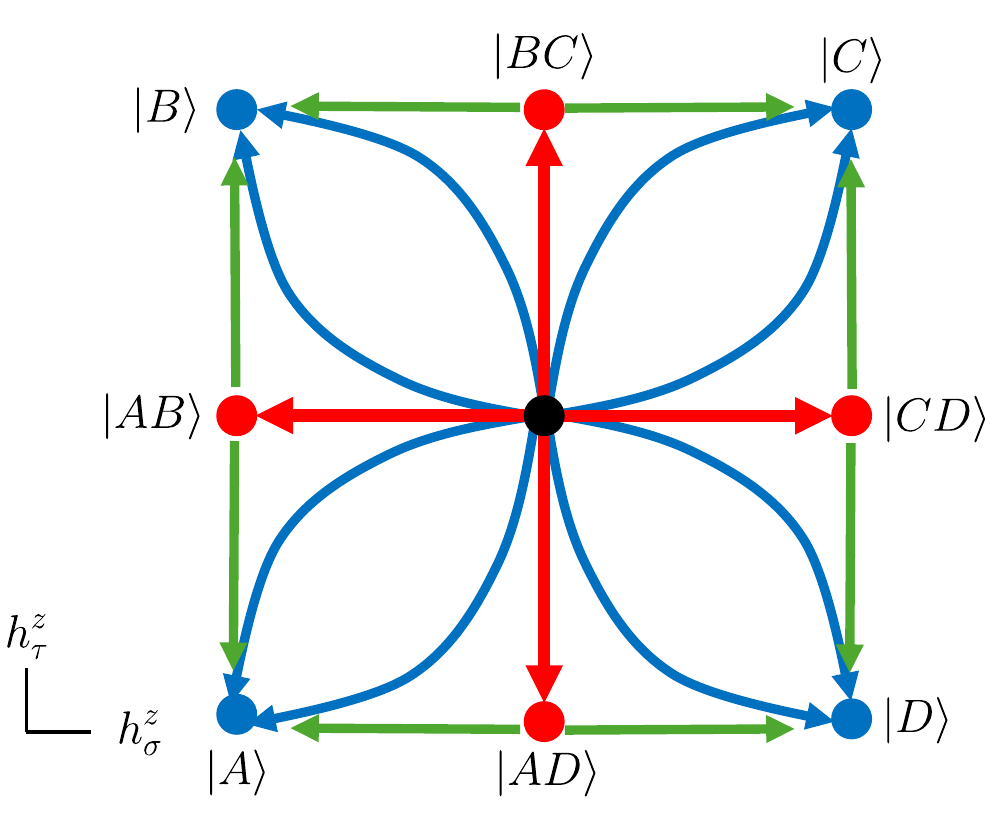}
  \caption{Illustration of boundary RG from free boundary condition to two-state mixed and fixed boundary conditions. The boundary fields $h_{\sigma}^z$ and $h_\tau^z$ break the corresponding $\mathbb{Z}_2$ spin-flip symmetry and induce a relevant boundary RG flow.}
  \label{fig:RG_1}
\end{figure}

The flow is then dictated by the interplay between $h^z_\sigma$ and $h^z_\tau$, governed by the breaking of the explicit $D_4$ symmetry of the Free BC.
When either of $h^z_\sigma$ and $h^z_\tau$ vanishes, the non-vanishing term breaks one of the $\mathbb{Z}_2$ spin-flip symmetries. So the boundary condition should flow to the corresponding
adjacent two-state mixed boundary condition (red lines in Fig.~\ref{fig:RG_1}).
These two directions should correspond to the two relevant perturbations due to the $\mathbb{Z}_2$ symmetry switching $\sigma$ and $\tau$ spins. When both of them are non-zero, the full $D_4$ symmetry is completely broken,
and the boundary condition flows to the fixed one corresponds to the magnetic field (blue lines in Fig.~\ref{fig:RG_1}). 

One can also discuss the BRG flow successively: first turn on one of the magnetic fields (say $h^z_\sigma$)
and then turn on the other one (say $h^z_\tau$) afterwards.
As discussed above, the first step drives the boundary condition to the adjacent two-state mixed boundary
condition (AB). 
We see from Eq.~\eqref{eq:AB-AB} that there is only one relevant boundary operator with scaling dimension $\frac{1}{2r^2}$ on a two-state mixed boundary condition, which has scaling dimension $\frac{1}{2r^2}$.
This can be identified with the boundary magnetic field $h^z_\tau$, which induces the BRG flow in the
second step (green lines in Fig.~\ref{fig:RG_1}). Since all the above perturbations are relevant, we see that the boundary entropy decreases strictly in this successive BRG flow:
\begin{align}
    g_\text{Free}>g_{AB}>g_{A}.
\end{align}

\subsubsection{Exactly marginal perturbation and continuous Dirichlet family of boundary conditions}
\label{sec:Dirichlet-family}

Now let us turn to the field $h^z_{\sigma\tau}$. 
As discussed for the decoupling point $\lambda=0$ ($r=1$), the operator $\sigma^z \tau^z$ corresponds
to the truly marginal perturbation which shifts the boundary value $\theta$.
Therefore, in the presence of the boundary field $h^z_{\sigma\tau}$ (only), 
the boundary state is the continous Dirichlet $\ket{D_O(\theta)}$ where
the boundary value $\theta$ is a function of $h^z_{\sigma\tau}$.

While we can deform the Dirichlet boundary condition by the marginal perturbation 
to tune $\theta$ to any value in $[0,\pi]$ in the BCFT,
we found that only a certain range of $\theta$ can be realized
by adding the boundary field $h^z_{\sigma\tau}\sigma^z_L\tau^z_L$
in the model~\eqref{eq:H_boundary-fields}, for $\lambda \neq 0$.

This is because the limits $\theta \to 0, \pi$ correspond to the spontaneous
symmetry breaking: the boundary states in these limits are given by a superposition of two
fixed boundary states, 
as we can see from Eqs.~\eqref{eq:DO_0_decomp} and \eqref{eq:fixed_DO}.
In other words, the limits $\theta \to 0, \pi$ are reachable only when
the quantum fluctuations at the boundary are completely suppressed.
This can be attained at the decoupling point $\lambda =0$ in the limit $h^z_{\sigma\tau}\to \pm \infty$,
as discussed in Ref.~\cite{oshikawaBoundaryConformalField1996}.
However, once $\lambda$ is turned on, the quantum fluctuations between the two low-energy
spin states chosen by the boundary field $h^z_{\sigma\tau}$ are induced,
even in the limit $h^z_{\sigma\tau}\to \pm \infty$. 
Our finding is that, in the model~\eqref{eq:H_boundary-fields} with $h^z_\sigma=h^z_\tau=0$,
the boundary condition is described by the Dirichlet boundary state $\ket{D_O(\theta)}$ of the
$\mathbb{Z}_2$ orbifold  with
\begin{align}
    \theta_c < \theta < \pi - \theta_c,
    \label{eq:theta-in-theta_c}
\end{align}
where
$\theta_c$ is defined in Eq.~\eqref{eq:theta_c-lambda} or \eqref{eq:theta_c-r},
for $-\infty < h^z_{\sigma\tau} < +\infty$.

In order to reach the region $\theta < \theta_c$ or $\pi-\theta_c < \theta$,
especially the limits $\theta \to 0, \pi$, we need to consider a different
lattice model in which we can completely suppress the quantum fluctuations
between the two spin states at the boundary.
We will discuss such a model in Sec.~\ref{sec:RG_AT_Neu}. Since $h_{\sigma\tau}^z$ introduces a truly marginal perturbation, the boundary entropy remains the same as the free boundary condition:
\begin{align}
    g_\text{Free}=g_{D_O(\theta)}.
\end{align}

In the case of the free boson, the boundary operator content of the Dirichlet boundary condition 
is independent of the boundary value, reflecting the U(1) symmetry of the theory.
In contrast, in the $\mathbb{Z}_2$ orbifold, the boundary operator content does depend on
the boundary value $\theta$, as the U(1) symmetry is broken in the orbifolding.
The boundary operator contents of the Dirichlet boundary condition can be read off from Eq.~\eqref{eq:AC-AC}.
We note that for all $r\in(1,\sqrt{2})$ there is always a marginal operator that shifts the boundary value $\theta$ along this Dirichlet line, and also a relevant operator with scaling dimension
$\min\left\{ \frac{2r^2\theta^2}{\pi^2}, \, 2r^2\left(1-\frac{\theta}{\pi}\right)^2 \right\}$.
The next leading boundary operator has the scaling dimension
$\max\left\{ \frac{2r^2\theta^2}{\pi^2}, \, 2r^2\left(1-\frac{\theta}{\pi}\right)^2 \right\}$.
This operator is relevant (scaling dimension smaller than $1$) when 
$\theta_p < \theta < \pi - \theta_p$, 
where 
\begin{align}
    \theta_p := \left(1-\frac{1}{\sqrt{2}r}\right)\pi .
\end{align}
This implies an interesting change of the BRG around the Dirichlet fixed line. 

As discussed in Sec.~\ref{sec:RG_Free_AT},
at the Free boundary condition $\theta=\pi/2$, there are two relevant boundary operators
with the same scaling dimension,
which are identified with the boundary fields $h^z_\sigma$ and $h^z_\tau$.
Once the field $h^z_{\sigma\tau}$ is turned on, we favour an antipodal pair of spin states,
say $AC$, over the other antipodal pair $BD$.
Now the two linear combinations of the boundary fields
$h^z_\sigma + h^z_\tau$ and $h^z_\sigma - h^z_\tau$ have different effects.
The former, if positive (negative), favors
the spin state $A$ over $C$ (or vice versa) between the two states already selected
by the field $h^z_{\sigma \tau}$,
while the latter favors $B$ or $D$ which were unfavored by the field $h^z_{\sigma\tau}$. 
We thus identify the former combination $h^z_\sigma + h^z_\tau$ with the most
relevant operator with the scaling dimension
$\min\left\{ \frac{2r^2\theta^2}{\pi^2}, \, 2r^2\left(1-\frac{\theta}{\pi}\right)^2 \right\}$.
The second relevant operator with the scaling dimension 
$\max\left\{ \frac{2r^2\theta^2}{\pi^2}, \, 2r^2\left(1-\frac{\theta}{\pi}\right)^2 \right\}$
is identified with the other combination $h^z_\sigma - h^z_\tau$.
When $h^z_{\sigma\tau}$ is sufficiently small, the boundary value $\theta$ is close to $\pi/2$.
In this regime, the second boundary field $h^z_\sigma-h^z_\tau$ is still relevant, driving the boundary condition to
a fixed spin state ($B$ or $D$), even though it was disfavored by the initial $h^z_{\sigma\tau}$ field.
However, when $h^z_{\sigma\tau}$ is large and $\theta < \theta_p$ (or $\theta > \pi - \theta_p$),
the second field $h^z_\sigma - h^z_\tau$ is irrelevant, implying that
an infinitesimal amount of such a field cannot override the preference imposed by
$h^z_{\sigma\tau}$.
It is natural to expect that, in the limit $h^z_{\sigma\tau} \to \pm \infty$, the boundary
spin is limited to an antipodal pair of states $AC$ or $BD$.
Indeed we find $\theta_c < \theta_p$, which is consistent with the physical expectation.
We note that our analysis here is consistent with that in Sec.~5.3 of Ref.~\cite{oshikawaBoundaryConformalField1996}
for the decoupling point $\lambda=0$.
The change in the BRG flow at $\theta=\theta_p$, which also occurs for the decoupled model with $\lambda=0$, 
was, however, not discussed in Ref.~\cite{oshikawaBoundaryConformalField1996}.

\begin{figure}[tb] 
    \centering 
    
    \begin{subfigure}{0.6\textwidth}
        \caption{}
        \includegraphics[width=\linewidth]{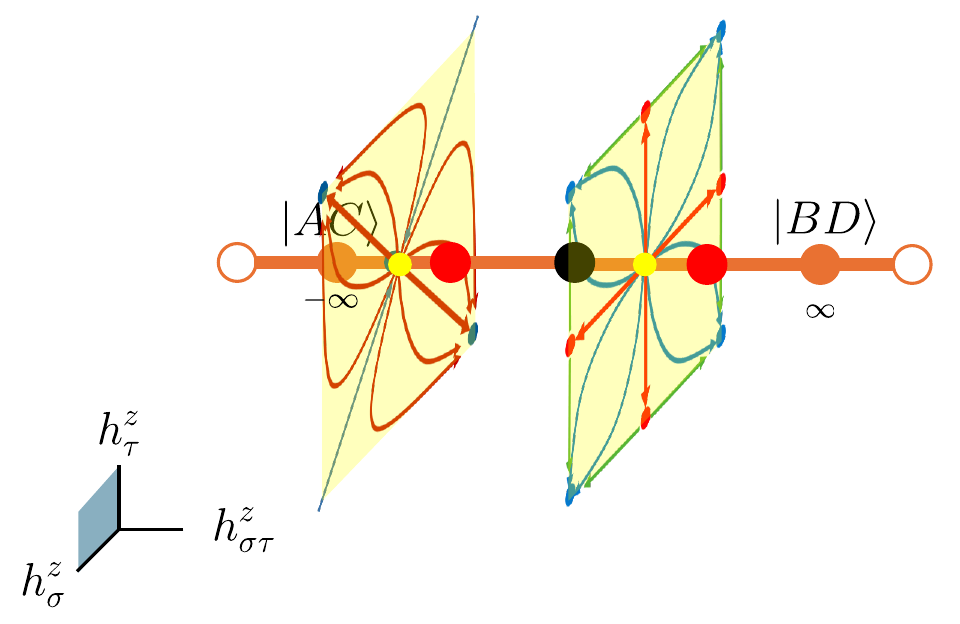} 
    \end{subfigure}
     \hfill
    \begin{subfigure}{0.38\textwidth} 
        \caption{}
        \includegraphics[width=\linewidth]{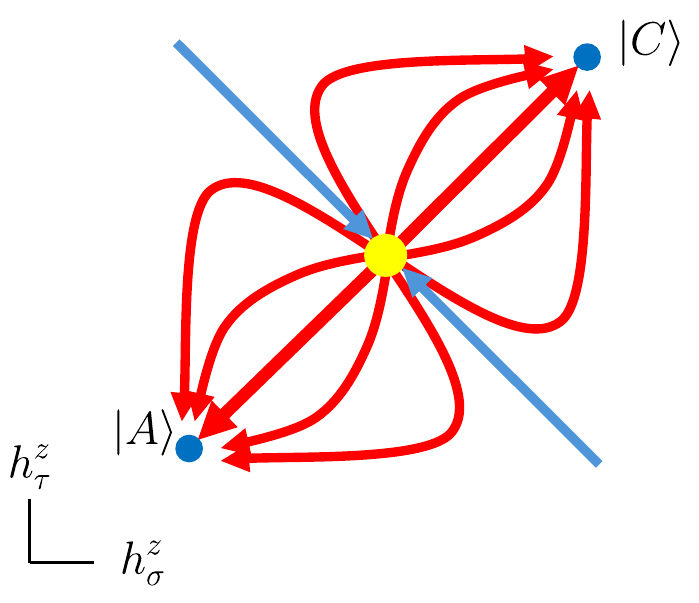} 
    \end{subfigure}

    \caption{Illustration of boundary RG from free boundary condition to antipodal two-state mixed and blob boundary conditions. The boundary field $h_{\sigma\tau}^z$ induces an integrable boundary perturbation towards $AC/BD$ boundary conditions. When this perturbation is small, the effect of the boundary fields $h_{\sigma}^z$ and $h_\tau^z$ resembles the case without this marginal perturbation (right yellow slice in (a)). Such behavior ends at a specific value (the red dots in (a)), as one of the relevant operators becomes irrelevant. Therefore, the boundary condition in the rest part should only flow to fixed boundary conditions, like shown in the left yellow slice in (a) and (b).}
    \label{fig:RG_2} 

\end{figure}

Summarizing our discussions, the BRG flows around the continuous Dirichlet family of the
boundary conditions realized by changing the value of the boundary field $h^z_{\sigma\tau}$
continuously, are picturized in Fig.~\ref{fig:RG_2}.

\subsubsection{Stability of the four-state SSB boundary condition}\label{sec:RG_AT_Neu}

As we have discussed, in the model~\eqref{eq:H_boundary-fields} with $h^z_\sigma=h^z_\tau=0$,
the Dirichlet boundary condition can be realized with the boundary value $\theta$
in the range $\theta_c < \theta < \pi - \theta_c$, owing to the quantum
fluctuations induced by the term $-\lambda \sigma_L^x \tau_L^x$.
In order to reach the other regions of $\theta$ in the Dirichlet boundary conditions,
we need to suppress the quantum fluctuations at the boundary.
In particular, at $\theta=0$ or $\theta=\pi$, the quantum fluctuation between
$A$ and $C$ (or $B$ and $D$) should be completely suppressed, and 
the corresponding boundary state is given by the superposition
$\ket{A} + \ket{C}$ (or $\ket{B}+\ket{D}$).

Construction of a lattice model that can realize such boundary conditions
is also hinted by the KW duality.
Eq.~\eqref{eq:KW_DO_theta} shows that (a superposition of) the Dirichlet boundary states with
$0 < \theta < \theta_p$ and $\pi - \theta_p < \theta < \pi$ can be obtained
by the KW transformation of the Dirichlet boundary condition with $\theta_p < \theta < \pi - \theta_p$.

In the lattice model, this corresponds to the KW dual
of the Hamiltonian~\eqref{eq:H_boundary-fields} at $h^z_\sigma = h^z_\tau=0$ is given by
\begin{align}
    \widetilde{H^{\text{Free-}h_{\sigma\tau}^z}}=&- \sum_{j=1}^{L-1}\left(\tilde{\sigma}_j^x+\tilde{\tau}_j^x+\lambda \tilde{\sigma}_j^x \tilde{\tau}_j^x\right)- \sum_{j=1}^{L-1}\left(\tilde{\sigma}_j^z \tilde{\sigma}_{j+1}^z+\tilde{\tau}_j^z \tilde{\tau}_{j+1}^z+\lambda \tilde{\sigma}_j^z \tilde{\tau}_j^z \tilde{\sigma}_{j+1}^z \tilde{\tau}_{j+1}^z\right)\notag\\
    &-(\tilde{\sigma}_1^z+\tilde{\tau}_1^z+\lambda\tilde{\sigma}_1^z\tilde{\tau}_1^z)
    -h_{\sigma\tau}^z \tilde{\sigma}_L^x \tilde{\tau}_L^x .
\label{eq:dual_H_Free_hsigmatau}
\end{align}

In particular, at $h^z_{\sigma\tau}=0$, this is reduced to the KW dual~\eqref{eq:dual_FreeFree} of the
Hamiltonian with the Free boundary condition on both ends.
In this limit, the boundary state on the right end is the four-state SSB state, which
is the superposition of the 4 fixed boundary states~\eqref{eq:KW_free} or equivalently \eqref{eq:KW_free_DO}.
Turning on $h^z_{\sigma\tau}$ shifts the boundary value $\theta$ of the Dirichlet boundary condition,
which also shifts the dual side via Eq.~\eqref{eq:KW_DO_theta}.

So far, the boundary state of the dual model~\eqref{eq:dual_H_Free_hsigmatau}
is still a superposition of the two Dirichlet boundary states.
We can choose one of them by adding the field
$ - \tilde{h}^z_{\sigma\tau} \tilde{\sigma}_L^z \tilde{\tau}_L^z$ 
at the right end to the Hamiltonian~\eqref{eq:dual_H_Free_hsigmatau},
or applying a projection operator.

To summarize the above discussion, the continuous Dirichlet boundary condition, including the
SSB limits $\theta \to 0, \pi$, can be realized by the Hamiltonian
\begin{align}
    H_\text{AT}^{A-h_{\sigma\tau}^x}=&- \sum_{j=2}^{L-1}\left(\sigma_j^x+\tau_j^x+\lambda \sigma_j^x \tau_j^x\right)- \sum_{j=1}^{L-1}\left(\sigma_j^z \sigma_{j+1}^z+\tau_j^z \tau_{j+1}^z+\lambda \sigma_j^z \tau_j^z \sigma_{j+1}^z \tau_{j+1}^z\right)\notag\\
    &-(\sigma_1^z+\tau_1^z+\lambda\sigma_1^z\tau_1^z)
    -h^x_{\sigma\tau}\sigma^x_L\tau^x_L - \epsilon \sigma^z_L \tau^z_L,
       \label{eq:H_hx_sigmatau}
\end{align}
where $\epsilon$ is an arbitrary non-zero field which lifts the degeneracy.
The transverse field $h^x_{\sigma\tau}$ introduces the quantum fluctuation.
At $h^x_{\sigma\tau}=0$, the boundary exhibits a 2-fold SSB
represented by the boundary state $\ket{D_O(0)} = \ket{A}+\ket{C}$ or 
$\ket{D_O(\pi)} = \ket{B}+\ket{D}$, depending on the sign of $\epsilon$.
Turning on $h^x_{\sigma\tau}$ shifts the boundary value $\theta$ of the Dirichlet boundary condition
away from $0$, $\pi$.

In addition to the continuous Dirichlet family $\ket{D_O(\theta)}$,
there is another continuous family of conformally invariant boundary conditions
for the $\mathbb{Z}_2$ orbifold, namely the continuous Neumann family $\ket{N_O(\theta)}$.
It can also be realized by perturbing the four-state SSB boundary condition~\eqref{eq:KW_free}. 

We claim that the continuous Neumann family of boundary conditions is realized in the Hamiltonian:
\begin{align}
    H^{\text{Free-}N_O(\theta)}_\text{AT}=-& \sum_{j=1}^{L-1}\left(\sigma_j^x+\tau_j^x+\lambda \sigma_j^x \tau_j^x\right)\notag- \sum_{j=1}^{L-1}\left(\sigma_j^z \sigma_{j+1}^z+\tau_j^z \tau_{j+1}^z+\lambda \sigma_j^z \tau_j^z \sigma_{j+1}^z \tau_{j+1}^z\right)\\
    -&\sigma^x_L-h^N_{\theta}\sigma^z_L\tau^x_L,
    \label{eq:H_Free_Neumann}
\end{align}
At the decoupling point $\lambda=0$, this Hamiltonian is 
equivalent to Eq.~(196) of Ref.~\cite{oshikawaBoundaryConformalField1996}, after identifying
$h^N_{\theta} \leftrightarrow b$.
In Ref.~\cite{oshikawaBoundaryConformalField1996}, it was constructed by applying
the KW transformation to one side of the standard defect model in the transverse-field Ising chain.
While the standard defect is mapped to the continuous Dirichlet boundary condition of the doubled
Ising model (AT model at the decoupling point) by folding, the KW transformation on one side
changes the boundary condition to the continuous Neumann one.
In modern terminology, this corresponds to fusion with a KW duality defect~\cite{Aasen2016}.

The correspondence between the model~\eqref{eq:H_Free_Neumann} and the continuous family of
Neumann boundary conditions remain valid for the generic AT model, generalizing
Ref.~\cite{oshikawaBoundaryConformalField1996} on the decoupling point $\lambda=0$.
Without the last two terms in the Hamiltonian,
the boundary condition on the right is the four-state SSB one.
The boundary transverse field $-\sigma^x_L$ introduces quantum fluctuations between the two states $AD$ and $BC$,
which correspond to $\sigma^z=\pm 1$.
Thus, once the transverse field $-\sigma^x_L$ is turned on, the boundary condition flows to the superposition
\begin{align}
\ket{AD} + \ket{BC} = \ket{N_O(\pi)} .
\end{align}
Furthermore, the last term $-h^N_{\theta} \sigma^z_L \tau^x_L$ introduces a marginal perturbation
which continuously shifts the boundary value $\theta$ from $\pi$. We see that the BRG flows under this successive operation are also consistent with the g-theorem:
\begin{align}
    g_{A+B+C+D}\equiv 4g_A>g_{AD}+g_{BC}=g_{N_O(\theta)},
\end{align}
where the greater-than symbol corresponds to the relevant perturbation by adding the quantum fluctuation $\sigma^x$.
\subsection{Four-state Potts model}\label{sec:RGflow_four-state}
\subsubsection{Stability of the free boundary condition}\label{sec:RGflow_four-state_3-state}
We now turn to the specific case of the four-state Potts point. Again, start with the free boundary condition. As discussed in the previous subsection, the scaling dimensions of the boundary operators depend on the compactification radius. At the four-state Potts point, the two previously relevant operators in the generic AT model become truly marginal (with scaling dimension $\Delta=1$). Consequently, instead of triggering an immediate flow to a stable fixed point, these operators generate continuous deformations of the boundary condition. This marginality implies the existence of three one-dimensional continuous families of boundary states connected to the Free boundary condition. In the context of the global phase diagram, the Free boundary condition sits at the intersection (or midpoint) of these lines. We can identify these three lines with the fixed point lines $F_{1,2,3}$ in the boundary state space of the $\mathrm{SU}(2)_1/D_2$ WZW model, as detailed in Sec.~\ref{sec:SU(2)/D_2}. Physically, these three marginal operators correspond to the (anti-)holomorphic components of the three bulk chiral primaries $J^a$ (for $a=1,2,3$).The resulting phase diagram along these directions resembles the structure shown in Fig.~\ref{fig:RG_2}. It is straightforward to see that once the system deviates from the high-symmetry Free point along these marginal lines, the stability changes. Specifically, as one moves away from the Free point, a single relevant operator emerges (corresponding to $\theta_p=\frac{\pi}{2}$ for the four-state Potts model), which governs the subsequent flow.

Since the symmetry of the four-state Potts model is lifted to $S_4$, one can further realize the three-state mixed boundary conditions in addition to those blob boundary conditions we defined in the generic Ashkin-Teller model. To study such boundary conditions, it turns out we can again start from the most general case:
\begin{align}\label{eq:H_Potts_pert}
     H_\text{Potts}=&- \sum_{j=1}^L\left(\sigma_j^x+\tau_j^x+\sigma_j^x \tau_j^x\right)\notag- \sum_{j=1}^{L-1}\left(\sigma_j^z \sigma_{j+1}^z+\tau_j^z \tau_{j+1}^z+\sigma_j^z \tau_j^z \sigma_{j+1}^z \tau_{j+1}^z\right)\\
    &+h^z_\sigma\sigma^z_L+h^z_\tau\tau^z_L+h^{z}_{\sigma\tau}\sigma^z_L\tau^z_L.
\end{align}
In the scaling limit, we may study such a boundary perturbation through the conformal perturbation theory. The corresponding Hamiltonian reads
\begin{align}
    H=H_\text{Potts}^\text{Free}+\sum_{a=1,2,3}\lambda^aJ^a,
\end{align}
where $H_\text{Potts}^\text{Free}$ is the unperturbed BCFT Hamiltonian with free boundary conditions on both sides, and we used the facts that the three boundary perturbation corresponds to the three chiral primaries in the theory. The fate of the boundary condition with such a perturbation can be studied by the following boundary RG equations:
\begin{align}
    \beta(\lambda^a)=(1-h^a)\lambda^a+C_{abc}\lambda^b\lambda^c+O(\lambda^3),
\end{align}
where $\beta(\lambda^a)=\frac{\partial \lambda^a}{\partial l}$ is the beta function, $h^a$ is the conformal dimension of $J^a$ and $C_{abc}$ is the coefficient of the three-point function $\langle J^aJ^bJ^c \rangle$. Since all three operators are marginal, the first term vanishes. It turns out the only non-vanishing coefficient is $C:=C_{123}$ due to the fusion rule $J^a\times J^b=\varepsilon_{abc}J^c$. Therefore, to the leading order, the BRG equations are
\begin{align}
     \left\{\begin{matrix}\beta(\lambda^1)=C\lambda^2\lambda^3,
 \\\beta(\lambda^2)=C\lambda^3\lambda^1,
 \\\beta(\lambda^3)=C\lambda^1\lambda^2.
\end{matrix}\right.
\end{align}
We see that these BRG equations resemble the bulk RG flow of the $\mathrm{SU}(2)$ Heisenberg model due to the similar symmetry. The only fixed points of these equations are those with only one non-vanishing coupling constant, which corresponds to the truly marginal flow of boundary conditions discussed in the last paragraph. For most of the rest cases, the RG trajectories flow towards four stable fixed points at infinity, forming a tetrahedron structure in the parameter space. These correspond to the asymptotes:
\begin{align}
    \vec{\lambda} \propto (1, 1, 1), \quad (1, -1, -1), \quad (-1, 1, -1), \quad\text{and} \quad  (-1, -1, 1),\notag
\end{align}
which simply corresponds to the flow to the four fixed boundary conditions $A,B,C,D$. The only attractive flows to the free boundary conditions occur along four discrete rays in the parameter space, defined by the proportionality:
\begin{align}
    \vec{\lambda} \propto (-1, -1, -1), \quad (-1, 1, 1), \quad (1, -1, 1), \quad \text{and} \quad (1, 1, -1)\notag
\end{align}
which corresponds to the perturbation towards the three-state mixed boundary conditions $ABC,BCD,ACD$ and $ABD$. This suggests that at leading order, boundary perturbations toward the three-state mixed spin direction should be marginally irrelevant. 

Moreover, these three-state mixed boundary conditions, if they exist as different fixed points at IR, shall exhibit very exotic behavior. If one considers the boundary RG flow from the free boundary condition to the three-state mixed boundary condition and then the two-state mixed boundary condition (e.g., $\text{Free}\to ABC\to AB$), the $g$-theorem requires that the $g$-factor should monotonically decrease or remain unchanged. However, since the free boundary condition is connected to the AB boundary condition in the four-state Potts model through a marginal perturbation, they must have the same $g$-factor value, which requires that the three-state mixed boundary condition should also have the same $g$-factor value as an intermediate state! At first glance, this implies that the flows from the Free boundary condition to both the three-state and two-state mixed boundary conditions must be truly marginal. However, we have already shown that for a generic boundary condition that deviates from the free boundary condition on the continuous line, there is only one marginal operator, so this could not be the case. As we will also present in the numerical section, these three-state mixed BCs, if they are distinct from the other blob BCs, will break Cardy's consistency law and fail to be consistent conformally invariant boundary conditions. We therefore claim that such a perturbation, which breaks the $S_4$ symmetry to $S_3$, will not induce a flow to a new fixed point at the IR.

\subsubsection{Stability of the four-state SSB boundary condition}\label{sec:RGflow_four-state_neumann}
At the four-state Potts point, the 1D line of Neumann boundary conditions is enriched by the $S_4$ symmetry, acquiring a 3-fold structure analogous to the continuous family of two-state mixed boundary conditions. Through Kramers-Wannier duality, the three lines of Dirichlet-like boundary conditions map to three lines of SSB BCs, specifically $\ket{f_a(\alpha)}+\ket{f_a(2\pi-\alpha)}$, which all converge at the four-state SSB boundary condition $\ket{A}+\ket{B}+\ket{C}+\ket{D}$. As discussed in Sec.~\ref{sec:SU(2)/D_2}, the moduli space of boundary conditions connected to the Neumann boundary conditions via marginal perturbations expands into an open ball. To realize this full manifold on the lattice, we propose the following Hamiltonian:
\begin{align}
    H^{\text{Free-}N_O(\theta)}_\text{AT}=-& \sum_{j=1}^{L-1}\left(\sigma_j^x+\tau_j^x+\lambda \sigma_j^x \tau_j^x\right)\notag- \sum_{j=1}^{L-1}\left(\sigma_j^z \sigma_{j+1}^z+\tau_j^z \tau_{j+1}^z+\lambda \sigma_j^z \tau_j^z \sigma_{j+1}^z \tau_{j+1}^z\right)\\
    -&h_1\sigma^x_L-h_2\sigma^z_L\tau^x_L-h_3\sigma^z_L\sigma^x_L\tau^x_L,
\end{align}
where the three perturbations are tuned by $h_{1,2,3}>0$. The first two terms are identical to those used for the Neumann boundary conditions in the generic Ashkin-Teller model; however, at the Potts point, $\sigma^x$ becomes a marginal perturbation. Consequently, its coefficient $h_1$ now parameterizes the position within the moduli space rather than driving an RG flow. These two terms allow one to traverse the line of SSB boundary conditions towards $\ket{AB}+\ket{CD}$. At any point along this line, the second term generates the extension into the generic $\mathrm{SU}(2)/D_2$ boundary conditions. The third term is produced by a product of the first two terms, in analogy to the $\mathfrak{su}(2)$ algebra, which we argue is essential to cover the full 3D structure.

\section{Numerical results and discussions}\label{sec:FSS}
In this section, we test our theoretical predictions through numerical calculations on the finite-size spectra. We use an extended version of DMRG\cite{chepigaExcitationSpectrumDensity2017} by keeping track of the excitation spectra of the effective Hamiltonian as a function of DMRG iterations. We extract 6-15 low-lying states, in chains ranging from $L=24\sim 100$ sites. We use a two-site routine, limit the bond dimension to 1500, truncate singular values below $10^{-8}$, and perform up to 7 sweeps. 

Following Eq.~\eqref{eq:FSS},
the finite-size low-energy spectra after rescaling should reveal a specific structure called conformal towers:
\begin{align}\label{eq:FSS_num}
    (E_n^{ab}-E_0^{ab})\frac{L}{\pi v}=(x^{ab}_n-x^{ab}_0)+o(1),
\end{align}
which contains the full operator contents of BCCOs between $a$ and $b$. These scaling dimensions $x^{ab}_n$ should match those appearing in the corresponding partition functions we obtained in Sec.~\ref{sec:boundary_states}.

It is noted that, to obtain the best estimation of the scaling dimension $x_n$ by Eq.~\eqref{eq:FSS_num}, a finite-size extrapolation is essential. In our work, we assumed the expression of the leading-order finite-size correction to be

\begin{align}\label{eq:extrapo}
 (E_n^{ab}-E_0^{ab})\frac{L}{\pi v}=(x^{ab}_n-x^{ab}_0)+
 \begin{cases}
  \frac{A}{L} & \quad r\leq \frac{2}{\sqrt{3}} \\[6pt]
  \frac{A}{L^{(4-r^2)}} & \quad \frac{2}{\sqrt{3}}<r<\sqrt{2}\\[6pt]
  \frac{A}{\log L} & \quad r=\sqrt{2}
 \end{cases}\ ,
\end{align}
where $A$ is a non-universal coefficient of the finite-size corrections. The first two cases follow from the competition between the shift to the system size and first-order perturbation from the bulk irrelevant operator with scaling dimension $4-r^2$~\cite{10.21468/SciPostPhys.17.4.099}. And the last case corresponds to the logarithmic correction from the marginally irrelevant operator\cite{affleckLogarithmicCorrectionsAntiferromagnetic1990,affleckLogarithmicCorrectionsQuantum1999}. We will see that such extrapolation greatly enhanced the collapse of numerical data and theoretical predictions, and it is crucial in exploring exotic critical phenomena in the four-state Potts model.

\subsection{Generic Ashkin-Teller model}
In this subsection, to demonstrate the generality of our identification, we present calculations for a wide range of coupling constants: $\lambda \in \{0, 0.2, 0.4, 0.6, 0.7071, 0.8, 1\}$. However, for the detailed analysis in the subsequent sections (e.g., marginal RG flow), we will restrict our focus to three representative values: $\lambda=0.2$ (generic irrational), $\lambda \approx 0.7071$ ($\mathbb{Z}_4$ parafermion), and $\lambda=1$ (four-state Potts). These three points are sufficient to capture the distinct physics of both the generic and rational regimes.
\subsubsection{Blob boundary conditions}\label{sec:num_blob}
Blob boundary conditions are realized by the corresponding projected Hamiltonian defined in Sec.~\ref{sec:bs_orbifold}. For simplicity, we only present the cases where one of the boundaries is fixed to an $A$ boundary condition. To be explicit, we extract the conformal towers from the following Hamiltonian:
\begin{align}
    H_\text{AT}^{A-\text{blob}} =& - \sum_{j=2}^{L-1}\left(\sigma_j^x+\tau_j^x+\lambda \sigma_j^x \tau_j^x\right) - \sum_{j=2}^{L-2}\left(\sigma_j^z \sigma_{j+1}^z+\tau_j^z \tau_{j+1}^z+\lambda \sigma_j^z \tau_j^z \sigma_{j+1}^z \tau_{j+1}^z\right)\notag\\
    & - (\sigma_{L-1}^z+\tau_{L-1}^z+\sigma_{L-1}^z\tau_{L-1}^z) + h_1,
\end{align}
where $h_1$ represents the projected interaction involving the first site, defined as:
\begin{equation}
    h_1 := -P \left( \sigma^z_1\sigma_2^z+\tau^z_1\tau_2^z+\lambda\sigma^z_1\sigma_2^z\tau^z_1\tau_2^z + \sigma^x_1+\tau_1^x+\lambda\sigma_1^x\tau_1^x \right) P.
\end{equation}
The explicit forms of $h_1$ for all blob boundary conditions are provided in App.~\ref{app:con}.

\begin{figure}[p]
    \centering 
    
    \begin{subfigure}{0.32\textwidth}
        \caption{}
        \includegraphics[width=\linewidth,trim=10 25 5 15, clip]{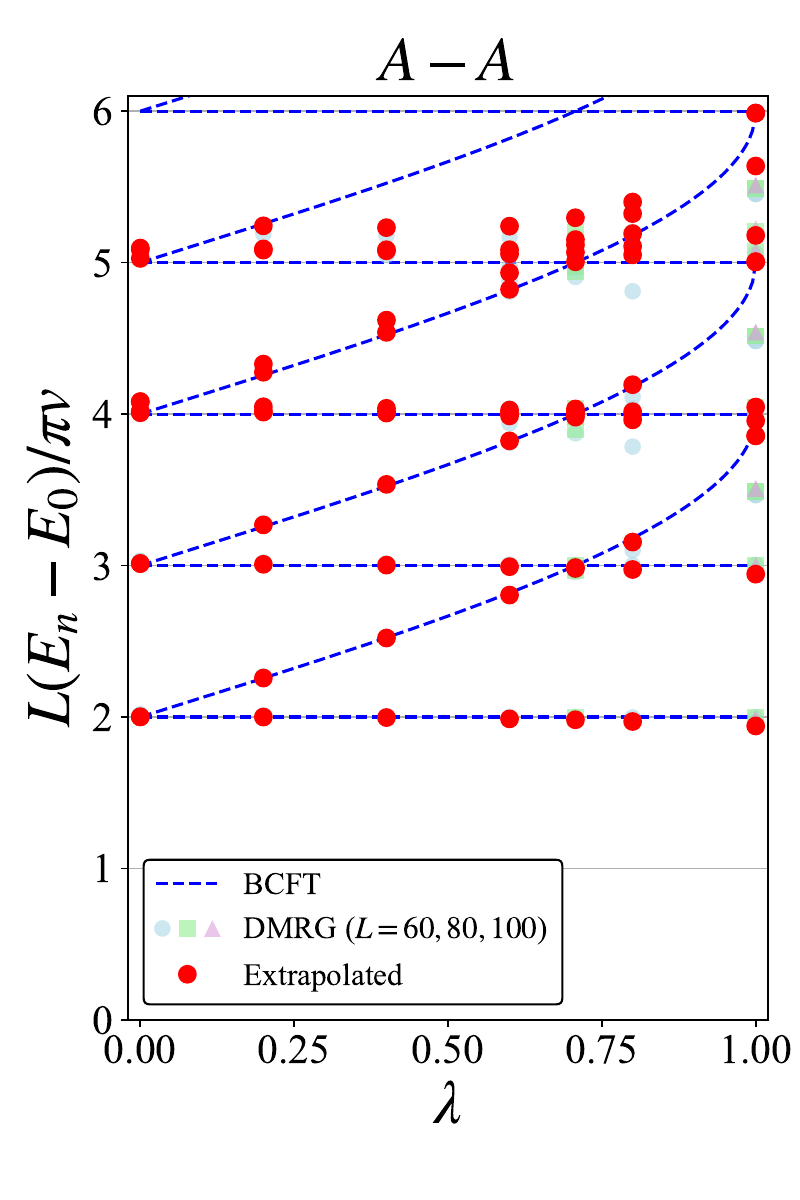} 
    \end{subfigure}
    \begin{subfigure}{0.32\textwidth} 
        \caption{}
        \includegraphics[width=\linewidth,trim=10 25 5 15, clip]{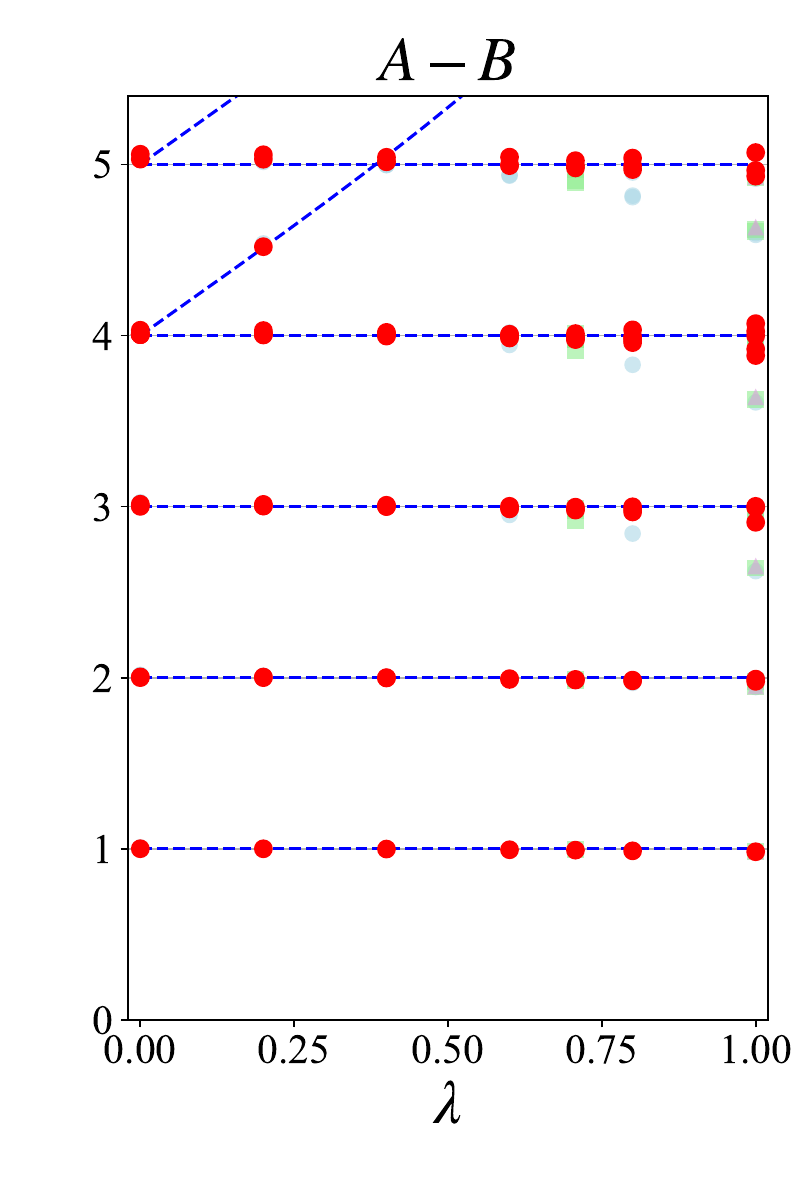} 
    \end{subfigure}
    \begin{subfigure}{0.32\textwidth} 
        \caption{}
        \includegraphics[width=\linewidth,trim=10 25 5 15, clip]{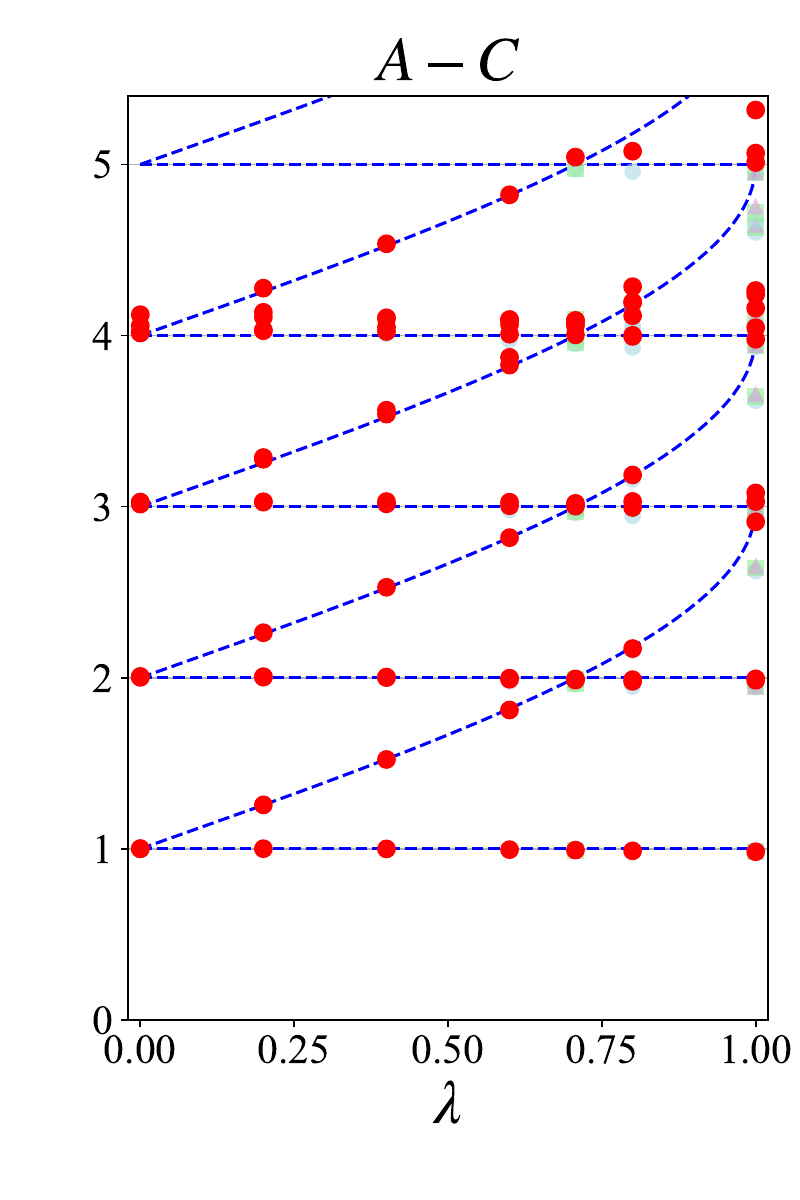} 
    \end{subfigure}

    \caption{Conformal towers for the Ashkin-Teller model under $A-A$ (a), $A-B$ (b) and $A-C$ (c) boundary conditions.
Theoretical predictions (blue dashed lines) are compared with $L\to\infty$ extrapolated data (red dots) and DMRG data (purple, green, and blue symbols).
The extrapolation (Eq.~\eqref{eq:extrapo}) uses system sizes $L=\{24,32,40,60\}$, with larger sets included for accuracy at $\lambda=0.7071$ ($L=80$) and $\lambda=1$ ($L=80,100$).
}
    \label{fig:blob_fixed} 
\end{figure}

\begin{figure}[p] 
    \centering 
    
    \begin{subfigure}{0.32\textwidth}
        \caption{}
        \includegraphics[width=\linewidth,trim=10 25 5 15, clip]{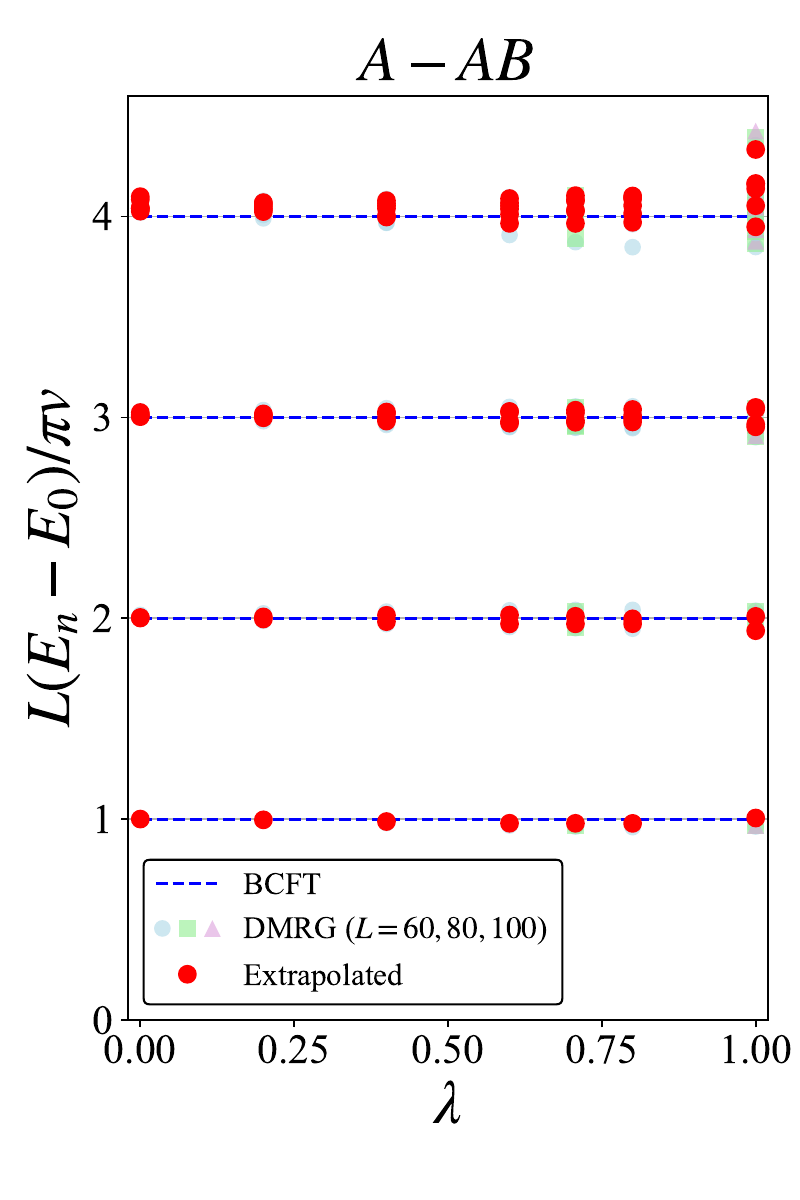} 
    \end{subfigure}
    \begin{subfigure}{0.32\textwidth} 
        \caption{}
        \includegraphics[width=\linewidth,trim=10 25 5 15, clip]{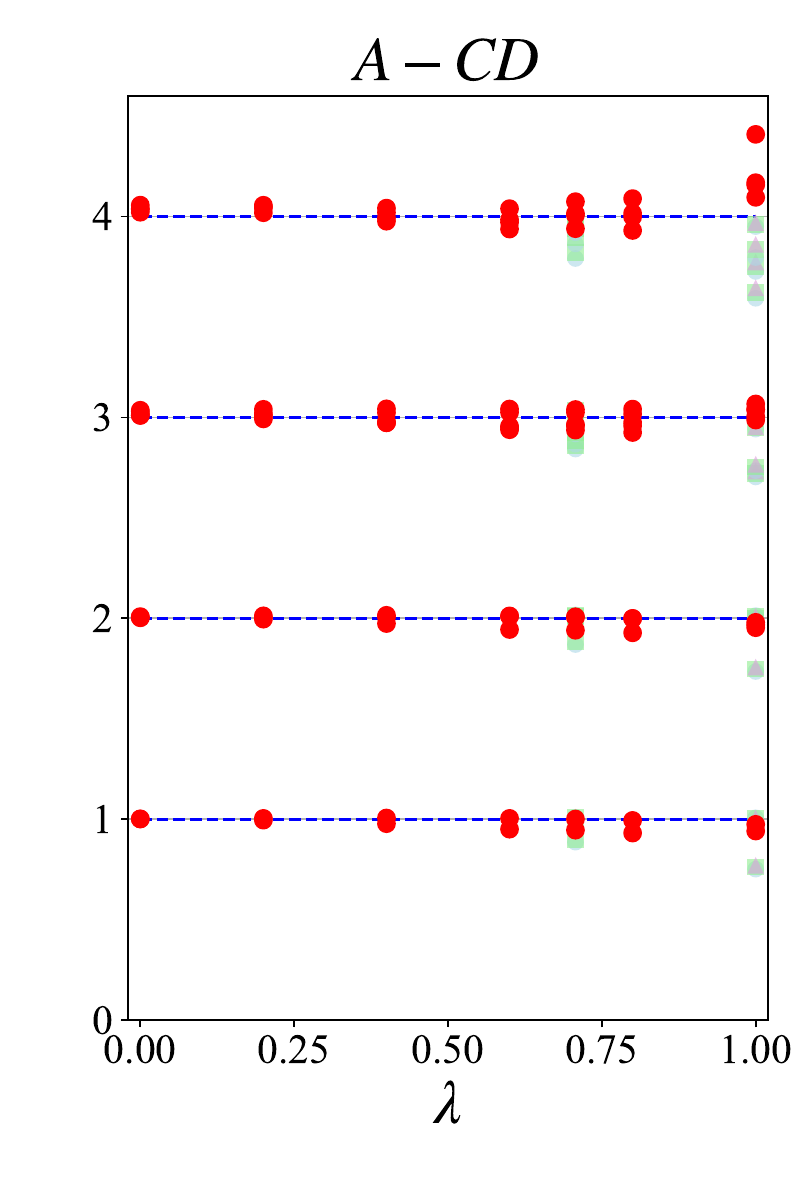} 
    \end{subfigure}
    \begin{subfigure}{0.32\textwidth} 
        \caption{}
        \includegraphics[width=\linewidth,trim=10 25 5 15, clip]{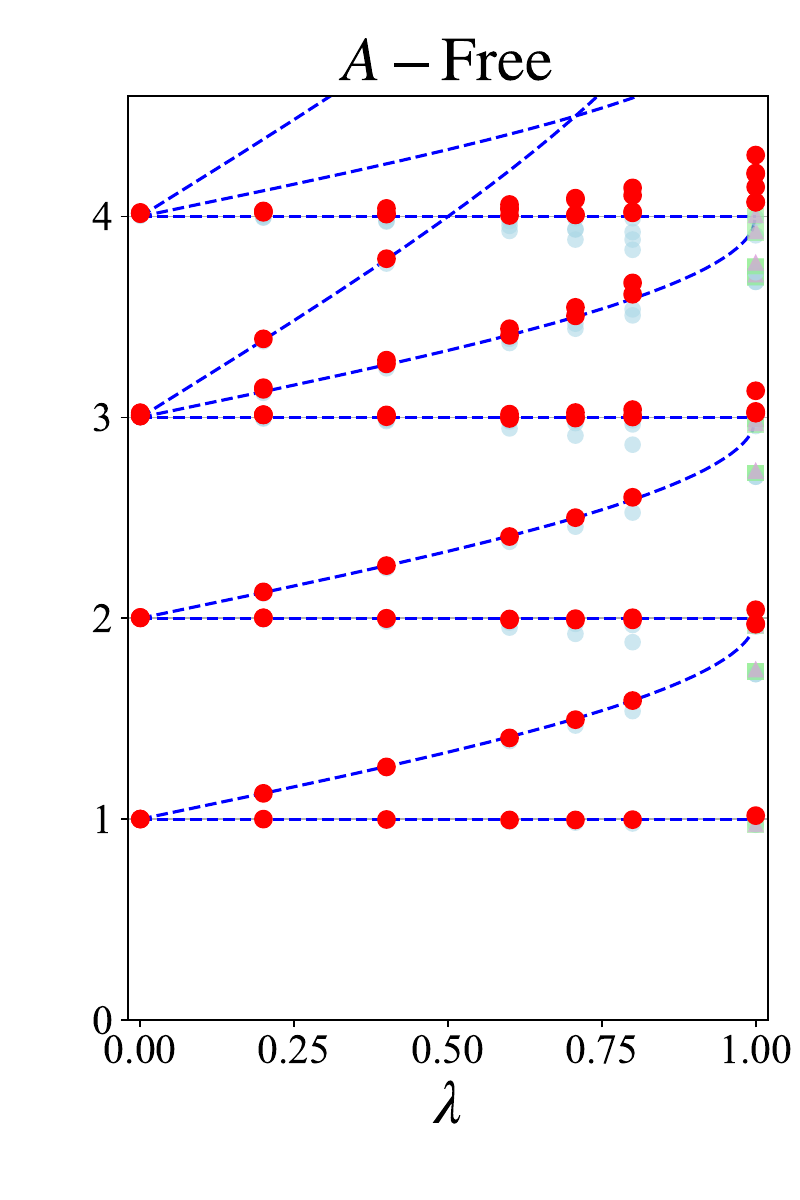} 
    \end{subfigure}
    \caption{Conformal towers for the Ashkin-Teller model under $A-AB$ (a), $A-CD$ (b) and $A-\text{Free}$ (c) boundary conditions.
Theoretical predictions (blue dashed lines) are compared with $L\to\infty$ extrapolated data (red dots) and DMRG data (purple, green, and blue symbols).
The extrapolation (Eq.~\eqref{eq:extrapo}) uses system sizes $L=\{24,32,40,60\}$, with larger sets included for accuracy at $\lambda=0.7071$ ($L=80$) and $\lambda=1$ ($L=80,100$).}
\label{fig:blob_two-state}
\end{figure}

\begin{figure}[tb] 
    \centering 
    \begin{subfigure}{0.32\textwidth}
        \caption{}
        \includegraphics[width=\linewidth,trim=10 25 5 15, clip]{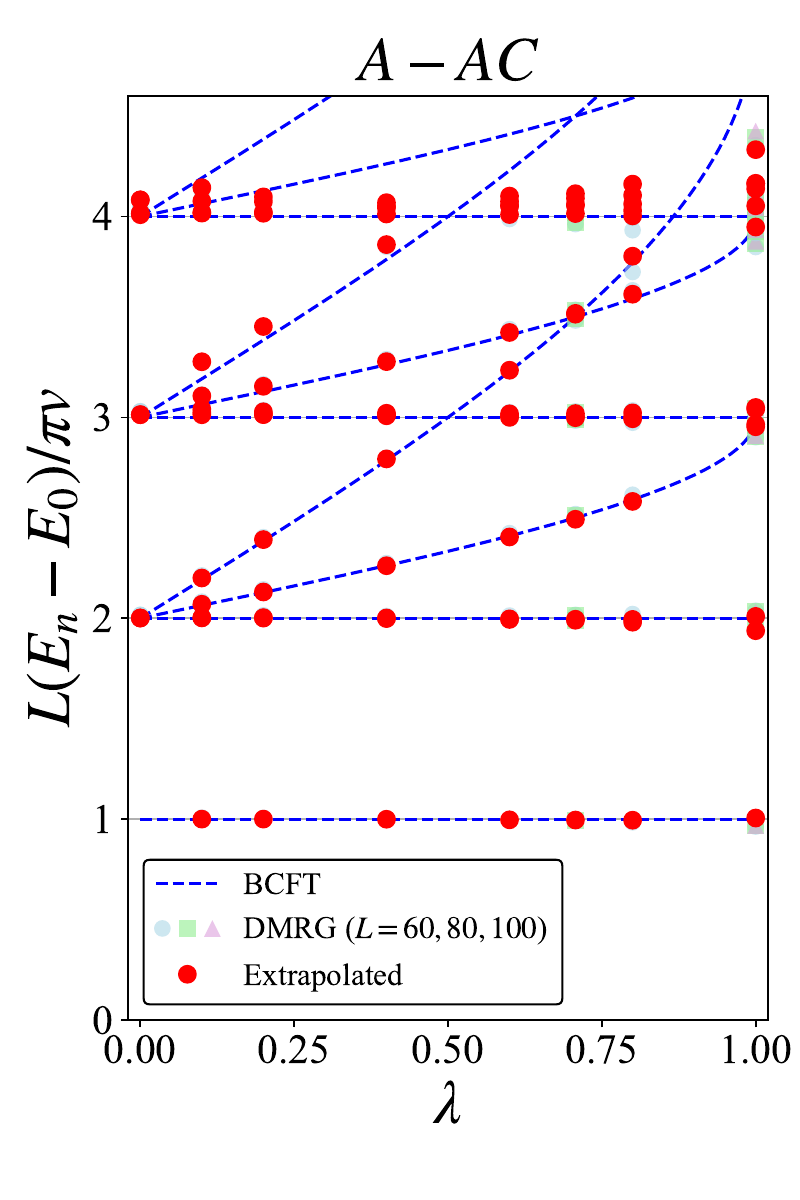} 
    \end{subfigure}
    \begin{subfigure}{0.32\textwidth} 
        \caption{}
        \includegraphics[width=\linewidth,trim=10 25 5 15, clip]{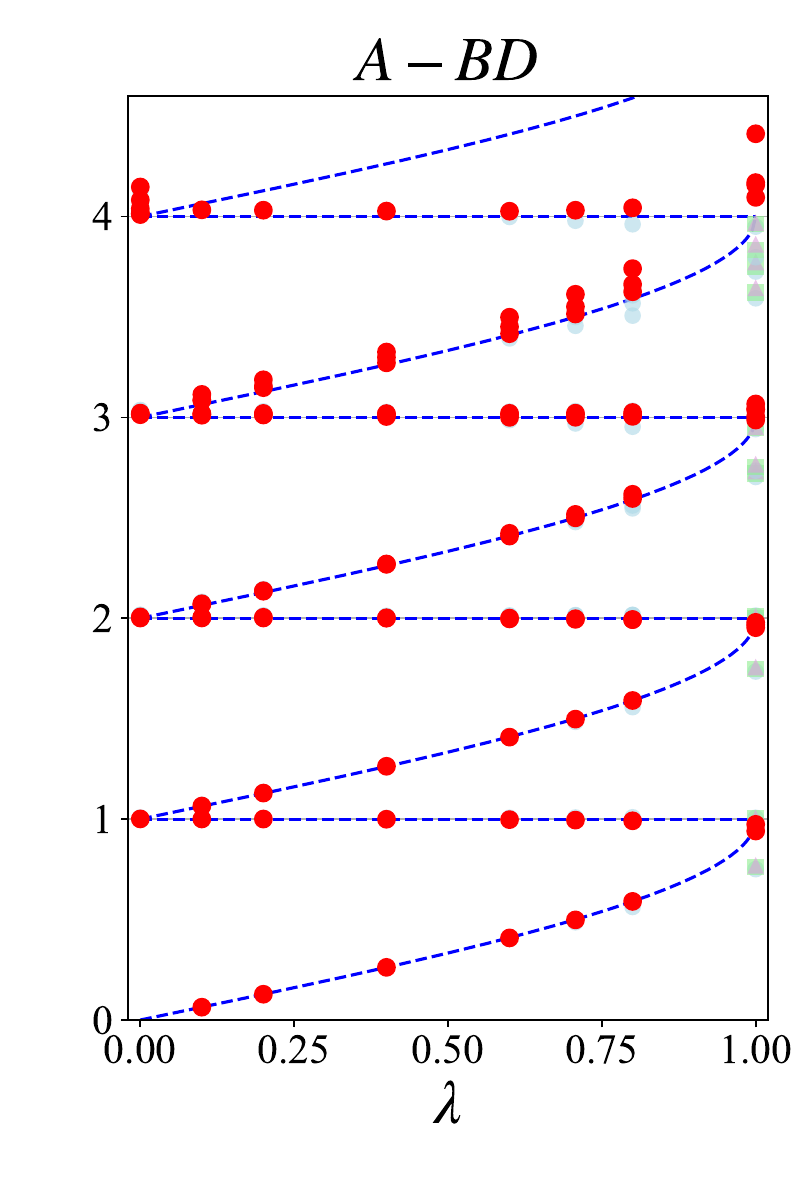} 
    \end{subfigure}
    \caption{Conformal towers for the Ashkin-Teller model under $A-AC$, and $A-BD$ boundary conditions.
Theoretical predictions (blue dashed lines) are compared with $L\to\infty$ extrapolated data (red dots) and DMRG data (purple, green, and blue symbols).
The extrapolation (Eq.~\eqref{eq:extrapo}) uses system sizes $L=\{24,32,40,60\}$, with larger sets included for accuracy at $\lambda=0.7071$ ($L=80$) and $\lambda=1$ ($L=80,100$).}
    \label{fig:blob_two-state-diag} 

\end{figure}

We present the results for fixed boundary conditions in Fig.~\ref{fig:blob_fixed}, two-state mixed and free boundary conditions in Fig.~\ref{fig:blob_two-state}, and antipodal two-state mixed boundary conditions in Fig.~\ref{fig:blob_two-state-diag}. We observe an excellent agreement between the BCFT predictions (blue dashed lines) and the extrapolated numerical results (red dots) for all blob boundary conditions defined in the generic Ashkin-Teller model ($\lambda\in[0,1]$, including the four-state Potts point). Note that for the region close to the four-state Potts point ($\lambda=1$), the conformal towers extracted from the maximal system size $L_\text{max}=100$ deviate significantly from the extrapolated data due to strong finite-size effects. This underscores the necessity of the extrapolation scheme~\eqref{eq:FSS_num} employed to demonstrate this agreement.

\subsubsection{Marginal deformation towards \texorpdfstring{$AC/BD$}{AC/BD} boundary conditions}

As we discussed in Sec.~\ref{sec:RGflow_AT},
the free boundary condition and the antipodal two-state mixed boundary conditions ($AC$ and $BD$)
belong to the continuous Dirichlet family of boundary conditions, connected by the exact
marginal deformation corresponding to the shift of the boundary value $\theta$.

We verify this numerically by studying the energy spectrum of the following Hamiltonian:
\begin{align}
     H_\text{AT}^{A-h_{\sigma\tau}^z}=&- \sum_{j=2}^{L}\left(\sigma_j^x+\tau_j^x+\lambda \sigma_j^x \tau_j^x\right)- \sum_{j=2}^{L-1}\left(\sigma_j^z \sigma_{j+1}^z+\tau_j^z \tau_{j+1}^z+\lambda \sigma_j^z \tau_j^z \sigma_{j+1}^z \tau_{j+1}^z\right)\notag\\
    &- (\sigma_{2}^z+\tau_{2}^z+\sigma_{2}^z\tau_{2}^z)-h_{\sigma\tau}^z \sigma_{L}^z \tau_{L}^z .
\end{align}
The left boundary condition is fixed to $A$ as in the previous section,
and the right boundary condition is the Dirichlet one $\ket{D_O(\theta)}$ 
with the boundary value $\theta$ tuned by the boundary field $h_{\sigma\tau}^z$.

\begin{figure}[tb]
    \centering 
    \begin{subfigure}{0.48\textwidth}
        \caption{}
        \includegraphics[width=\linewidth,trim=75 25 20 25, clip]{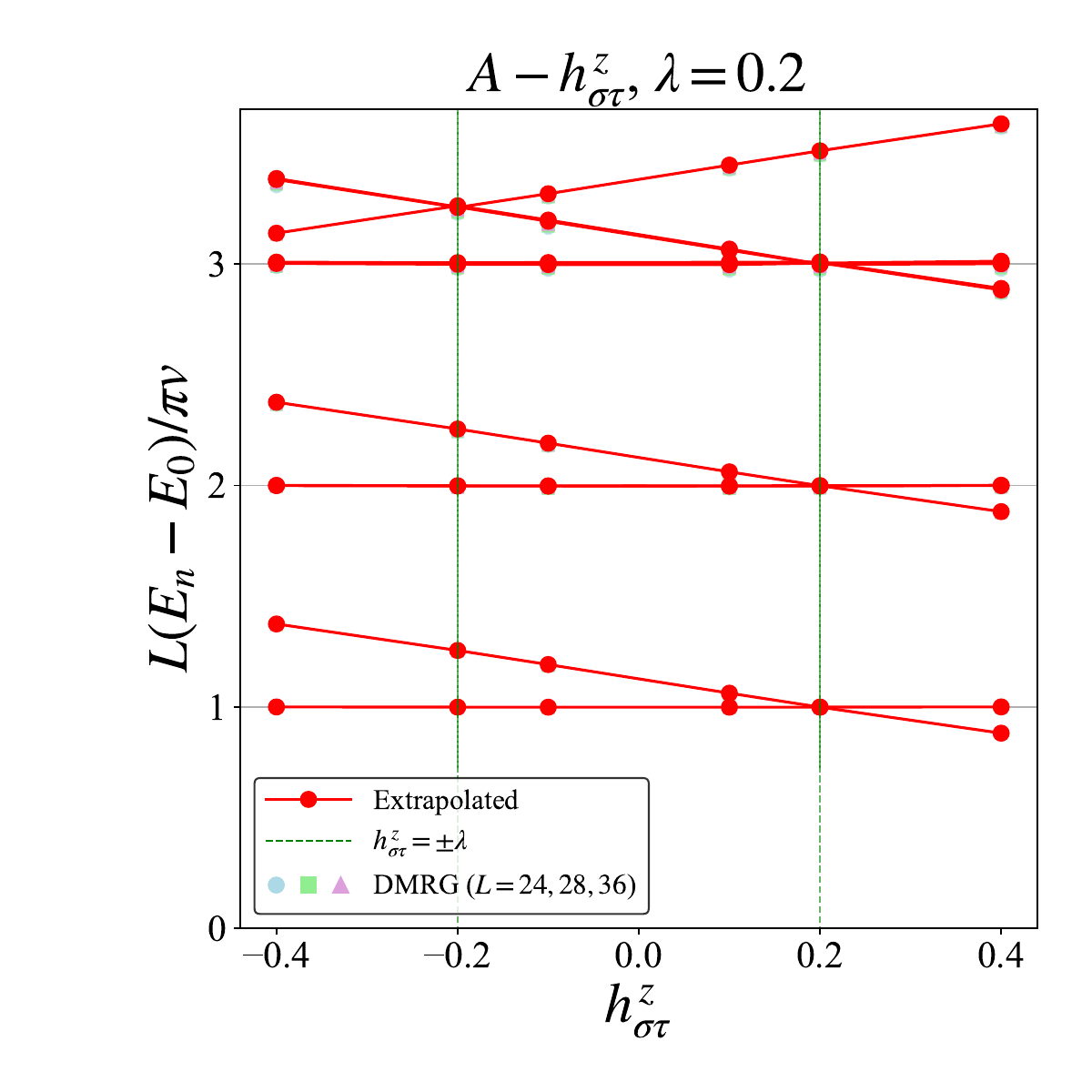} 
    \end{subfigure}
     \hfill
    \begin{subfigure}{0.48\textwidth} 
        \caption{}
        \includegraphics[width=\linewidth,trim=75 25 20 25, clip]{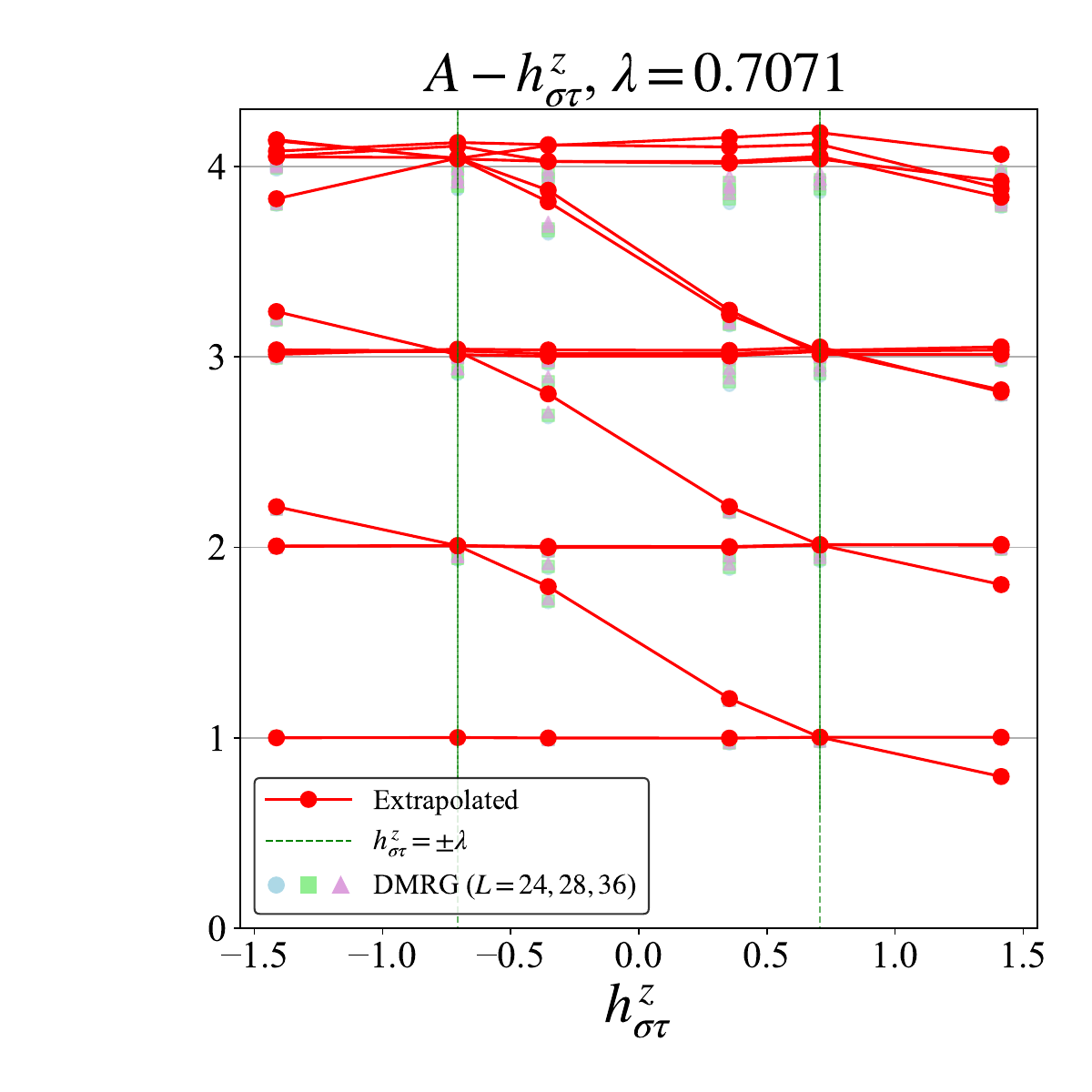} 
    \end{subfigure}
    \caption{Conformal towers for the Ashkin-Teller model under $A-h_{\sigma\tau}^z$ boundary conditions, shown for (a) $\lambda=0.2$ and (b) $\lambda=0.7071$. Finite-size DMRG data (purple, green and blue symbols) are compared with the extrapolated results in the thermodynamic limit (red circles).
    The vertical green lines mark the points $h_{\sigma\tau}^z=\pm \lambda$ where the level crossing occurs, corresponding to $\theta=\theta_d$.}
    \label{fig:AC-A}
\end{figure}

The BCFT partition function can be computed from the boundary states for general values of $\theta$,
generalizing Eq.~\eqref{eq:AC-A}.
In the open string channel, it reads
\begin{align}\label{eq:num_A-AC}
    Z_{\theta-A}(q,\theta)=q^{2r^2(\frac{\theta}{2\pi})^2}(1+q+q^{2r^2(1-\frac{\theta}{\pi})}+2q^2+\cdots) .
\end{align}
This predicts the low-energy spectrum corresponding to the differences in the conformal dimensions
given by integers and those depend on the boundary value $\theta$, via Eq.~\eqref{eq:FSS_num}.
In particular, the smallest integer difference is $1$, and the smallest ``floating''
difference is $2r^2(1-\frac{\theta}{\pi})$, where $\theta$ is a function of $h^z_{\sigma\tau}$.
This can indeed be seen in the numerical data presented in Fig.~\ref{fig:AC-A}.

We do not know the exact dependence of $\theta$ on $h^z_{\sigma\tau}$, which is model dependent
but could be read off from the numerical data in Fig.~\ref{fig:AC-A}.
Nevertheless, we have an exact prediction on the crossing of the two lowest excited states
energies. This happens when $\theta=\theta_d$, where
\begin{align}
    2r^2 (1 - \frac{\theta_d}{\pi}) = 1 .
\end{align}
Comparing with Eq.~\eqref{eq:theta_c-r} we find that
\begin{align} 
 \theta_d = \frac{\pi}{2} - \theta_c .
\end{align}
That is, the Dirichlet boundary condition at $\theta=\theta_d$
is a KW dual~\eqref{eq:KW} of that at $\theta=\theta_c$, which
corresponds to the limit of $h^z_{\sigma\tau}\to \infty$ where the
boundary spin is projected to the antipodal pair of spin states $AC$.
In terms of the lattice model, we found that the KW dual of the model
with $h_{\sigma\tau}^z\to \infty$ is effectively given by Eq.~\eqref{eq:H_dual_Free_AC-2}
which has $h_{\sigma\tau}^z = \lambda$. 
Therefore, the KW duality predicts that the crossing of the two lowest excited levels
occurs at  $h_{\sigma\tau}^z = \lambda$.
This is indeed observed in the numerical data in Fig.~\ref{fig:AC-A}.

\begin{figure}[tb]
\centering
        \includegraphics[width=0.5\linewidth]{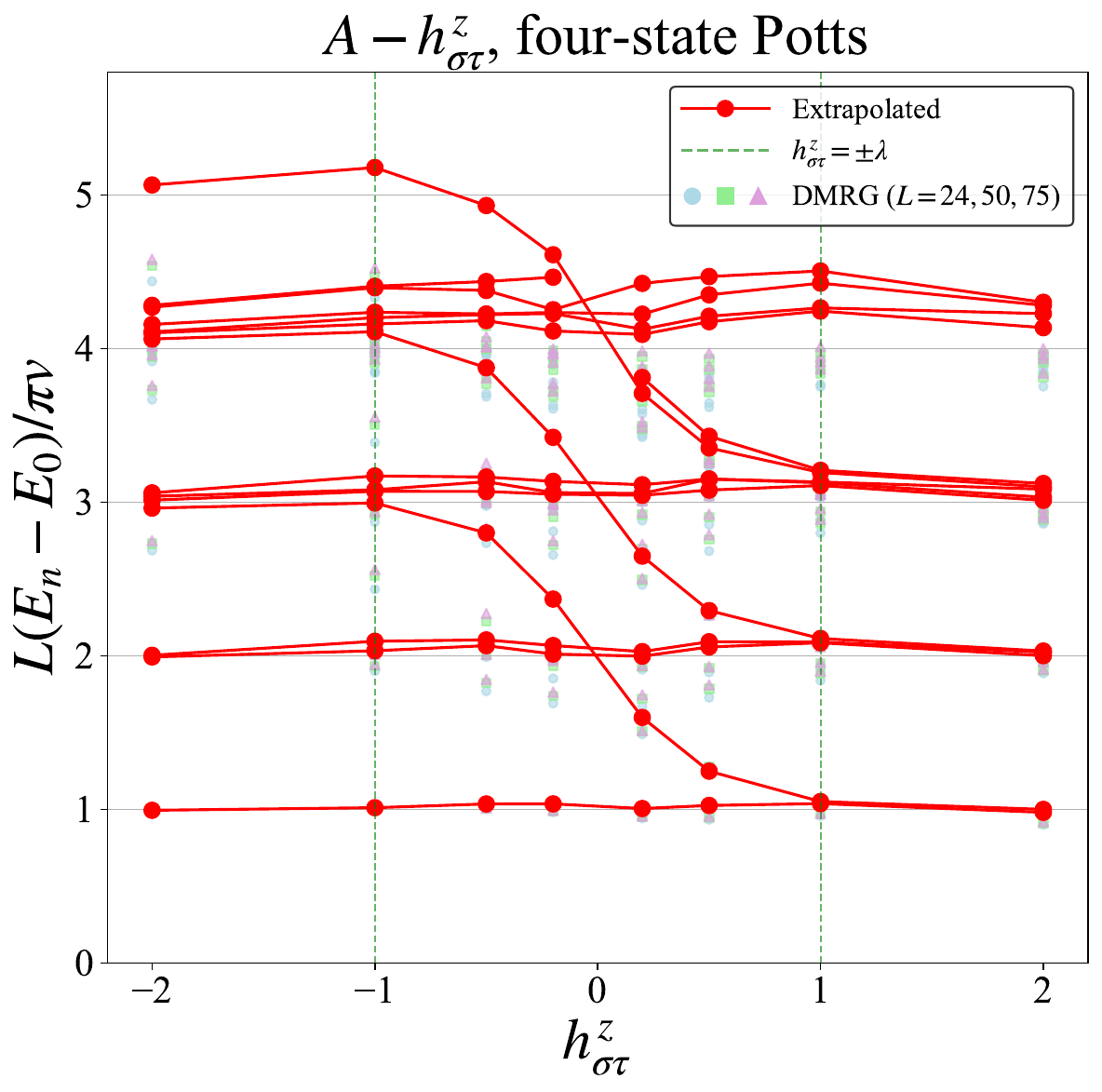} 
    \caption{Conformal towers for the four-state Potts model under $A$-$h_{\sigma\tau}^z$ boundary conditions. Finite-size DMRG data (purple, green, and blue symbols) are compared with the extrapolated results in the thermodynamic limit (red circles). The vertical green lines mark the points $h_{\sigma\tau}^z=\pm 1$ where the marginal flow should end. Red lines are connected by hands to indicate the predicted conformal towers data points should belong to. We see that extrapolated data points differ dramatically from the original data points and fit well with the predicted conformal towers \eqref{eq:num_A-AC}.}
    \label{fig:Potts_A-AC}
\end{figure}

At four-state Potts point $\lambda=1$, $\theta_d = \theta_c$, reflecting 
the self-duality of the blob boundary condition
\begin{align}
    D_{KW}\ket{AC/BD}=\ket{AC}+\ket{BD}.
\end{align}
This implies that, at $h_{\sigma\tau}^z = \pm 1$, the spectrum should be already identical to
that at $h_{\sigma\tau}^z \to \pm \infty$.
This is indeed observed in the numerical data presented in Fig.~\ref{fig:Potts_A-AC}.
Beyond $h_{\sigma\tau}^z = \pm 1$, the spectrum is basically unchanged by $h_{\sigma\tau}^z$,
indicating that the marginal deformation in the model reached its limit.

\subsubsection{Near-SSB region of Dirichlet boundary conditions}

\begin{figure}[tbp] 
    \centering 
    \begin{subfigure}{0.272\textwidth}
    \caption{}
        \includegraphics[width=\linewidth,trim=5 15 0 15, clip]{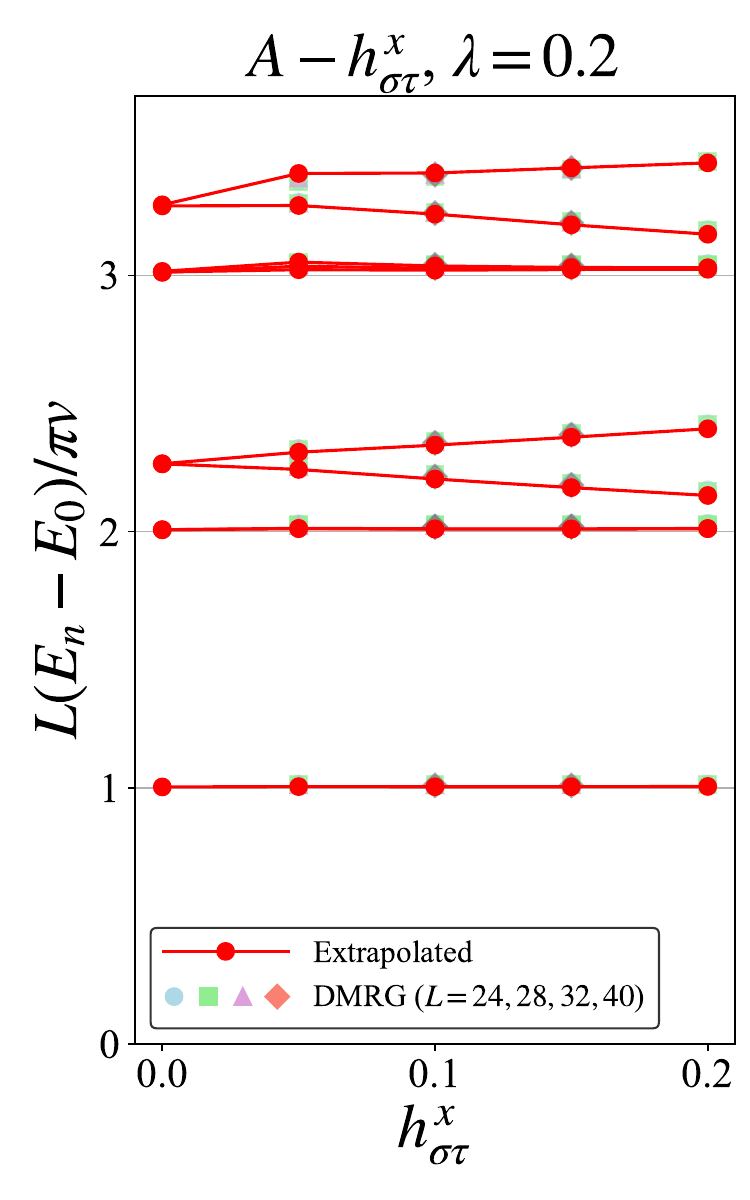} 
    \end{subfigure}\hfill
    \begin{subfigure}{0.434\textwidth}
    \caption{}
        \includegraphics[width=\linewidth,trim=8 15 0 15, clip]{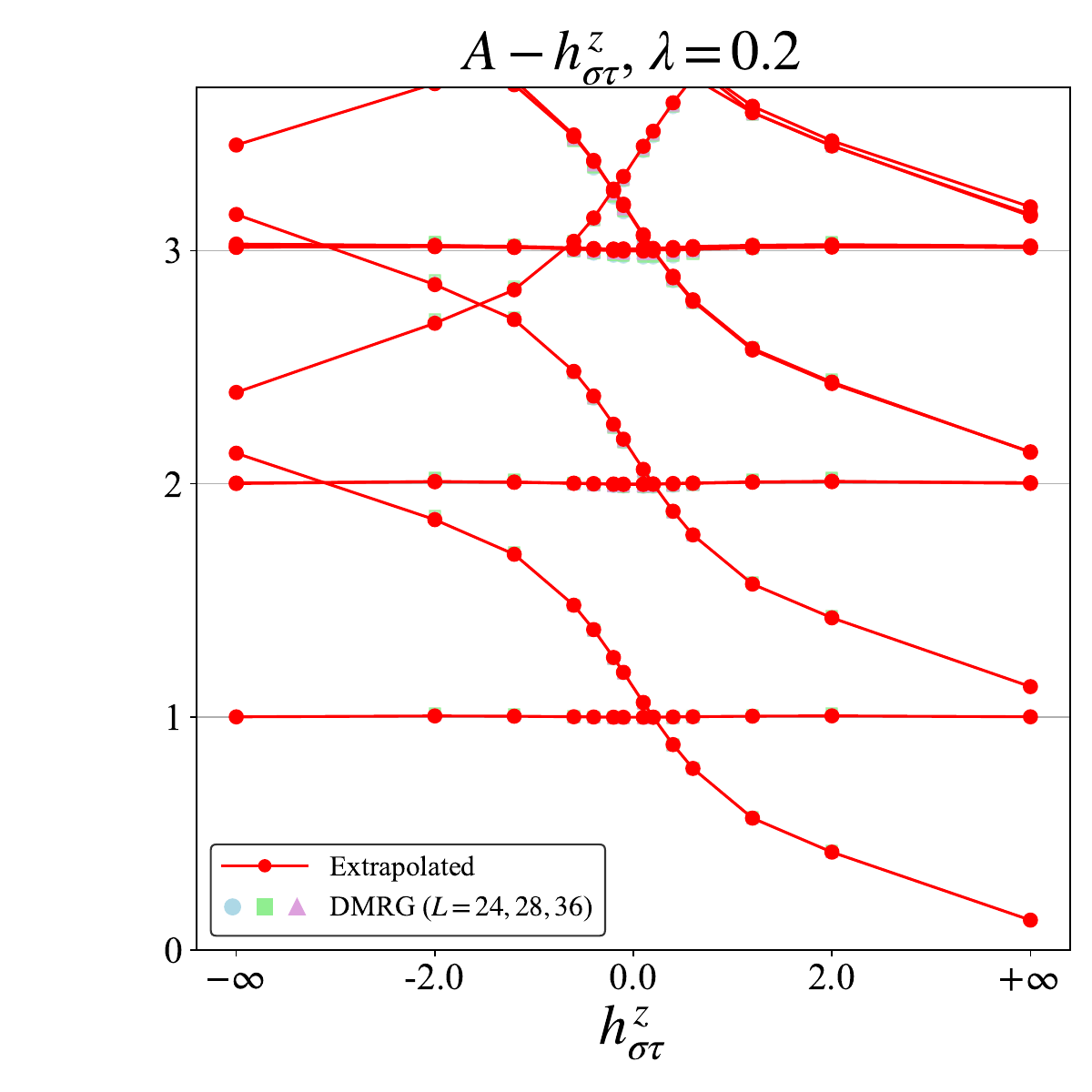} 
    \end{subfigure}\hfill
    \begin{subfigure}{0.272\textwidth}
    \caption{}
        \includegraphics[width=\linewidth,trim=5 15 0 15, clip]{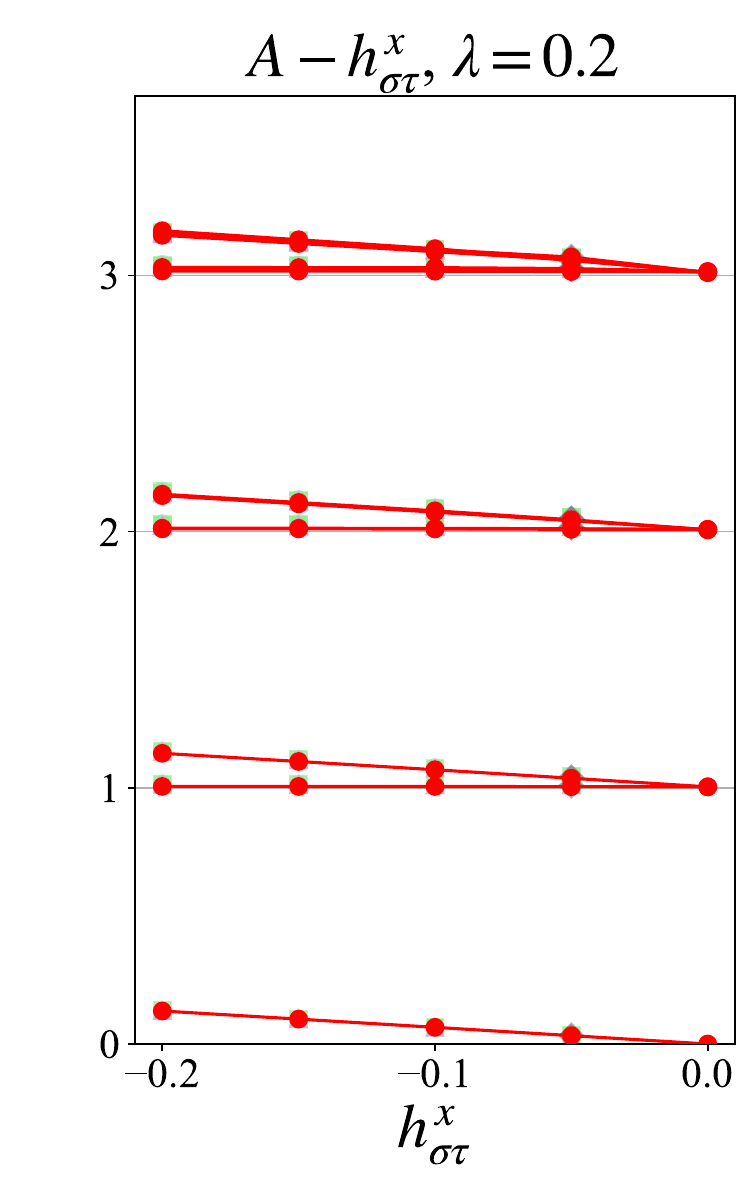} 
    \end{subfigure}
    \begin{subfigure}{0.272\textwidth}
    \caption{}
        \includegraphics[width=\linewidth,trim=5 15 0 15, clip]{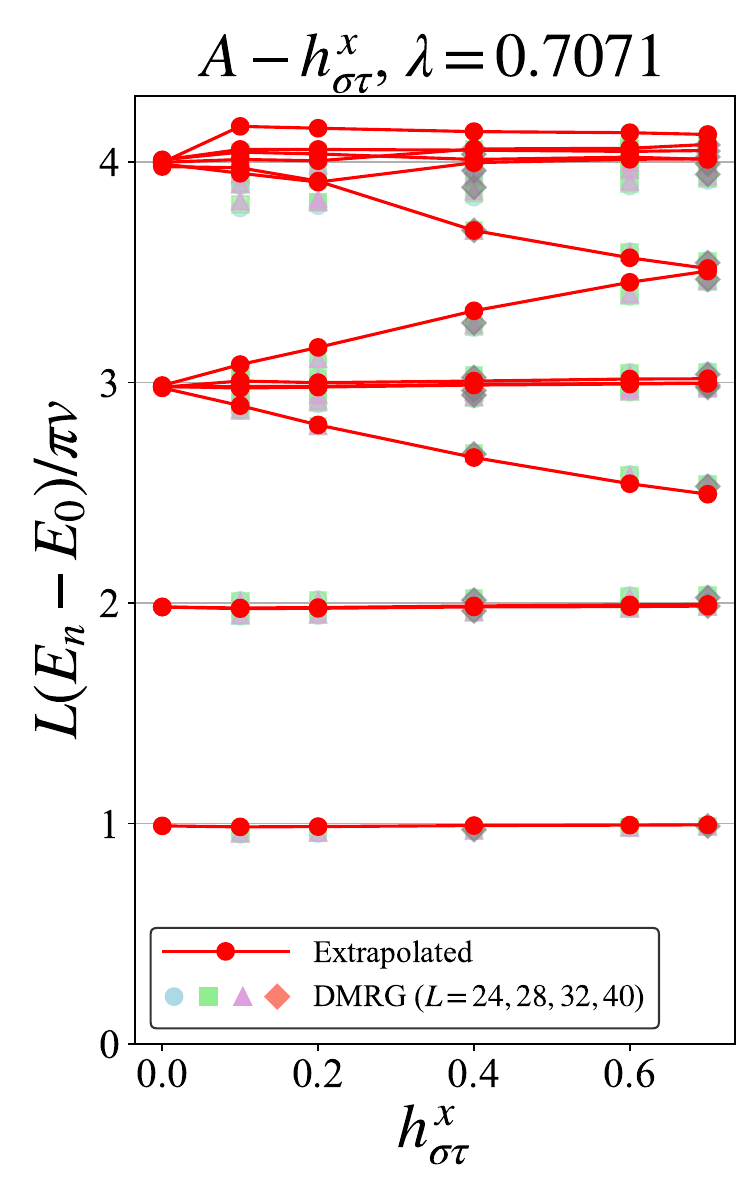} 
    \end{subfigure}\hfill
    \begin{subfigure}{0.434\textwidth}
    \caption{}
        \includegraphics[width=\linewidth,trim=8 15 0 15, clip]{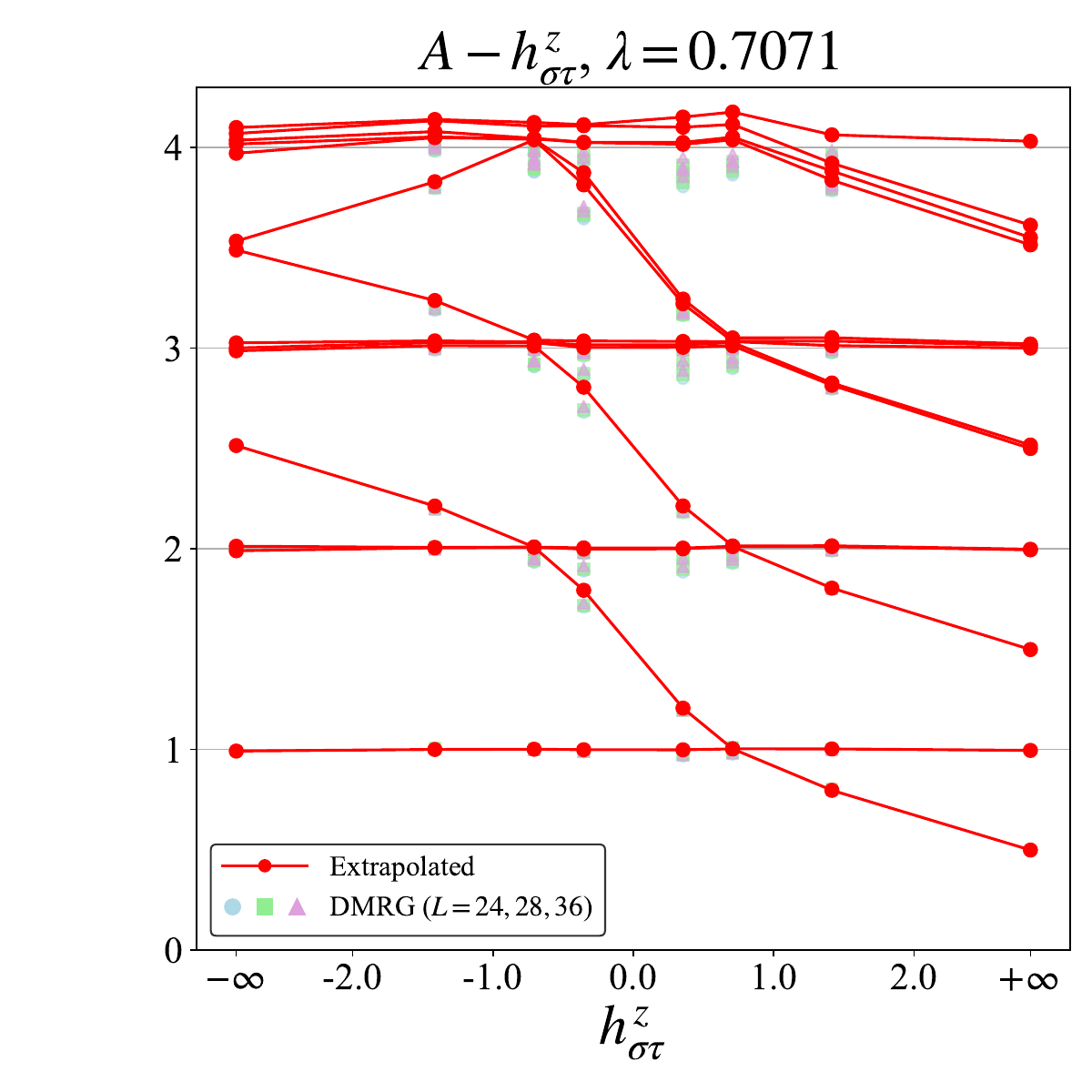} 
    \end{subfigure}\hfill
    \begin{subfigure}{0.272\textwidth}
    \caption{}
        \includegraphics[width=\linewidth,trim=5 15 0 15, clip]{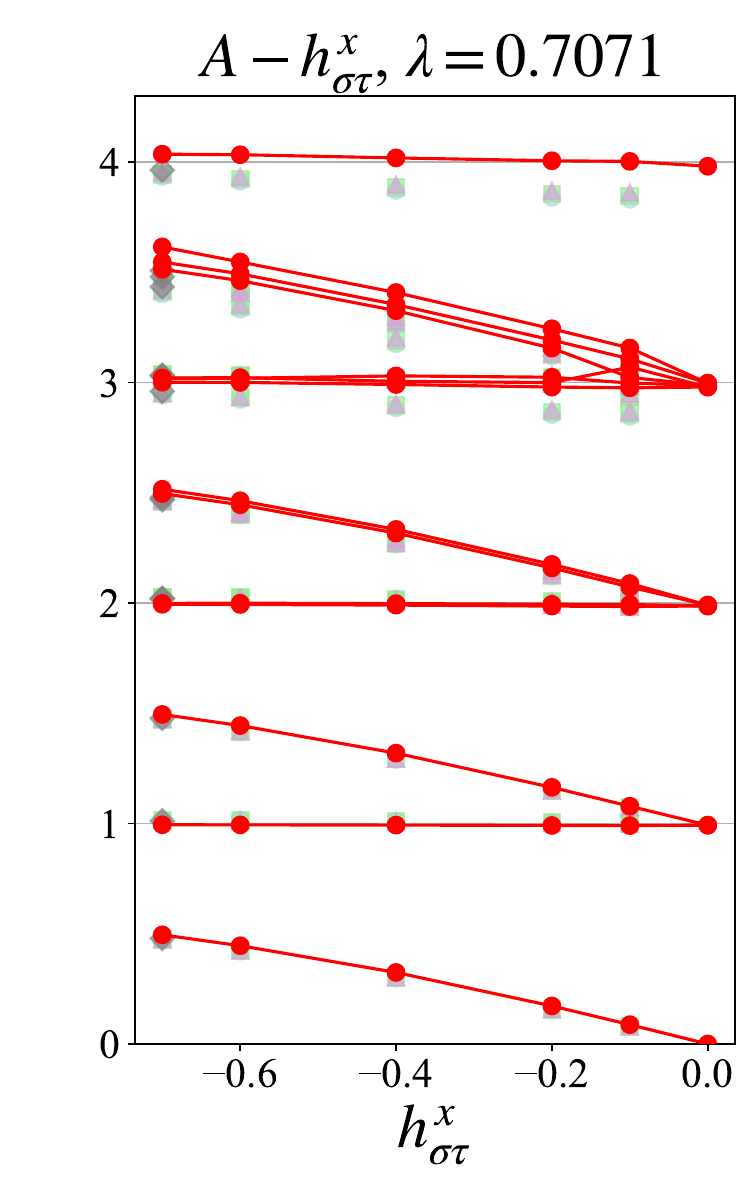} 
    \end{subfigure}
\caption{Conformal towers for the Ashkin-Teller model under $A$-$h_{\sigma\tau}^x$ and $A$-$h_{\sigma\tau}^z$ boundary conditions, shown for (a-c) $\lambda=0.2$, (d-f) $\lambda=0.7071$. Finite-size DMRG data (blue, green, purple, and pink symbols) are compared with the extrapolated results in the thermodynamic limit (red circles). For special points $h_{\sigma\tau}^x=0$ and $h_{\sigma\tau}^z=\pm\infty$, only extrapolated data are presented, and the original data come directly from the blob boundary condition case in Sec.~\ref{sec:num_blob}. Data points at $h_{\sigma\tau}^x=0$ and $h_{\sigma\tau}^z=\pm\infty$ correspond to the blob boundary conditions cases, so we directly adopt the extrapolated result without including the finite-size data.}
\label{fig:rest_Diri}
\end{figure}

In Sec.~\ref{sec:RGflow_AT}, we claimed that the boundary perturbation $h^z_{\sigma\tau}\sigma^z\tau^z$ in the standard model~\eqref{eq:H_boundary-fields}
could only realize a limited range of the Dirichlet boundary conditions in $\mathbb{Z}_2$-orbifold BCFT,
corresponding to the boundary value $\theta$ in the range~\eqref{eq:theta-in-theta_c}.

As discussed earlier in Sec.~\ref{sec:Dirichlet-family},
the Dirichlet boundary conditions outside the range~\eqref{eq:theta-in-theta_c}
include the SSB limits $\theta \to 0, \pi$ and can be realized in the model~\eqref{eq:H_hx_sigmatau}
in which the transverse fields are turned off at the end of the chain.
Here we test this claim by numerically studying the model~\eqref{eq:H_hx_sigmatau}.

The result for $\lambda=0.2$ and $\mathbb{Z}_4$-parafermion point is presented in Fig.~\ref{fig:rest_Diri}. We see that a nice continuous flow of the conformal tower
is presented from the left-hand side to the right-hand side, crossing three different lattice Hamiltonian regions. One can easily check that this flow of the conformal towers is consistent with the partition function \eqref{eq:AC-A}, with $\theta$ varying from $0$ to $\pi$. Therefore, by including
the $h^x_{\sigma\tau}$ boundary perturbation to the SSB boundary condition in the absence of the
transverse field, we successfully realized the entire range of the boundary value $0 \leq \theta \leq \pi$
in the Dirichlet boundary conditions
for generic Ashkin-Teller criticality.

\subsubsection{Neumann boundary condition}

In Sec.~\ref{sec:RG_AT_Neu}, we proposed that for the generic Ashkin-Teller model, the Neumann boundary condition can be realized by adding a marginal perturbation to the SSB boundary condition $\ket{AB}+\ket{CD}$ or $\ket{AD}+\ket{BC}$,
for example in the model Hamiltonian~\eqref{eq:H_Free_Neumann}.
In this section, we provide evidence for this construction with the following Hamiltonian with the
same Neumann boundary condition on both sides
\begin{align} 
\label{eq:H_Neu-Neu}
H^{N_O(\theta)-N_O(\theta)}_\text{AT}=-& \sum_{j=2}^{L-1}\left(\sigma_j^x+\tau_j^x+\lambda \sigma_j^x \tau_j^x\right)\notag- \sum_{j=1}^{L-1}\left(\sigma_j^z \sigma_{j+1}^z+\tau_j^z \tau_{j+1}^z+\lambda \sigma_j^z \tau_j^z \sigma_{j+1}^z \tau_{j+1}^z\right)\\
    -&\sigma^x_1-h^N_{\theta}\sigma^z_1\tau^x_1-\sigma^x_L-h^N_{\theta}\sigma^z_L\tau^x_L,
\end{align}
where the $\sigma^x$ term ensures that the unperturbed case corresponds to
$\ket{AB}+\ket{CD}=\ket{N_O(0)}$,
and $h_\theta^N$ is the marginal perturbation causing the shift of the boundary value $\theta$ of the
Neumann boundary condition $\ket{N_O(\theta)}$.
To compare our numerical results with theoretical predictions, the diagonal partition function
with Neumann boundary conditions is given by
\begin{align}\label{eq:Z_N}
Z_{N_O(\theta)-N_O(\theta)}=\frac{1}{\eta(q)}\left[\sum_{n=-\infty}^\infty q^{\frac{n^2}{2r^2}}+\sum_{n=0}^\infty q^{\frac{(n+\frac{\theta}{\pi})^2}{2r^2}}\right].
\end{align}
As in the Dirichlet case, the relation between $h_{\theta}^N$ and $\theta$ is non-universal,
and we do not know the exact functional form of $\theta(h_{\theta}^N)$.
Therefore, we focus on the characteristic features of the conformal tower flow. We observe that the first term in Eq.~\eqref{eq:Z_N} remains invariant as $\theta$ varies, containing a two-fold degenerate operator with the lowest conformal dimension $h=\frac{1}{2r^2}$ and a marginal operator responsible for the $\theta$-deformation.
The second term contains a ``floating'' set of boundary operators, whose
scaling dimensions continuously depend on $\theta$ (and thus $h_{\theta}^N$).

\begin{figure}[tb] 
    \centering 
    
    \begin{subfigure}{0.32\textwidth}
        \caption{}
        \includegraphics[width=\linewidth,trim=10 25 5 15, clip]{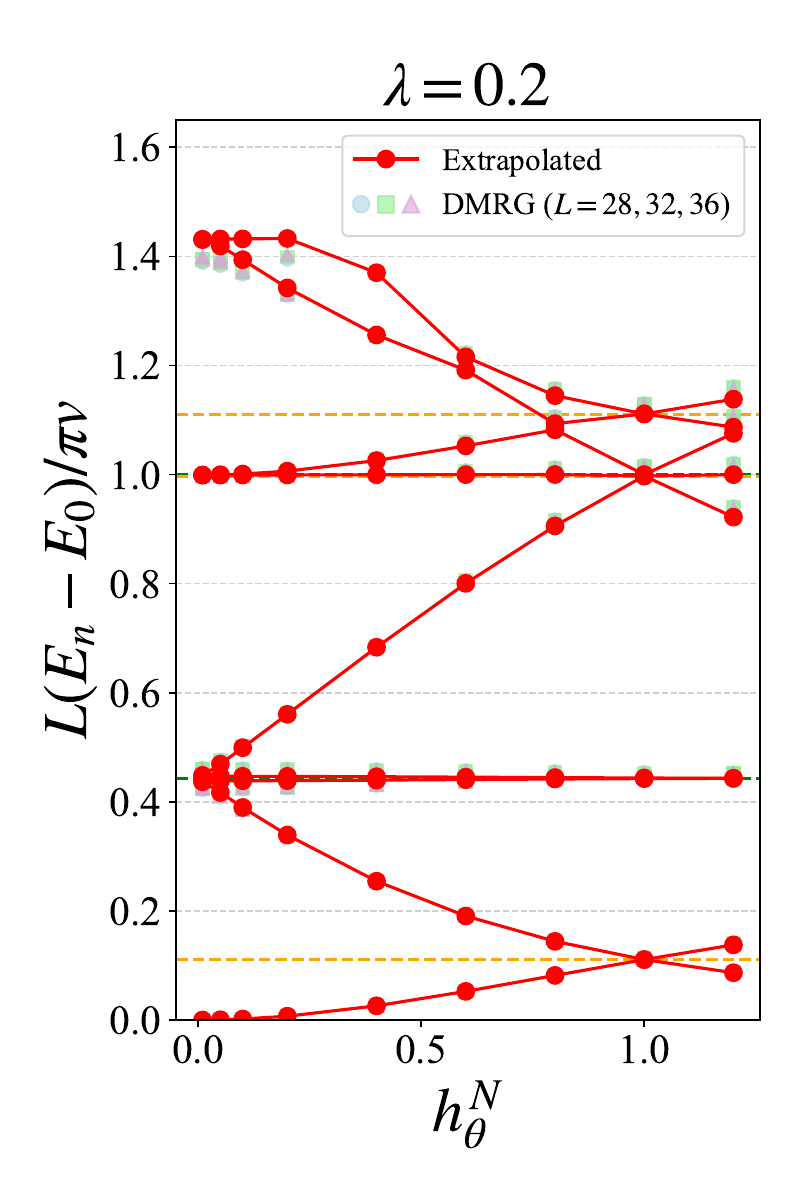} 
    \end{subfigure}
    \begin{subfigure}{0.32\textwidth} 
        \caption{}
        \includegraphics[width=\linewidth,trim=10 25 5 15, clip]{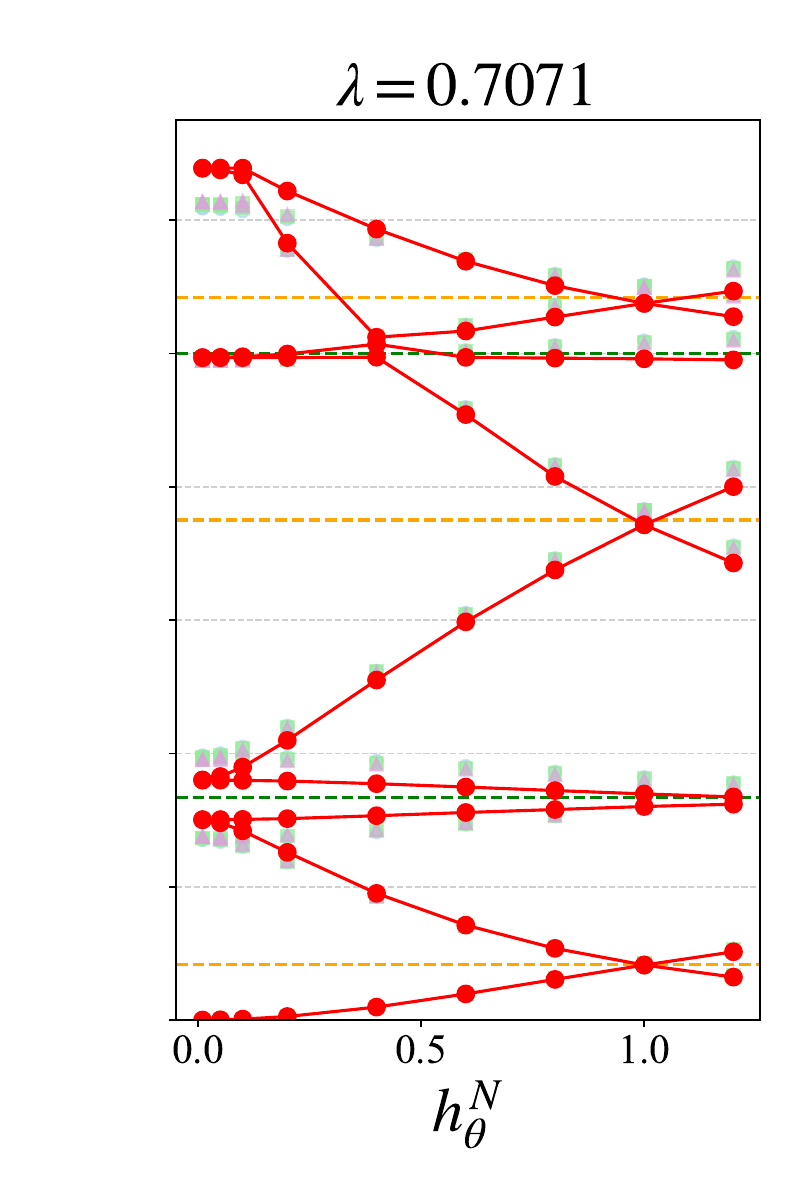} 
    \end{subfigure}
    \begin{subfigure}{0.32\textwidth} 
        \caption{}
        \includegraphics[width=\linewidth,trim=10 25 5 15, clip]{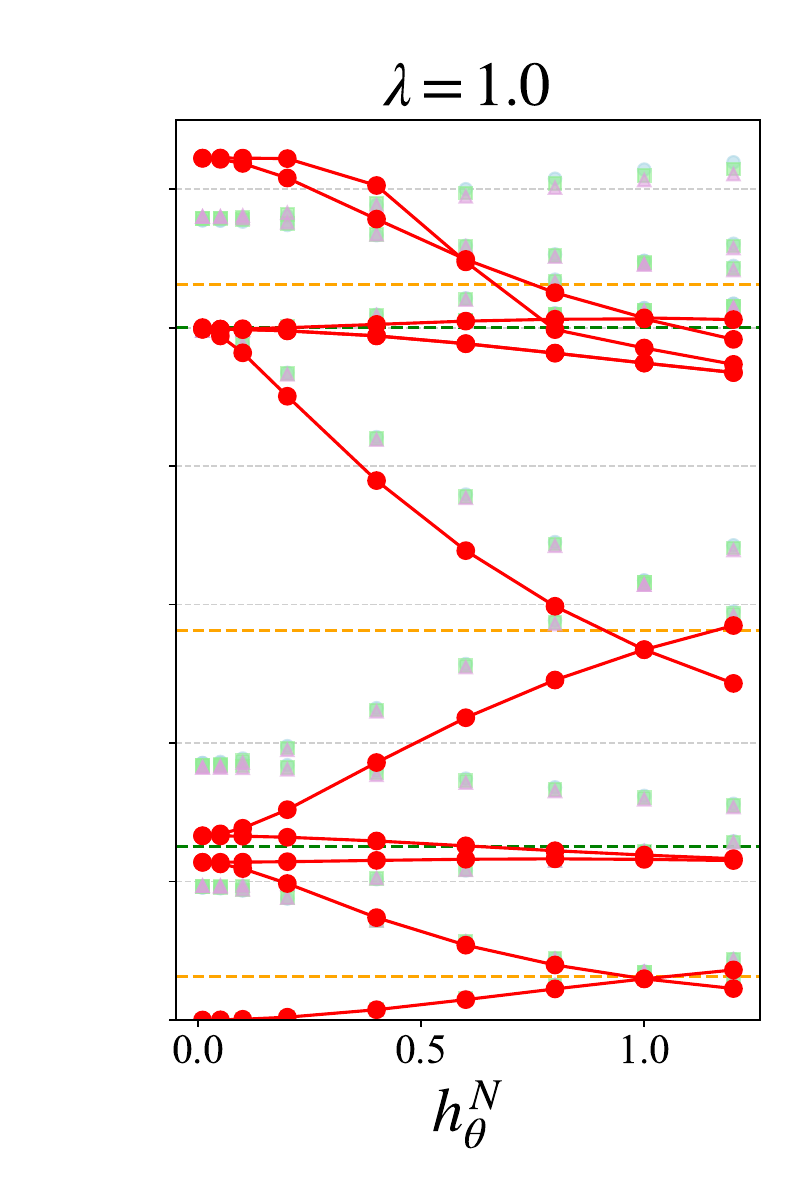} 
    \end{subfigure}
    \caption{Conformal towers for the Ashkin-Teller model under symmetric $N_O(\theta)$ boundary conditions~\eqref{eq:H_Neu-Neu}, shown for (a) $\lambda=0.2$ (b) $\lambda=0.7071$ and (c) $\lambda=1$.
Theoretical predictions (blue dashed lines) are compared with $L\to\infty$ extrapolated data (red dots) and DMRG data (purple, green, and blue symbols).
The extrapolation (Eq.~\eqref{eq:extrapo}) uses system sizes $L=\{24,28,32,36\}$ (giving $L_{\text{max}}=36$), with larger sets included for accuracy at $\lambda=0.2$ ($L=40$) and $\lambda=0.7071$ ($L=40$). Orange dashed lines }
\label{fig:Neumann}
\end{figure}

The numerical data shown in Fig.~\ref{fig:Neumann} are consistent with the theoretical expectations, reproducing the
constant part of the spectrum independent of $\theta$ and also the floating part.
The scaling dimensions of the constant part of the spectrum also agree with
the theoretical predictions (green dashed line).
We note that, at $\theta=\pi/2$, the scaling dimensions in the floating set become degenerate
(between $n=0,-1$, $n=1,-2$, etc.).
The high-degeneracy point $\theta=\pi/2$ is a self-dual point under the KW
transformation, according to Eq.~\eqref{eq:KW-Neumann}.
At the decoupling point $\lambda=0$, it was shown that $\theta=\pi/2$
occurs for $h^N_{\theta}=1$, which corresponds to applying KW transformation
to half of the infinite critical Ising chain~\cite{oshikawaBoundaryConformalField1996}.
While we do not have a proof for the lattice model, it appears that
the self-dual point $\theta=\pi/2$ of Neumann boundary condition is
realized for $h^N_{\theta}=1$ even for a non-vanishing $\lambda$.
This can be seen numerically in Fig.~\ref{fig:Neumann}, where the crossing of levels
indeed occurs at $h^N_{\theta}=1$.
We speculate that the crossing at $h^N_{\theta}=1$ is protected by a semi-duality
of the model.
In any case, at the self-dual point $\theta=\pi/2$ where the crossings occur,
the scaling dimension of the degenerate boundary operators can be read off
from the partition function~\eqref{eq:Z_N} as $\frac{1}{8r^2},\frac{9}{8r^2}, \ldots$,
which are shown as orange dashed lines in Fig.~\ref{fig:Neumann}.
Indeed we see that they agree with the numerically obtained scaling dimensions where
the crossings occur.

These confirm our picture that the boundary condition is described by the
Neumann boundary states $\ket{N_O(\theta)}$.

\subsection{Three-state mixed boundary perturbations in the four-state Potts model}
\label{sec:ABC}
In Sec.~\ref{sec:RGflow_four-state_3-state}, our theoretical analysis predicted that the three-state mixed boundary conditions (e.g., $BCD$) are not new stable fixed points. Instead, perturbations in these directions correspond to marginally irrelevant operators, implying a slow flow back to the Free boundary condition. In this section, we verify this picture numerically, starting with the symmetric setup.

\begin{figure}[tb] 
    \centering 
    
    \begin{subfigure}{0.66\textwidth}
        \caption{}
        \includegraphics[width=\linewidth]{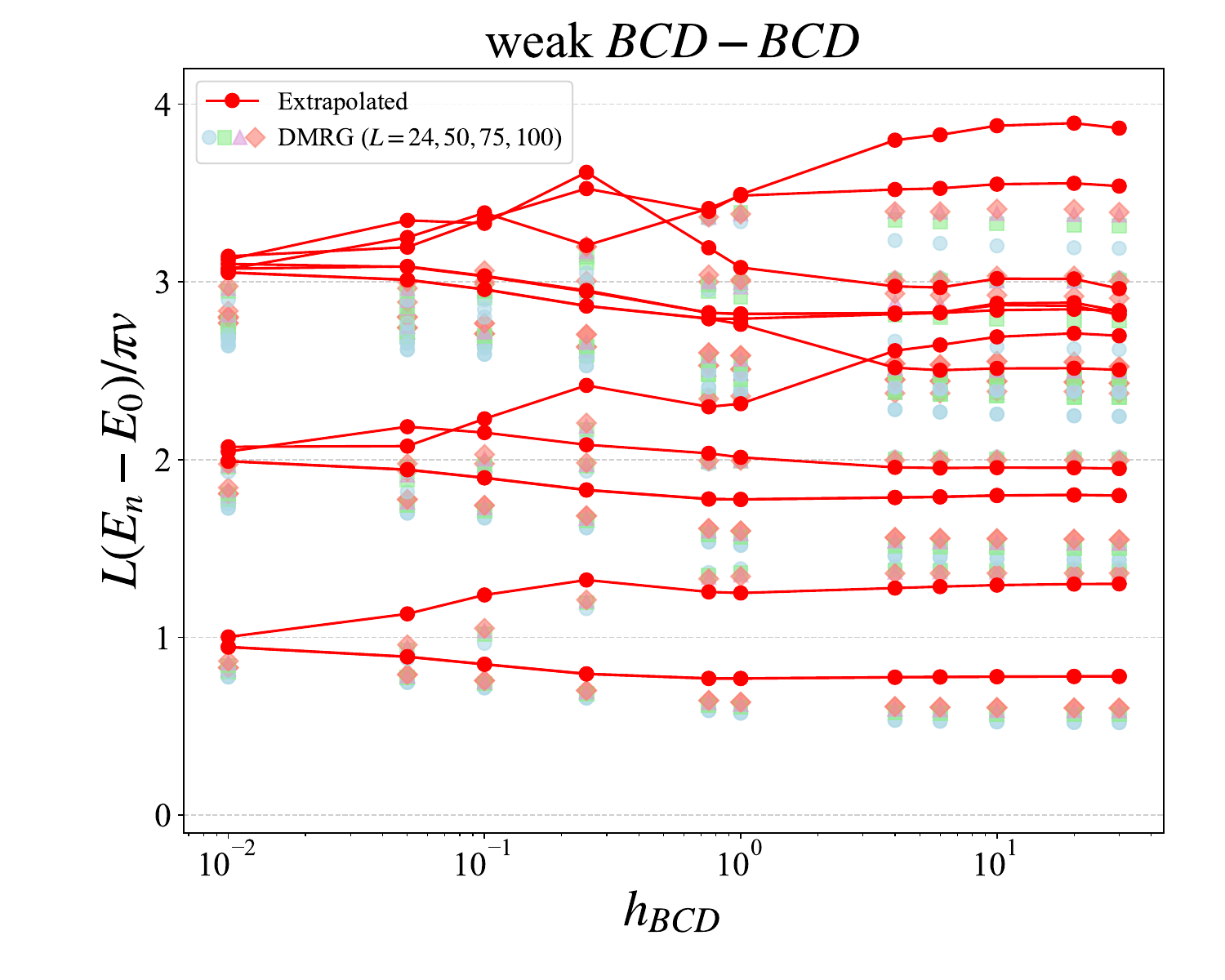} 
    \end{subfigure}
    \hfill
    \begin{subfigure}{0.33\textwidth} 
        \caption{}
        \includegraphics[width=\linewidth]{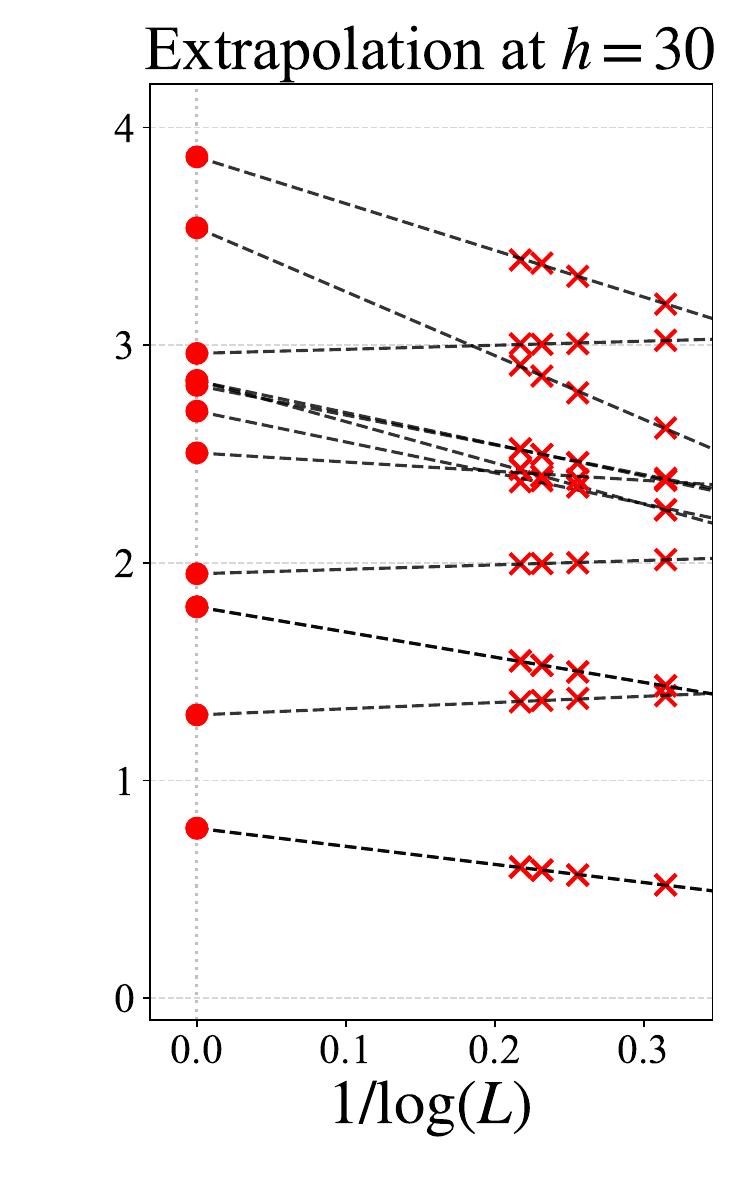} 
    \end{subfigure}

\caption{(a) Conformal towers for the Ashkin-Teller model under weak $BCD$-$BCD$ boundary conditions (defined as in Eq.~\eqref{eq:H_BCD}) with $h_{BCD}=0.01,0.05,...,30$. Finite-size DMRG data (grey, purple, green, and blue symbols) are compared with the extrapolated results in the thermodynamic limit (red circles). (b) Finite-size extrapolation of conformal towers with $h_{BCD}=30$. DMRG data (red cross marks) are extrapolated (black dashed lines) to the thermodynamic limit $L\to\infty$ (red circles).}
    \label{fig:BCD-BCD} 

\end{figure}

To study the stability of the $BCD$ direction, we consider the Hamiltonian with symmetric boundary terms polarizing the spins against the A state:
\begin{align}\label{eq:H_BCD}H_\text{AT}^{h_{BCD}-h_{BCD}}=&- \sum_{j=1}^{L}\left(\sigma_j^x+\tau_j^x+ \sigma_j^x \tau_j^x\right)- \sum_{j=1}^{L-1}\left(\sigma_j^z \sigma_{j+1}^z+\tau_j^z \tau_{j+1}^z+\sigma_j^z \tau_j^z \sigma_{j+1}^z \tau_{j+1}^z\right)\notag\\&+h_{BCD}(\sigma_L^z+\tau_L^z+\sigma_L^z\tau_L^z)+h_{BCD}(\sigma_1^z+\tau_1^z+\sigma_1^z\tau_1^z)\end{align}
with $h_{BCD}\geq 0$. The conformal towers extracted from this Hamiltonian are presented in Fig.~\ref{fig:BCD-BCD}. We observe that as the magnitude of the perturbation $h_{BCD}$ increases, the spectrum eventually stabilizes, forming a plateau in the strong magnetic field regime.

However, a careful examination reveals that the scaling dimensions at this plateau exhibit a persistent deviation from the expected values of the Free boundary condition (or any other rational fixed point). Even after performing the standard finite-size extrapolation (which accounts for the bulk logarithmic corrections), this discrepancy remains unresolved. This behavior is a direct consequence of the marginally irrelevant nature of the flow, combined with the strength of the perturbation. Since we manually tune the boundary coupling $h_{BCD}$ to be of order $O(1)$ to reach the plateau, the system is driven into a strong-coupling crossover regime rather than the perturbative infrared limit. Due to the extremely slow, logarithmic decay of the effective boundary coupling ($g_{\text{eff}} \sim 1/\ln L$), the system cannot flow back to the Free fixed point within accessible system sizes. Consequently, the standard extrapolation fails to bridge the gap between the pinned boundary state and the asymptotic Free BC. Thus, the observed plateau and its deviation should be understood as a finite-size artifact of this slow crossover, rather than evidence of a new exotic fixed point.

\begin{figure}[tb] 
    \centering 
    
    \begin{subfigure}{0.66\textwidth}
        \caption{}
        \includegraphics[width=\linewidth]{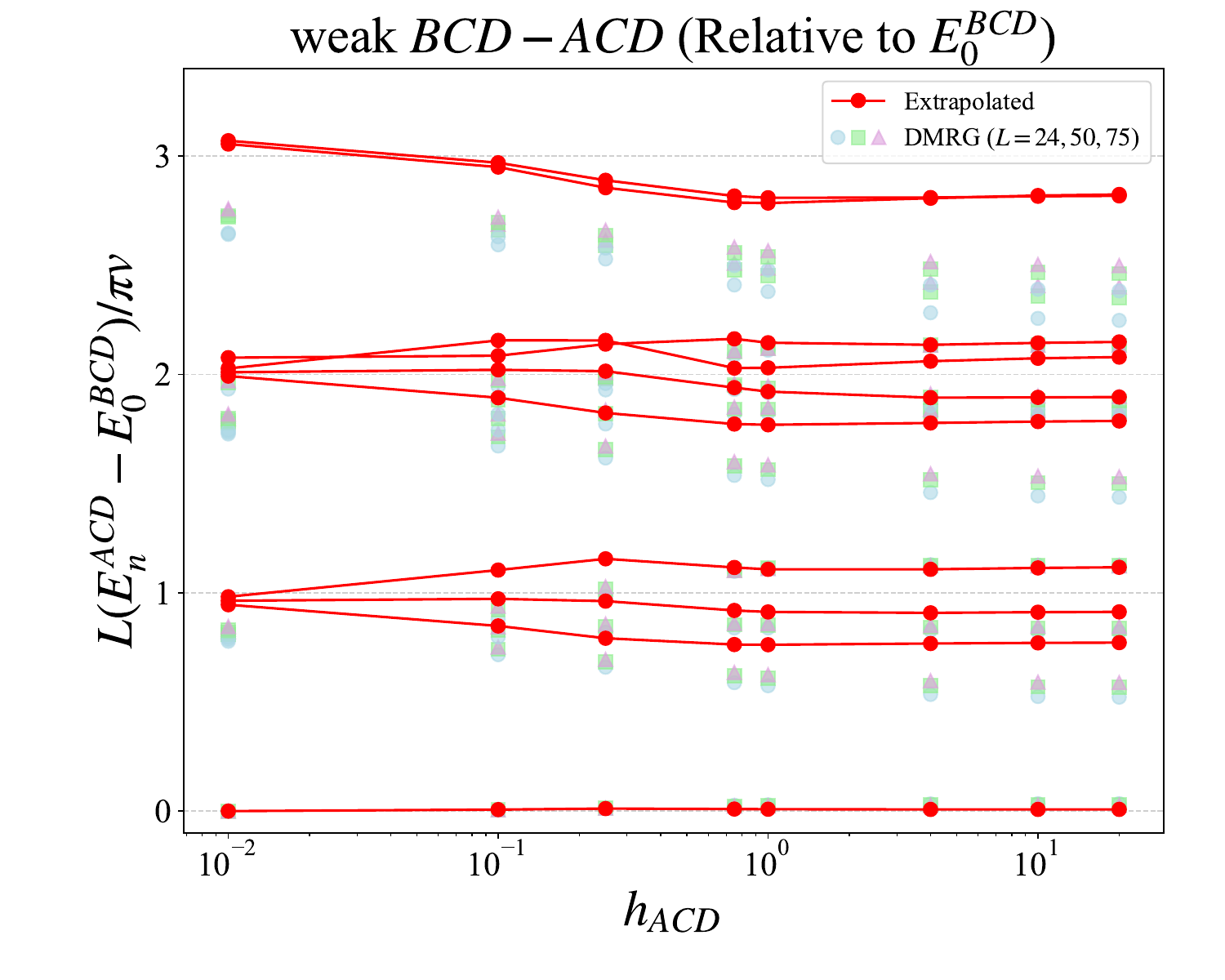} 
    \end{subfigure}
    \hfill
    \begin{subfigure}{0.33\textwidth} 
        \caption{}
        \includegraphics[width=\linewidth]{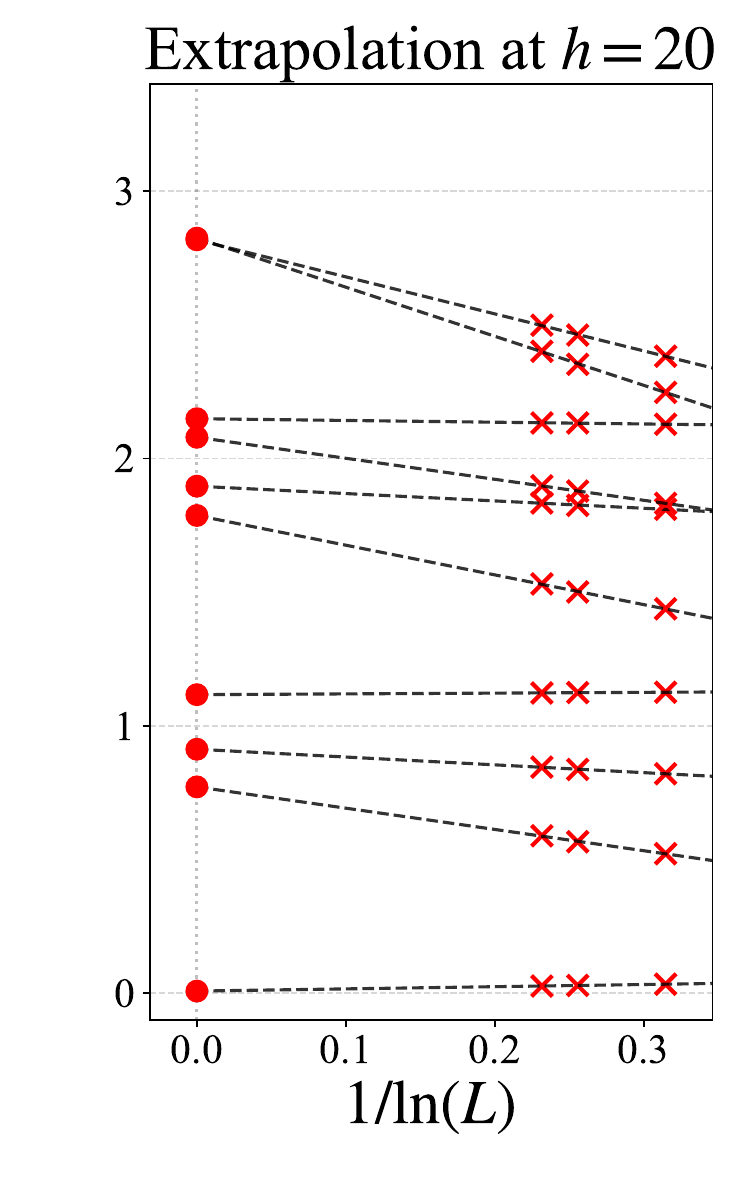} 
    \end{subfigure}

\caption{(a) Conformal towers for the Ashkin-Teller model under weak BCD-ACD boundary conditions (defined as in Eq.~\eqref{eq:H_BCD_ACD}) with $h_{ACD}=0.01,0.05,...,20$. Finite-size DMRG data (purple, green, and blue symbols) are compared with the extrapolated results in the thermodynamic limit (red circles). Note that here we are comparing the excitation energy to the ground state of the Hamiltonian with symmetric three-state mixed BCs~\eqref{eq:H_BCD}. The lowest conformal tower with vanishing conformal dimension then signifies an identity tower. We attribute the deviation of the second lowest operator ($h_{AC}\in[0.3,2]$) to the breakdown of the extrapolation, caused by the small magnitude of the first-order finite-size correction. (b) Finite-size extrapolation of conformal towers with $h_{ACD}=20$. DMRG data (red cross marks) are extrapolated (black dashed lines) to the thermodynamic limit $L\to\infty$ (red circles).}
    \label{fig:BCD-ACD} 

\end{figure}

To conclusively show that the $BCD$ direction does not define a distinct boundary condition, we examine the overlap between different boundary states using an asymmetric setup. We apply $h_{BCD}$ and $h_{ACD}$ fields at the left and right boundaries, respectively:
\begin{align}\label{eq:H_BCD_ACD}H_\text{AT}^{h_{ACD}-h_{BCD}}=&- \sum_{j=1}^{L}\left(\sigma_j^x+\tau_j^x+ \sigma_j^x \tau_j^x\right)- \sum_{j=1}^{L-1}\left(\sigma_j^z \sigma_{j+1}^z+\tau_j^z \tau_{j+1}^z+\sigma_j^z \tau_j^z \sigma_{j+1}^z \tau_{j+1}^z\right)\notag\\&+h_{BCD}(\sigma_1^z+\tau_1^z+\sigma_1^z\tau_1^z)+h_{ACD}(\sigma_L^z-\tau_L^z-\sigma_L^z\tau_L^z),\end{align}
where for simplicity we take $h_{BCD}=h_{ACD}$.
The extracted spectrum is shown in Fig.~\ref{fig:BCD-ACD}. If $BCD$ and $ACD$ were distinct elementary boundary conditions (analogous to the orthogonal $A$ and $B$ states), the open-string partition function $Z_{BCD-ACD}$ should be dominated by non-trivial boundary condition changing operators. However, Fig.~\ref{fig:BCD-ACD} clearly reveals the presence of an identity tower (approaching dimension 0). The existence of the identity channel proves that the $BCD$ and $ACD$ boundary states are not mutually orthogonal. In fact, this is also observed in the conformal loop model realization of the four-state Potts model, where they found the lowest conformal dimension of BCCOs between free and three-state mixed boundary conditions is zero~\cite{jacobsenConformalBoundaryLoop2008}. This violation of distinctness implies that these states have a non-zero overlap with the Free boundary condition (and thus with each other). This numerical evidence conclusively supports our claim that the three-state mixed boundary conditions are not new intrinsic Cardy states.

In the generic AT model $0 \leq \lambda < 1$, the leading boundary operator for
the perturbation on the Free boundary condition towards the three-state mixed boundary condition, say $BCD$,
is identical to the exactly marginal operator which shifts the boundary value $\theta$
of the Dirichlet boundary state $\ket{D_O(\theta)}$ from free boundary $\theta=\pi/2$
towards the antipodal two-state mixed boundary condition $BD$, $\theta=\theta_c$.
Thus, such a perturbation just produces a marginal deformation along
$\ket{D_O(\theta)}$.
Again, the ``three-state mixed boundary conditions'' do not give a new stable fixed point
of BRG flows, or equivalently, a new conformally invariant boundary condition.

\section{Conclusion}\label{sec:conclusion}In this work, we have performed a comprehensive study of the boundary conformal field theory (BCFT) for the Ashkin-Teller model, combining theoretical derivation with large-scale numerical simulations. We have successfully established the identification between the stable lattice boundary conditions and the BCFT boundary states along this $c=1$ critical line. In particular, we focused on the four-state Potts point as a special case, where we verified the consistency of this identification against the model's enhanced $S_4$ symmetry and Kramers-Wannier duality. Furthermore, supported by DMRG results, we have mapped out the boundary renormalization group flows between these fixed points, providing a clear picture of the boundary critical phenomena in this system.

A central theme of our theoretical analysis, as emphasized in the introduction, is the nontrivial identification between the generic $c=1$ free boson CFT (which describes the general Ashkin-Teller line and possesses an infinite number of primary fields) and the CFT with extended symmetry (which emerges at special points like the four-state Potts model and organizes into a finite number of primary fields). This perspective is fundamentally important beyond the specific models studied here. The same strategy—analyzing the simple current extension of the chiral algebra~\cite{fukusumiFusionRuleConformal2025,fukusumiGaugingExtendingBulk2025,fukusumiExtendingFusionRules2025}—should apply to a wider class of systems, such as generic Narain CFTs~\cite{Narain:1985jj,Narain:1986am}, their deformations~\cite{Koh:1989cv}, and the corresponding lattice realizations~\cite{Alavirad:2019iea}. We also note that the resultant extended chiral algebra corresponds to a $\mathbb{Z}_{N}$-graded symmetry topological field theory (SymTFT)~\cite{chatterjeeSymmetryShadowTopological2023}, which should describe the (symmetry-enriched) topological orders. In this context, revisiting the finite group symmetry analysis of characters (or modular forms) becomes crucial. For future theoretical developments, it would be particularly interesting to revisit the studies of Rogers-Ramanujan type identities~\cite{Andrews:1984af}. These identities share a deep connection with such integrable lattice models and fractional quantum Hall systems~\cite{Bernevig2008PropertiesON,Bernevig_2008}\footnote{To some extent, the study of Rogers-Ramanujan type identities can be viewed as the study of simple current extension at the level of chiral characters.}, and we refer interested readers to a recent review~\cite{Campbell_2024}.

While this work has concentrated on theoretical and numerical aspects, we also emphasize the relevance of these results to experimental settings. The manipulation of boundaries and defects plays a vital role in understanding critical systems~\cite{Batchelor:2015osa}, and the realization of CFT physics in cold atom experiments, such as Rydberg atom arrays, has seen significant progress~\cite{Slagle:2021ene}. We expect that our explicit construction of boundary conditions and RG flows in the Ashkin-Teller universality class can give a useful theoretical basis for detecting signatures of integer spin simple currents or ``fractional supersymmetry" in these experimental platforms.

\section*{Acknowledgements}
YL acknowledges Atsushi Ueda for helpful discussions on the boundary renormalization group flow and suggestions on the manuscript. NC is indebted to Ian Affleck for teaching her the basics of boundary conformal field theory. YF thanks helpful discussions with Takamasa Ando on gauging operations in lattice models. YF acknowledges Masahiro Hoshino for useful references on numerical investigation of the Kramers-Wannier transformation. 
Many parts of the present work are based on Refs.~\cite{oshikawaIsingDefectPRL,oshikawaBoundaryConformalField1996}, which were the outcome of the first project of MO as a postdoc of Ian Affleck. It opened up many later applications of boundary CFT by the present authors and other scientists, and was a pioneering work on what are now called ``conformal defects''.

\paragraph{Funding information}
YL is supported by the Global Science Graduate Course (GSGC) program of the University of Tokyo. NC acknowledges the financial support by Delft Technology Fellowship. Numerical simulations were performed at the DelftBlue HPC and at the Dutch national e-infrastructure with the support of the SURF Cooperative. YF thanks the support of NCTS and CTC. 
The work of MO is partially supported by JSPS KAKENHI Grant No. JP23K25791 and No. JP24H00946, and by
JST CREST Grant No. JPMJCR19T2. This research was supported in part by grant NSF PHY-2309135 to the Kavli Institute for Theoretical Physics (KITP).

\begin{appendix}
\numberwithin{equation}{section}

\section{Conventions}\label{app:con}
In this work, blob boundary conditions are realized on the lattice through the projection of the Hamiltonian. Here, we provide all the boundary Hamiltonians realizing all the blob boundary conditions used in the numerical simulations. As the projector commutes with most parts of the Hamiltonian, we only consider the terms involving site 1:
\begin{align}
    H^{L}=-(\sigma^z_1\sigma_2^z+\tau^z_1\tau_2^z+\lambda\sigma^z_1\sigma_2^z\tau^z_1\tau_2^z+\sigma^x_1+\tau_1^x+\lambda\sigma_1^x\tau_1^x),
\end{align}
where $H^L$ denotes the left-hand side of the Hamiltonian. By substituting $1$ by $L$ and $2$ by $L-1$, one can obtain the expression for the right-hand side.

For the fixed and usual two-state mixed BC, one can simply remove the degrees of freedom for the boundary spins. For instance, for fixed BCs:
\begin{align}
A&:-\sigma_2^z-\tau_2^z-\lambda\sigma_2^z\tau_2^z,\\
B&:-\sigma_2^z+\tau_2^z+\lambda\sigma_2^z\tau_2^z,\\
C&:+\sigma_2^z+\tau_2^z-\lambda\sigma_2^z\tau_2^z,\\
D&:+\sigma_2^z-\tau_2^z+\lambda\sigma_2^z\tau_2^z.
\end{align}
Note that spin operators $\{\sigma_1\}$ and $\{\tau_1\}$ are simply removed by taking corresponding expectation values. Similarly, for two-state mixed BCs:
\begin{align}
AB&:-\tau_1^x-\sigma_2^z-\tau_1^z\tau_2^z-\lambda\tau_1^z\sigma_2^z\tau_2^z,\\
CD&:-\tau_1^x+\sigma_2^z-\tau_1^z\tau_2^z+\lambda\tau_1^z\sigma_2^z\tau_2^z,\\
AD&:-\sigma_1^x-\tau_2^z-\sigma_1^z\sigma_2^z-\lambda\sigma_1^z\sigma_2^z\tau_2^z,\\
BC&:-\sigma_1^x+\tau_2^z-\sigma_1^z\sigma_2^z+\lambda\sigma_1^z\sigma_2^z\tau_2^z.
\end{align}

For the rest cases, we cannot perform further simplification by directly eliminating degrees of freedom. For antipodal two-state mixed BCs:
\begin{align}
    AC&:-\frac
{1}{2}(\sigma_1^z+\tau_1^z)(\sigma_2^z+\tau_2^z)-\frac{\lambda}{2}(\sigma_2^z\tau_2^z+\sigma^x_1\tau^x_1-\sigma_1^y\tau_1^y+\sigma_1^z\tau_1^z\sigma_2^z\tau_2^z),\\
BD&:-\frac
{1}{2}(\sigma_1^z-\tau_1^z)(\sigma_2^z-\tau_2^z)-\frac{\lambda}{2}(-\sigma_2^z\tau_2^z+\sigma^x_1\tau^x_1+\sigma_1^y\tau_1^y+\sigma_1^z\tau_1^z\sigma_2^z\tau_2^z).
\end{align}
And for three-state mixed BCs:
\begin{align}
BCD&:P_{BCD}H(z)-\frac{1}{2}(1-\tau_1^z)\sigma_1^x-\frac{1}{2}(1-\sigma_1^z)\tau_1^x-\frac{a}{2}(\sigma_1^x\tau_1^x+\sigma_1^y\tau_1^y),\\
ABC&:P_{ABC}H(z)-\frac{1}{2}(1-\tau_1^z)\sigma_1^x-\frac{1}{2}(1+\sigma_1^z)\tau_1^x-\frac{a}{2}(\sigma_1^x\tau_1^x-\sigma_1^y\tau_1^y),\\
ACD&:P_{ACD}H(z)-\frac{1}{2}(1+\tau_1^z)\sigma_1^x-\frac{1}{2}(1-\sigma_1^z)\tau_1^x-\frac{a}{2}(\sigma_1^x\tau_1^x-\sigma_1^y\tau_1^y),\\
ABD&:P_{ABD}H(z)-\frac{1}{2}(1+\tau_1^z)\sigma_1^x-\frac{1}{2}(1+\sigma_1^z)\tau_1^x-\frac{a}{2}(\sigma_1^x\tau_1^x+\sigma_1^y\tau_1^y),
\end{align}

where $H(z):=-(\sigma_1^z\sigma_2^z+\tau_1^z\tau_2^z+\lambda\sigma^z_1\sigma^z_2\tau^z_1\tau^z_2)$ is the part that commutes with the projector.

\section{Boundary conditions at the \texorpdfstring{$Z_4$}{Z4}-parafermion point}
In this section, we apply our identification obtained in Sec.~\ref{sec:bs_orbifold} to another rational point of the AT criticality, the $Z_4$-parafermion theory. Like the four-state Potts model, the rationality allows us to quickly identify boundary states with the primaries of the extended theory. The identification of the spin representation is then proposed and partially accomplished in Ref.~\cite{Picco_2010}, where free, fixed, and four of the two-state mixed boundary conditions are identified. To compare these with our results, we will only consider the partition function $Z_{A-\beta}$ where the boundary condition on the left-hand side is fixed to the $A$ boundary condition. 

By simple substitution with the corresponding compactification radius $r=\frac{\sqrt{6}}{2}$, we obtain the following results:
\begin{align}
    Z_{A-A}^\text{pf}&=\frac{1}{\eta(q)}\left[1+\sum_{n=1}^\infty q^{3n^2}+(-1)^nq^{n^2}\right],\\
    Z_{A-C}^\text{pf}&=\frac{1}{\eta(q)}\left[\sum_{n=1}^\infty q^{3n^2}-(-1)^nq^{n^2}\right],\\
    Z_{A-B}^\text{pf}=Z_{A-D}^\text{pf}&=\frac{1}{\eta(q)}\sum_{n=1}^\infty q^{\frac{3(2n-1)^2}{4}},\\
    Z_{A-AB}^\text{pf}=Z_{A-AD}^\text{pf}&=\frac{1}{ \eta(q)}\sum_{n=-\infty}^{\infty}q^{\frac{(8n+1)^2}{16}},\\
    Z_{A-CD}^\text{pf}=Z_{A-BC}^\text{pf}&=\frac{1}{ \eta(q)}\sum_{n=-\infty}^{\infty}q^{\frac{(8n+3)^2}{16}},\\
    Z_{A-AC}^\text{pf}&=\frac{1}{ \eta(q)}\sum_{n=-\infty}^{\infty}q^{\frac{3(8n+1)^2}{64}},\\
    Z_{A-BD}^\text{pf}&=\frac{1}{ \eta(q)}\sum_{n=-\infty}^{\infty}q^{\frac{3(8n+3)^2}{64}},\\
    Z_{A-\text{Free}}^\text{pf}&=\frac{1}{ \eta(q)}\sum_{n=-\infty}^{\infty}q^{\frac{3(4n+1)^2}{16}}.
\end{align}
We found that our results agree with the identification made in Ref.~\cite{Picco_2010}. We see that the free and antipodal two-state mixed boundary conditions are not W-symmetry preserving boundary conditions, which is consistent with the fact that the $\mathbb{Z}_4$-parafermion model is anomalous~\cite{PhysRevB.110.045118}. The remaining two W-symmetry preserving boundary conditions can be identified with the $\mathbb{Z}_2$-orbifold boundary states $\ket{D_O(\frac{\pi}{3}/\frac{2\pi}{3})}$. Surprisingly, these are the boundary conditions that are mapped to the antipodal two-state mixed boundary conditions $AC/BD$ under the KW transformation.

\end{appendix}





\bibliography{SciPost_Example_BiBTeX_File.bib}


\end{document}